\documentclass{elsart}

\usepackage{graphicx}
\usepackage{epstopdf}
\usepackage{amssymb}
\usepackage{enumerate}
\usepackage{amsmath}
\usepackage{psfrag}
\usepackage{amsbsy}
\usepackage{amsfonts}
\usepackage{setspace}
\usepackage[usenames, dvipsnames]{color}
\usepackage{float}
\usepackage{tabularx}
\usepackage{epsfig}
\usepackage{bm}
\usepackage{booktabs,array,dcolumn}
\usepackage{arydshln}
\usepackage{dashrule}
\usepackage{wasysym}
\usepackage{multicol}
\usepackage{multirow}
\usepackage{ulem}
\usepackage{framed} 
\usepackage{nomencl}
\usepackage{multicol}
\usepackage{xpatch}

\usepackage{subfig}
\usepackage{epstopdf}
\usepackage{color, soul}
\sethlcolor{SkyBlue}

\graphicspath{{figures/}}

\makeatletter

\long\def\@maketablecaption#1#2{\@tablecaptionsize
	\global \@minipagefalse
	\hbox to \hsize{\parbox[t]{\hsize}{\centering #1 \\ #2}}}

\makeatother

\def\vec#1{\mbox{\boldmath $#1$}}

\makeatletter

\begin{document}
\begin{frontmatter}
\title{A novel 3D variational aeroelastic framework for flexible multibody dynamics: Application to bat-like flapping dynamics}
\author{G. Li},
\author{Y. Z. Law}, and
\author{R. K. Jaiman\corauthref{cor}}
\address{Department of Mechanical Engineering, National University of Singapore, Singapore 119077}
\corauth[cor]{Corresponding author}
\ead{mperkj@nus.edu.sg}

\begin{abstract}
We present a novel three-dimensional (3D) variational aeroelastic framework for flapping wing with a flexible multibody system subjected to an external incompressible turbulent flow. The proposed aeroelastic framework consists of a three-dimensional fluid solver with a hybrid RANS/LES model based on the delayed detached eddy simulation (DDES) treatment and a nonlinear monolithic elastic structural solver for the flexible multibody system with constraints. Radial basis function (RBF) is applied in this framework to transfer the aerodynamic forces and structural displacements across the discrete non-matching interface meshes while satisfying a global energy conservation. For the consistency of the interface data transfer process, the mesh motion of the fluid domain with large elastic deformation due to high-amplitude flapping motion is also performed via the standard radial basis functions. 
The fluid equations are discretized using a stabilized Petrov-Galerkin method in space and the generalized-$\alpha$ approach is employed to integrate the solution in time. The flexible multibody system is solved by using geometrically exact co-rotational finite element method and an energy decaying scheme is used to achieve numerical stability of the multibody solver with constraints. A nonlinear iterative force correction (NIFC) scheme is applied in a staggered partitioned iterative manner to maintain the numerical stability of aeroelastic coupling with strong added mass effect. 
An isotropic aluminum wing with flapping motion is simulated via the proposed aeroelastic framework and the accuracy of the coupled solution is validated with the available experimental data. We next study the robustness and reliability of the 3D flexible multibody aeroelastic framework for an anisotropic flapping wing flight involving battens and membranes with composite material and compare against the experimental results. Finally, we demonstrate the aeroelastic framework for a bat-like wing and examine the effects of flexibility on the flapping wing dynamics.
\end{abstract}


\end{frontmatter}

\section{Introduction}
\subsection{Background on biologically-inspired flapping flight}
Biologically-inspired flapping flight has intrigued human-kind over the past several centuries. Recently, there is a growing trend in aeronautical engineering applications to incorporate 
the understanding of flapping flight dynamics of birds, insects and bats. 
A lot of remarkable research work in this area has been carried out during the past decades, including experiments \cite{jones1999experimental,Singh2008Insect}, computational fluid dynamics (CFD) simulations \cite{heathcote2004flexible,hamamoto2007application} and real flapping robots \cite{bahlman2013design}. 
A number of recent reviews on several aspects of flapping flight have been documented \cite{Wei1999Flapping,Rozhdestvensky2003Aerohydrodynamics,Triantafyllou2004Review,platzer2008flapping,Shyy2010Recent}.
Compared with a fixed-wing flight vehicle, 
the flapping flight of birds and bats involve active morphing and adaptive flexible wing configuration, which may offer some unique benefits with regard 
to efficiency, noise and manoeuvrability \cite{Proctor1994Manual}. Therefore, the understanding of flapping wing mechanisms can be useful to incorporate in the designs of micro air vehicles (MAVs) and unmanned aerial vehicles (UAVs).
An unsteady turbulent aerodynamics interacting with nonlinear flexible multibody dynamics poses a serious challenge in the study of such flapping flight dynamics of animals. 
The emerging engineering requirements and the fundamental understanding of 
bio-inspired flapping flight have motivated the present computational development focusing 
on the flexible multibody solver interacting with a separated turbulent flow.

During the past decades, numerous researchers \cite{platzer2008flapping,shyy2007aerodynamics,dai2013computational} have explored the biologically-inspired flapping flight for different flying species (e.g., insects, birds and bats). 
The aerodynamic characteristics around rigid and flexible flapping wings as well as the coupled nonlinear structural responses were investigated in detail. 
However, several challenges  are still remained 
for a more comprehensive and complete understanding of flapping mechanism and the role of flexibility during flight \cite{platzer2008flapping}. 
For example, previous researches are limited to some specific airfoils and simplified insect-like wings, which lack the generalization to a real flapping flight. 
Biological structures are quite different for various species, which 
lead to their unique flapping patterns with specific mechanisms. 
Most of the insects have single or several pairs of wings consisted of the vein and membrane components with widely varying distributions of flexibility  \cite{shyy2007aerodynamics}. Specifically, a wing of the bird is made of bones and muscles to control its flight attitude and the surface is covered by different types of feathers. 
In particular, the unique skeletal anatomical structure of a bat has more degrees of freedom, compared with insects and birds and the membrane-like wing skin has high flexibility with anisotropic material properties \cite{swartz1998skin}. It is noticed that a physical structure of different wings can be considered as a generic flexible multibody system with kinematic constraints. Recently, several work focus on the study of fluid-flexible multibody interaction, 
and a realistic flapping wing including the vein and membrane components is incorporated into the simulation. For instance, Gogulapati et al. \cite{gogulapati2013approximate} developed an approximate aerodynamic model coupled with shell elements to simulate a flexible anisotropic flapping wing. Farhat et al. \cite{farhat2014ale} built an ALE-embedded computational framework to deal with aeroelastic problems with large structural deformation and demonstrated this framework for the anisotropic flapping wing. The underlying structural models in these two work are discretized only with shell elements. In Cho et al. \cite{cho2017improved}, the co-rotational beam elements are used for veins and the co-rotational shell elements are employed for the anisotropic flapping wing simulation. 
All the numerical methods in the above mentioned research work were developed to analyze and to provide a physical understanding of the aeroelastic phenomena around insect-like or bird-mimicking wings.
Only a handful of publications on 3D fully coupled aeroelastic analysis on a bat flapping dynamics can be found in the literature, compared with a large number of work on the flight of birds and insects \cite{wang2015lift}. 
The main objectives of the present work are (i) to develop a fully-coupled aeroelastic framework for flexible multibody analysis of flapping flight dynamics, and  (ii) to demonstrate the proposed aeroelastic framework for a bat-like wing.  

\subsection{Aspects of computational modeling}
Generally, a typical numerical simulation of the aeroelastic problem involves the coupling of the governing equations of fluid and structural dynamical systems in two different domains. Currently, two main aeroelastic schemes are considered to couple the fluid equations and the structural equations namely, monolithic  \cite{blom1998monolithical,liu2014stable} and partitioned \cite{felippa2001partitioned,yenduri2017new}. A monolithic approach can achieve good numerical stability for the aeroelastic problem with strong added mass effect, which assembles the fluid and structural equations into a single block then solves the coupled equations in a unified manner. However, it is difficult to take a full advantage of the existing stable and advanced fluid and structural solvers, which restricts the scalability and flexibility of an aeroelastic framework \cite{liu2014stable,hron2006monolithic,jaiman2015fully}. Considering the drawbacks of a monolithic scheme, a partitioned approach is developed to employ the existing suitable fluid and structural solvers to solve the complex and generic aeroelastic problem. The fluid and structural equations are solved in a sequential manner, and the traction and velocity continuities are satisfied along the fluid-structure interface to achieve numerical stability and accuracy \cite{felippa2001partitioned,jaiman2011combined,jaiman2011transient}. However, the added mass effect associated with acceleration of the flapping wing during flight and geometry of the wing with thin structures may result to numerical instability in a partitioned approach \cite{chin2016flapping,causin2005added,olivier2014fluid}. For the purpose of generality of flexible multibody aeroelastic analysis, we adopt a partitioned approach in the present study. In what follows, some backgrounds and relevant literature are associated with the partitioned aeroelastic framework for flexible multibody simulations.

In a typical partitioned-based aeroelastic scheme, the surface boundary data  must be transferred along the interface between the fluid and structural domains to satisfy a Dirichlet-type interface conditon (displacements or velocity) and a Neumann-type (fluid momentum flux or traction) interface condition. A proper care during interpolation and projection process is required to transfer  the physical data accurately across non-matching meshes  between the partitioned fluid and structural domains \cite{jaiman2011transient,jaiman2005assessment,Yu2018A}.  Global and local energy conservation should be satisfied during the aeroelastic coupling while maintaining the accuracy of data transfer along the interface via Dirichlet-Neumann coupling. 
Recently, a high accuracy interface projection scheme with  the global and local energy conservation is successfully applied to 3D fluid-structure problems  \cite{Yu2018A}. 
In addition to the interface projection problem, a large structural deformation due to high-amplitude flapping motion poses a  challenge for mesh motion to maintain good mesh quality 　in a transient aeroelastic computation. 
%
%
In the recent years, the radial basis function (RBF) method has been demonstrated as a simplified approach to interpolate scattered data \cite{rendall2008unified,lombardi2013radial} while satisfying the property of the conservation of global energy transfer.  For several aeroelastic problems, the RBF-based interpolation has been employed for the data transfer across non-matching meshes \cite{beckert2001multivariate,de2007review} and the mesh motion with large deformation \cite{de2007mesh,bos2010numerical}. 
The marked advantage of the RBF method is that the connectivity of the nodes is not required, which provides an effective and convenient way to implement RBF in any existing framework. 
For the ease of implementation during the coupling of fluid domain with flexible multibody surfaces, we consider the radial basis functions for the interface coupling and the mesh motion in the present study.

In the context of the present aeroelastic work, the structure of a flexible wing consists of various components with different material properties and the relative inertia, which can be considered as a nonlinear elastic multibody system. The elastic behavior of such a system is inherently nonlinear, hence the elastic displacements and rotations cannot be assumed to be small \cite{bauchau2010flexible}. Geometrically exact elements are able to deal with arbitrarily large displacement and rotation components in a multibody system exactly and those components are defined in a common inertial frame.  
Energy preserving (EP) time integrators are generally employed to ensure the discrete conservation of the total mechanical energy while achieving nonlinear unconditional stability for multibody systems. However, the undesired high frequency oscillations lead to numerical divergence for the solution of the nonlinear equations of motion when solving some complex problems with large number of degrees of freedom \cite{bauchau1995numerical}. The nonlinearities of the system induce the energy to transfer from low to the high frequency modes. 
The energy decaying (ED) scheme \cite{bauchau1996energy,bottasso1997energy,bauchau1999design} is designed based on energy preserving (EP) scheme to deal with the high frequencies in a flexible multibody system, which can achieve nonlinear unconditional stability. The multibody interaction is solved by a time discontinuous Galerkin scheme based on the ED approach. The constraints in the system are typically enforced by the Lagrange multiplier technique. Such a  flexible multibody formulation has been successfully applied for the interaction of turbulent flow with a floater-mooring system comprising a rigid-floater body and a long flexible riser and mooring lines \cite{Gurugubelli2018A}. We adopt a similar flexible multibody formulation for the present aeroelastic framework.

In nature, the flight condition of flying animals varies from low Reynolds numbers to high Reynolds numbers, depending on the real size, flying speed and flight environment of these animals. Such a highly variable flight condition involves rich aerodynamic physics during flight, including leading edge vortex (LEV) generation, massively separated flow, laminar-turbulent transition and trailing edge vortex (TEV) shedding \cite{shyy2007aerodynamics}. Therefore, a practical turbulent model is desired to capture the complex flow dynamics around a flapping wing. Compared to direct numerical simulation (DNS) and large eddy simulation (LES) methods, the delayed detached eddy simulation (DDES) method has significant advantages to save the computational resources while keeping the reasonable accuracy for the separated turbulent flow, as demonstrated by \cite{joshi2017variationally} for large-scale fluid-structure simulations.  The DDES model offers an effective and reasonable way to simulate the vortex structures played during a flapping flight, hence we employ this model in our computational framework.

\subsection{Contributions and organization}
In this study, a novel 3D variational aeroelastic framework is developed to simulate the flapping flight with flexible multibody system (e.g., insect-like and bat-like wing) in a turbulent flow. While the Petrov-Galerkin finite element method is used to solve Navier-Stokes equations for the external fluid flow, geometrically nonlinear co-rotational finite element method is applied to the flexible multibody system in the structural domain. The DDES method is employed to simulate the turbulent separated flow during flapping motion. The ALE fluid-turbulent solver and the flexible multibody solver are coupled via a partitioned iterative scheme combined with the nonlinear iterative force correction (NIFC) approach, which achieve numerical stabilization for the coupled aeroelastic framework. The RBF-based data interpolation approach is implemented in our framework to transfer the aerodynamic forces and the structural displacements across the fluid-structure interface while satisfying the property of conservation of energy transfer. The interface force is corrected at the end of each fluid sub-iteration by means of the NIFC method. The RBF approach with a compact support is utilized to handle the mesh motion with a large deformation condition and keep the initial mesh quality in simulation. For the initial validation, an isotropic aluminum wing and an anisotropic wing with composite material in flapping flight condition are simulated by using the proposed novel 3D variational aeroelastic framework with flexible multibody dynamics. Results are compared against results obtained from experiments and literature. Finally, we demonstrate the proposed aeroelastic framework to simulate a bat-like flexible wing with supported skeletons and covered membranes. 

In the present paper, we address two important challenges associated with the variationally coupled aeroelastic framework with flexible multibody system for the flexible flapping wing simulation: (i) coupling of an incompressible turbulent flow and a flexible multibody system with geometrically nonlinear co-rotational finite elements, (ii) the interpolation of aerodynamic forces and structural displacements between fluid surface elements and structural finite elements in a flexible multibody environment. For the first challenge, the proposed partitioned iterative scheme referred above is used to achieve the coupling between the fluid solver and the multibody structural solver in a robust and generic manner. It is worth noting that two main problems should be considered when dealing with the second challenge. Firstly, the information on the multiple surfaces belonging to different components of an entire structural model need to be collected then exchanged with the information from the fluid domain in an energy conservation manner. While the aerodynamic tractions at the nodes of each fluid element are interpolated to the target multiple structural elements by using the RBF method, the collected nodal structural displacements and the velocity vectors of each structural element are interpolated to the corresponding fluid mesh nodes. To address the numerical instability caused by the added mass effect, the strongly-coupled NIFC implementation \cite{jaiman2016stable} is employed. The force equilibrium and the velocity continuity condition conservation on the interface are satisfied by evaluating the approximate interface force corrections in the nonlinear sequence transformation. The generalization of Aitken's $\Delta^2$ extrapolation technique is applied to the iterative sequence coupling, which achieves a stable and convergent force updating process. 

The remainder of this manuscript is organized as follows. In Section 2, the variational formulations for the fluid and the flexible multibody system with constraints are reviewed. The detailed procedures for aeroelastic interface interpolation and the NIFC-based fluid-flexible multibody coupling by using RBF method are presented. The implementation of RBF method in mesh motion with large deformation condition is then introduced. An isotropic aluminum wing and an anisotropic wing with composite material in flapping flight condition are simulated with the proposed aeroelastic framework and compared with the available simulation and experimental data for validation purpose in Section 3. Section 4 presents an application on a bat-like flexible wing with skeletons and membrane and explores the effects of flexibility and aerodynamic load. The key conclusions of the present work are summarized in Section 5.

\section{Partitioned aeroelastic framework for flexible multibody system}
The governing equations and the formulations for the present 3D variational aeroelastic framework are similar to those of  \cite{Gurugubelli2018A}. For the sake of completeness, we review the variational formulations for  the moving fluid and flexible multibody solvers, whereas the Navier-Stokes equations are solved using the Petrov-Galerkin finite element method in the ALE coordinates and the flexible multibody system is solved via nonlinear co-rotational finite element method. 
\subsection{Petrov-Galerkin finite element for turbulent flow}
Consistent with the work of \cite{Gurugubelli2018A}, the Navier-Stokes equations are discretized using a stabilized Petrov-Galerkin formulation. The gerneralized-$\alpha$ method  is implemented to integrate the ALE flow solution in time domain, which can achieve unconditionally stable as well as second-order accuracy for linear problem. Furthermore, user-controlled high frequency damping desired for a coarser discretization in space and time is enabled by this scheme. The solution updates for the flow variables with the generalized-$\alpha$ scheme can be written as
\begin{align}
\overline{\boldsymbol{u}}^{f,n+1}_h=\overline{\boldsymbol{u}}^{f,n}_h+\Delta t ((1-\gamma^f)\partial_t \overline{\boldsymbol{u}}^{f,n}_h+\gamma^f \partial_t \overline{\boldsymbol{u}}^{f,n+1}_h)    \label{eqGA1} \\
\overline{\boldsymbol{u}}^{f,n+\alpha^f}_h=\alpha^f \overline{\boldsymbol{u}}^{f,n+1}_h + (1-\alpha^f) \overline{\boldsymbol{u}}^{f,n}_h \label{eqGA2} \\
\partial_t \overline{\boldsymbol{u}}^{f,n+\alpha^f_m}_h=\alpha^f_m \partial_t\overline{\boldsymbol{u}}^{f,n+1}_h + (1-\alpha^f_m)
\partial_t \overline{\boldsymbol{u}}^{f,n}_h     \label{eqGA3} \\
\boldsymbol{u}^{m,n+\alpha^f}_h=\alpha^f \boldsymbol{u}^{m,n+1}_h+(1-\alpha^f)\boldsymbol{u}^{m,n}_h \label{eqGA4}
\end{align}
where $\overline{\boldsymbol{u}}^{f,n+1}_h$ and $\boldsymbol{u}^{m,n}_h$ represent the fluid and mesh velocities defined for each spatial fluid point $\boldsymbol{x}^f \in \Omega^f(t)$, respectively, whereas $\boldsymbol{x}^f$ and $t$ are the spatial and temporal coordinates. 
Here, $\alpha^f$, $\alpha^f_m$ and $\gamma^f$ represent the standard integration parameters given as
\begin{align}
\alpha^f=\frac{1}{1+\rho^f_{\infty}},  \quad \quad \alpha^f_m=\frac{1}{2} \left( \frac{3-\rho^f_{\infty}}{1+\rho^f_{\infty}} \right),   \quad \quad \gamma^f=\frac{1}{2}+\alpha^f_m-\alpha^f  \label{eqGA5} 
\end{align}
The fluid spatial domain $\Omega^f$ can be discretized into $n^f_{el}$ number of  non-overlapping finite elements and $\Omega^f=\bigcup^{n_{el}^f}_{e=1}\Omega^e$. While $\mathcal{S}^{f,h}$ represents the space of the trial solutions, $\mathcal{V}^{f,h}$ denotes the space of test function. The variational formulation of the fluid equations within the Petrov-Galerkin framework can be written as: find $[\overline{\boldsymbol{u}}^{f,n+\alpha^f}_h,\overline{p}^{f,n+1}_h]\in{\mathcal{S}^{f,h}}$ such that $\forall[\boldsymbol{\phi}^f_h,q_h]\in{\mathcal{V}^{f,h}} $
\begin{align}
\begin{split}
&\int_{\Omega^e} \rho^f (\partial_t \overline{\boldsymbol{u}}^{f,n+\alpha^f_m}_h+(\overline{\boldsymbol{u}}^{f,n+\alpha^f}_h-\boldsymbol{u}^{m,n+\alpha^f}_h)\cdot \nabla \overline{\boldsymbol{u}}^{f,n+\alpha^f}_h) \cdot \boldsymbol{\phi}^f_h {\rm{d}\Omega}  \\
&+\int_{\Omega^e} \overline{\boldsymbol{\sigma}}^{f,n+\alpha^f}_h:\nabla \boldsymbol{\phi}^f_h {\rm{d}\Omega} + \int_{\Omega^e} {\boldsymbol{\sigma}^{des}}^{f,n+\alpha^f}_h:\nabla \boldsymbol{\phi}^f_h \rm{d}\Omega \\
&-\int_{\Omega^e} \nabla q_h \cdot \overline{\boldsymbol{u}}^{f,n+\alpha^f}_h {\rm{d}\Omega} \\
&+ \sum_{e=1}^{n^f_{el}} \int_{\Omega^e} \tau_m (\rho^f (\overline{\boldsymbol{u}}^{f,n+\alpha^f}_h-\boldsymbol{u}^{m,n+\alpha^f}_h) \cdot \nabla \boldsymbol{\phi}^f_h+\nabla q_h) \cdot \boldsymbol{\mathcal{R}}_m(\overline{\boldsymbol{u}}^f,\overline{p}) {\rm{d}\Omega^e} \\
&+\sum_{e=1}^{n^f_{el}} \int_{\Omega^e} \nabla \cdot \boldsymbol{\phi}^f_h \tau_c \nabla \cdot \overline{\boldsymbol{u}}^{f,n+\alpha^f}_h {\rm{d}\Omega^e} \\
&-\sum_{e=1}^{n^f_{el}} \int_{\Omega^e} \tau_m \boldsymbol{\phi}^f_h \cdot (\boldsymbol{\mathcal{R}}_m(\overline{\boldsymbol{u}}^f,\overline{p}) \cdot \nabla \overline{\boldsymbol{u}}^{f,n+\alpha^f}_h) {\rm{d}\Omega^e} \\
&-\sum_{e=1}^{n^f_{el}} \int_{\Omega^e} \nabla \boldsymbol{\phi}^f_h : (\tau_m \boldsymbol{\mathcal{R}}_m(\overline{\boldsymbol{u}}^f,\overline{p})\otimes \tau_m \boldsymbol{\mathcal{R}}_m(\overline{\boldsymbol{u}}^f,\overline{p})) {\rm{d}\Omega^e} \\
&= \int_{\Omega^e} \boldsymbol{b}^f(t^{n+\alpha^f}) \cdot \boldsymbol{\phi}^f_h {\rm{d}\Omega} + \int_{\Gamma_h} \boldsymbol{h}^f \cdot \boldsymbol{\phi}^f_h {\rm{d} \Gamma} 
\label{eqGA6} 
\end{split}
\end{align}
where $\rho^f$ is the density of the fluid and $\boldsymbol{b}^f$ is the body force applied on the fluid, ${\boldsymbol{\sigma}^{des}}^{f,n+\alpha^f}_h$ represents the turbulent stress term, $\overline{\boldsymbol{\sigma}}^{f,n+\alpha^f}_h$ is the Cauchy stress tensor for a Newtonian fluid. Here,
$\boldsymbol{\phi}^f_h$ and $q_h$ denote the test functions of the fluid velocity $\overline{\boldsymbol{u}}^f$ and pressure $\overline{p}$. 
In Eq.~(\ref{eqGA6}), the Galerkin terms for the flow equations are shown in the first, second and third lines. The Petrov-Galerkin stabilization term for the momentum equation is shown in the fourth line and the term for the continuity is in the fifth line. The remaining terms represent the approximation of the fine scale velocity on the element interiors. The stabilization parameters $\tau_m$ and $\tau_c$ are added to the fully discretized formulation and $\boldsymbol{\mathcal{R}}_m(\overline{\boldsymbol{u}}^f,\overline{p})$ denotes the residual of the momentum equation \cite{brooks1982streamline}. A hybrid RANS/LES model based on the DDES treatment is applied to model the flow turbulence for high Reynolds numbers. The DDES model behaves as RANS model in the near wall region and switches to LES-mode in the separated turbulent region. Further details can be referred to \cite{spalart2009detached}.

\subsection{Flexible multibody solver with constraints}
The motion equations for a flexible structure are discretized using finite element method and can be written into a weak variational form using the virtual work principle
\begin{align}
\begin{split}
\int^{t^{n+1}}_{t^n} 
\left( \int_{\Omega^s_i} \rho^s \frac{\partial^2 \boldsymbol {d}^s_h}{\partial t^2} \cdot \boldsymbol{\phi}^s_h {\rm{d}\Omega} + \int_{\Omega^s_i} \boldsymbol{\sigma}^s (\boldsymbol{\tilde{E}}(\boldsymbol{d}^s_h)) : \nabla \boldsymbol{\phi}^s_h  {\rm{d}\Omega}
\right) {\rm{d}}t
= \\
\int^{t^{n+1}}_{t^n} 
\left(
\int_{\Omega^s_i} \boldsymbol{b}^s \cdot \boldsymbol{\phi}^s_h  {\rm{d}\Omega} + \int_{\Gamma_i} \boldsymbol{t}^s \cdot \boldsymbol{\phi}^s_h {\rm{d} \Gamma}
\right) {\rm{d}}t   \label{eqMB1} 
\end{split}
\end{align}
where $\Omega^s_i$ denotes the multibody domain, $\boldsymbol{\phi}^s_h$ and $\rho^s$ represent the test function for the structural displacements and the structural density, respectively. 
Here, $\boldsymbol{d}^s_h$ is the structural displacement, $\boldsymbol{\sigma}^s$ is defined as the first Piola-Kirchhoff stress tensor, $\boldsymbol{b}^s$ denotes the body force on the multibody $\Omega^s_i$ and $\boldsymbol{\tilde{E}}(\boldsymbol{d}^s_h)$ is the simplified Cauchy-Green Lagrangian strain tensor. The external force caused by flow on the interface $\Gamma_i$ is defined as $\boldsymbol{t}^s$. 

The kinematic joints, like revolution joint and sphere joint, are used to connect different components in the flexible multibody system and can be considered as multibody constrains with the following expression:
\begin{align}
\boldsymbol{c}(\boldsymbol{d}^s)=0.    \label{eqMB2} 
\end{align}
Generally, the discretized motion equations with constraints for flexible multibody system can be written into a matrix form
\begin{align}
\int^{t^{n+1}}_{t^n}  (\boldsymbol{M}^s\ddot{\boldsymbol{d}^s}(t)+\boldsymbol{C}^s{\boldsymbol{d}^s}(t)+\boldsymbol{K}^s\boldsymbol{d}^s(t)) {\rm{d}}t = \int^{t^{n+1}}_{t^n} (\boldsymbol{F}^s(t)) {\rm{d} }t \label{eqMB3}
\end{align}
where $\boldsymbol{M}^s$, $\boldsymbol{C}^s$ and $\boldsymbol{K}^s$ represents the mass, constrain and stiffness matrices for the flexible multibody system respectively. All the body forces and external forces caused by flow acting on the multibody system can be combined into a whole force matrix $\boldsymbol{F}^s$. The forces from constraints will not produce any work for the multibody system at the discrete solution level, which are not considered in the force matrix. 

The structural variables are updated via an unconditionally stable energy decaying scheme temporally, which is obtained by using a linear time discontinuous Galerkin approximation to Eq.~\eqref{eqMB3}. We briefly present the general discretized motion equations for flexible multibody system with different geometrically nonlinear co-rotational finite element models, including beam, cable, shell and membrane. The detailed description of geometrically exact shell model in multibody dynamics can be found in \cite{simo1989stress,bauchau2002time}.

\subsection{Aeroelastic coupling and mesh motion via radial basis functions}
In this section, we briefly introduce the RBF method and its applications to the multibody aeroelastic coupling and the mesh motion interpolation in our ALE formulation.
In the proposed partitioned iterative coupling scheme, the interface data is exchanged between the fluid domain and the multibody structural domain. Across the non-matching meshes along the aeroelastic interface, the surface data need to be transferred in a conservative manner with reasonable accuracy.
\subsubsection{Review of RBF interpolation process}
Assume that data are interpolated from a set of control points to a set of target points. Let $\boldsymbol{x}^c_{i}=(x^c_i,y^c_{i},z^c_{i})$ be the coordinate of the $i$-th control point, $\boldsymbol{x}^t_{j}=({x^t_{j},y^t_{j},z^t_{j}})$ be the coordinate of the $j$-th target point, $N_c$ and $N_t$ be the number of control points and target points. The global interpolation function defined by a radial basis interpolation is given by
\begin{align}
g(\boldsymbol{x})&=\sum_{i=1}^{i=N_c}\beta_i \phi_i + p(\boldsymbol{x}) \nonumber \\
				&=\sum_{i=1}^{i=N_c}\beta_i \phi(\left \| \boldsymbol{x}-\boldsymbol{x}^c_{i} \right \|) + p(\boldsymbol{x})
 \label{eqRBF}
\end{align}
where $g(\boldsymbol{x})$ is the global interpolation function and $\boldsymbol{x}$ is the coordinate of an arbitrary point in space and $\left \| \cdot \right \|$ denotes the Euclidean norm.
The coefficient $\beta_i$ represents the weights related to the $i$-th basis function $\phi_i$ and $p(\boldsymbol{x})$ is a linear polynomial to recover translation and rotation motion, where its expression is $p(\boldsymbol{x})=\lambda_0+\lambda_1 x + \lambda_2 y + \lambda_3 z$.

As shown in \cite{rendall2008unified,lombardi2013radial}, the interpolation relationship between the control vector $\boldsymbol{G}^c$ and  the target vector $\boldsymbol{G}^t$ can be written in the matrix form as
\begin{align}
\boldsymbol{G}^t
=\boldsymbol{A}^I (\boldsymbol{M}^I)^{-1} \boldsymbol{G}^c=\boldsymbol{H}^I \boldsymbol{G}^c   \label{eqRBF3}
\end{align}
where
{\setstretch{1.0}
\begin{align}
\boldsymbol{G}^t=\left [ \begin{matrix}
0 \\
0 \\
0 \\
0 \\
g(\boldsymbol{x}^t_{1}) \\
\vdots \\
g(\boldsymbol{x}^t_{N_t})
\end{matrix}
\right],   \quad \quad \quad
\boldsymbol{G}^c=\left [ \begin{matrix}
0 \\
0 \\
0 \\
0 \\
g(\boldsymbol{x}^c_{1}) \\
\vdots \\
g(\boldsymbol{x}^c_{N_c})
\end{matrix}
\right]  \label{eqRBF4}
\end{align}	
}
where $\boldsymbol{H}^I$ denotes the final interpolation matrix between the control vector and target vector. $\boldsymbol{A}^I$ and $\boldsymbol{M}^I$ are the interpolation matrices given by
{\setstretch{1.0}
\begin{align}
\boldsymbol{M}^I=\left [ \begin{matrix}
0 & 0 & 0 & 0 & 1 & 1 & \cdots & 1\\
0 & 0 & 0 & 0 & x^c_{1} & x^c_{2} & \cdots & x^c_{N_c}\\
0 & 0 & 0 & 0 & y^c_{1} & y^c_{2} & \cdots & y^c_{N_c}\\
0 & 0 & 0 & 0 & z^c_{1} & z^c_{2} & \cdots & z^c_{N_c}\\
1 & x^c_{1} & y^c_{1} & z^c_{1} & \phi^{c,c}_{11} & \phi^{c,c}_{1 2}  & \cdots & \phi^{c,c}_{1 N_c}\\
\vdots & \vdots & \vdots & \vdots & \vdots & \vdots  & \ddots & \vdots \\
1 & x^c_{N_c} & y^c_{N_c} & z^c_{N_c} & \phi^{c,c}_{N_c 1} & \phi^{c,c}_{N_c 2}   & \cdots & \phi^{c,c}_{N_c N_c}
\end{matrix}
\right]      \label{eqMatrix1} 
\end{align}
\begin{align}
\boldsymbol{A}^I= \left [ \begin{matrix}
1 & x^t_{1} & y^t_{1} & z^t_{1} & \phi^{t,c}_{11} & \phi^{t,c}_{1 2}  & \cdots & \phi^{t,c}_{1 N_c}\\
\vdots & \vdots & \vdots & \vdots & \vdots & \vdots  & \ddots & \vdots \\
1 & x^t_{N_t} & y^t_{N_t} & z^t_{N_t} & \phi^{t,c}_{N_t 1} & \phi^{t,c}_{N_t 2}   & \cdots & \phi^{t,c}_{N_t N_c}
\end{matrix}
\right]           \label{eqMatrix2}
\end{align}
}
with $\phi^{c,c}_{i j}=\phi(\left \| \boldsymbol{x}^c_{i}-\boldsymbol{x}^c_{j} \right \|)$ and $\phi^{t,c}_{i j}=\phi(\left \| \boldsymbol{x}^t_{i}-\boldsymbol{x}^c_{j} \right \|)$, respectively. A compactly supported Wendland's $C^2$ function \cite{rendall2008unified,beckert2001multivariate} has been proved as an effective basis function for the interpolation with improved accuracy, which is considered in the framework. The basis function with a compact support is defined as 
$\phi(\left \| \boldsymbol{x} \right \|)=(1-\left \| \boldsymbol{x} \right \| / r)^4(4\left \| \boldsymbol{x} \right \| / r+1)$, where $r$ denotes the support radius.  \\[-0.2cm]

\subsubsection{Multibody aeroelastic coupling}
\begin{figure}
	\centering
	\includegraphics[width=0.9\textwidth]{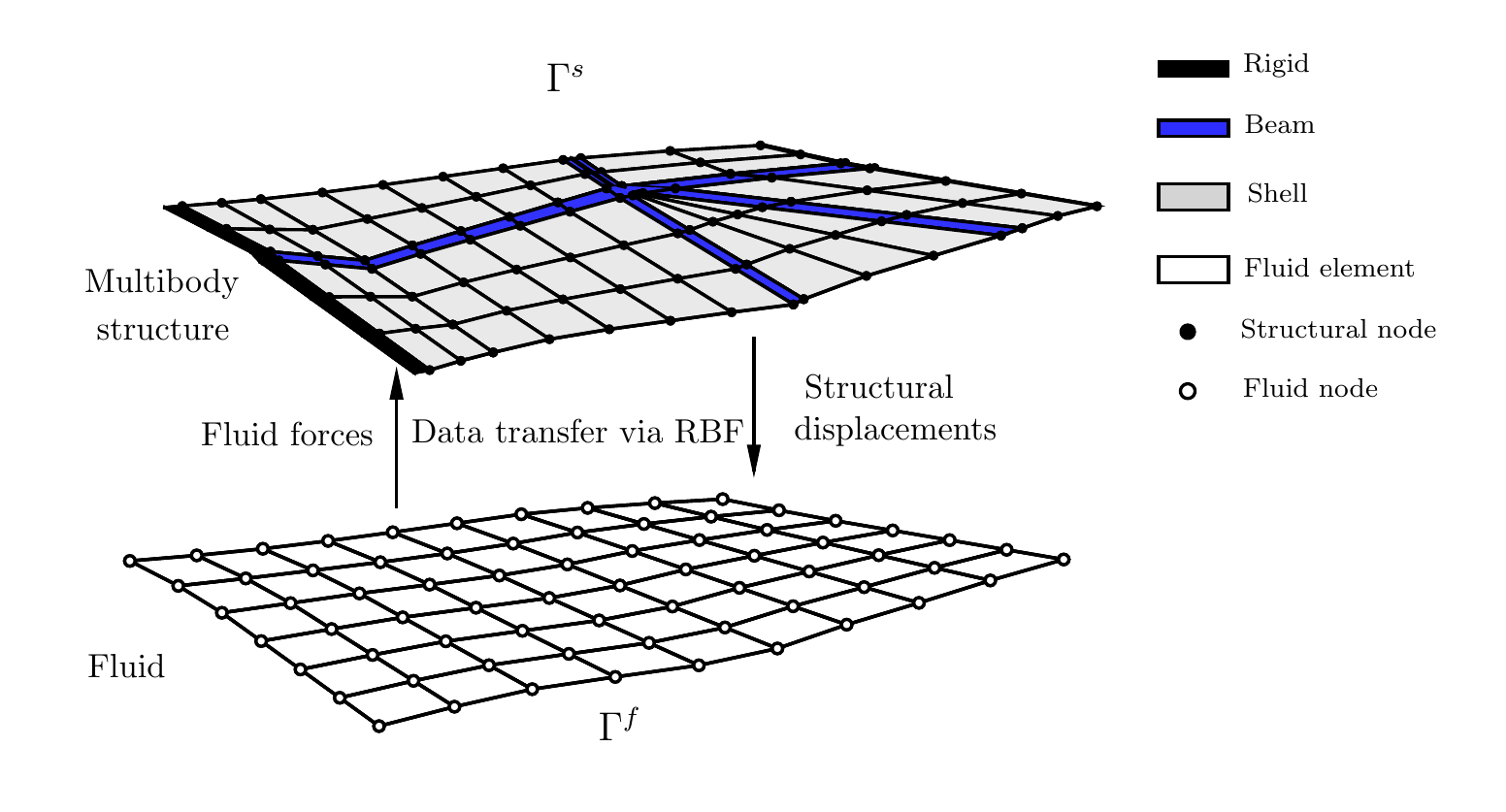}
	\caption{Schematic of interface data transfer via RBF method across non-matching meshes between fluid domain and multibody structure system involving rigid link, flexible beam and shell members.}
	\label{rbfinter}
\end{figure}

The aeroelastic coupling involves two data transfers across the interface: (i) forces from the fluid domain to the structural domain and (ii) displacements from the structural domain to the fluid domain. A schematic diagram of interface data interpolation via RBF method is depicted in Fig. \ref{rbfinter}.　
The energy conservation should be satisfied in the interpolation of forces and displacements along the aeroelastic interface. Generally, the work $W^s$ done by forces in structural domain on the interface equals to the work $W^f$ done by the aerodynamic forces according to the definition of energy conservation, which is given as
\begin{align}
W^s(\boldsymbol{d}^s)=\int_{\Gamma^{fs}} (\boldsymbol{\sigma}^s \boldsymbol{n}^s) \cdot \boldsymbol{d}^s {\text{ d} \Gamma}= \int_{\Gamma^{fs}} (\boldsymbol{\sigma}^f \boldsymbol{n}^f) \cdot \boldsymbol{d}^f {\text{ d} \Gamma} = W^f(\boldsymbol{d}^f) \label{eqEC}
\end{align}
where $\boldsymbol{d}^s$ and $\boldsymbol{d}^f$ are the displacements along interface in the structural and fluid domains,  respectively. 

For the aeroelastic interface coupling process, the interpolation equation of displacements along interface can be written as
\begin{align}
\boldsymbol{D}^{f, I} 
=\boldsymbol{H}^I_{ss} \boldsymbol{D}^{s,I} \label{eqdis}
\end{align}
where $\boldsymbol{D}^{f,I}$ and $\boldsymbol{D}^{s,I}$ are the displacement vectors along interface for the fluid domain and the structural domain, respectively. Both the vectors have the same vector form as shown in Eq.~\eqref{eqRBF4}.
The matrix $\boldsymbol{H}_{ss}^I$ interpolates the displacements from the structural surface nodes to the fluid surface nodes. 
As the balance of energy transfer at the interface must be achieved during the interpolation process, according to Eq.~\eqref{eqEC}
and Eq.~\eqref{eqdis}, the interpolation between the fluid forces  and the structural forces can be written as
\begin{align}
\boldsymbol{F}^{s,I} = (\boldsymbol{H}_{ss}^I)^T \boldsymbol{F}^{f,I} \label{eqfor}
\end{align}
where $\boldsymbol{F}^{f,I}$ and $\boldsymbol{F}^{s,I}$ are the force vectors along the interface for the fluid domain and the structural domain, respectively. We numerically evaluate the momentum flux through the boundary faces and provide the interpolated forces to the structural solver.
Such an interpolation process across aeroelastic interface is performed at each time step in the partitioned iterative scheme with the conservation of energy transfer at the fluid-structure interface. 

\subsubsection{Mesh motion interpolation}
The aeroelastic coupling involve a deformation of the fluid mesh in response to a movement of the structure.
The interpolation of the mesh motion using radial basis function can handle mesh motion with a relatively moderate deformation without distorting the mesh quality. Moreover, the mapping matrix used for the interface displacement interpolation can be easily extended to a three-dimensional fluid domain. In our framework, the mesh motion of the fluid nodes in the domain is interpolated from the fluid nodes on the interface.  The relationship between the displacement vector of fluid nodes in the domain $\boldsymbol{D}^f_v$ and the displacement vector of the fluid nodes on the interface $\boldsymbol{D}^{f,I}_s$ is given as 
\begin{align}
\boldsymbol{D}^f_v
=\boldsymbol{H}_{vs}^I \boldsymbol{D}^{f,I}_s   \label{eqINter}
\end{align}
where $\boldsymbol{H}_{vs}^I$ denotes the interpolation matrix for the fluid space nodes.
Note that, $\boldsymbol{H}_{vs}^I$ includes the polynomial term $p(\boldsymbol{x})$ shown in Eq. \eqref{eqRBF}, where its value increases linearly with $\boldsymbol{x}$. While these terms can preserve good mesh quality in the near wall region, an undesired translation and rotation motion will be interpolated to the fluid nodes at the boundary of the computational domain.
Hence, a smooth cut-off function \cite{lombardi2013radial} is implemented to adjust the deformation far away from the interface.

\subsection{Nonlinear interface force correction scheme}
Nonlinear interface force correction scheme \cite{jaiman2016stable} can provide a numerical stability for the partitioned aeroelastic coupling when significant added mass effect is encountered in an aeroelastic simulation. This scheme has been extended to the flexible multibody aeroelastic coupling simulation to achieve numerical stability and the iterative force correction procedure is briefly summarized herein. The formulation for a coupled linear system between the fluid domain and the structure domain which is discretized by the finite element method can be written into the matrix form $\boldsymbol{A}\boldsymbol{U}=\boldsymbol{R}$, where $\boldsymbol{U}$ denotes the vector of the unknowns for the coupled system and $\boldsymbol{R}$ represents the right-hand side. The abstract matrix form can be written as 

{\setstretch{1.0}
\begin{align}
\left [ \begin{matrix}
\boldsymbol{A}_{11} & 0 & 0 & \boldsymbol{A}_{14} \\
\boldsymbol{A}_{21} & \boldsymbol{A}_{22} & 0 & 0 \\
0 & \boldsymbol{A}_{32} & \boldsymbol{A}_{33} & 0 \\
0 & 0 & \boldsymbol{A}_{43} & \boldsymbol{A}_{44} 
\end{matrix}
\right]
\left \{ \begin{matrix}
\Delta \boldsymbol{d}^s \\
\Delta \boldsymbol{d}^I \\
\Delta \boldsymbol{q}^f \\
\Delta \boldsymbol{f}^I 
\end{matrix}
\right \}=
\left \{ \begin{matrix}
\boldsymbol{R}_1 \\
\boldsymbol{R}_2 \\
\boldsymbol{R}_3 \\
\boldsymbol{R}_4 
\end{matrix}
\right \} \label{eqNIFC1} 
\end{align}
}
where $\boldsymbol{A}_{11}$ denotes the mass and stiffness matrices of flexible multibody system and $\boldsymbol{A}_{14}$ is the load vector on the solid surface. $\boldsymbol{A}_{21}$ represents the transformation matrix which maps the structural displacements to aeroelastic interface and $\boldsymbol{A}_{43}$ denotes the force calculation and transform to the interface. $\boldsymbol{A}_{22}$ and $\boldsymbol{A}_{44}$ are the identity matrices. $\boldsymbol{A}_{32}$ gives the relationship between the displacement on interface and the ALE mapping in the fluid domain. $\boldsymbol{A}_{33}$ is the coupled fluid velocity and pressure linear system. $\boldsymbol{d}^s$ denotes the structural displacement for the flexible multibody system and $\boldsymbol{d}^I$ represents the displacement along the coupling interface. $\boldsymbol{q}^f$ is the unknown variables in fluid domain and $\boldsymbol{f}^I$ is the force on interface. The right-hand side vector $\boldsymbol{R}_i$ shows the residual of this linear system for the different equations. In this matrix form, the first equation is the flexible multibody system and the third equation is related to the fluid and turbulence equations. The second and fourth equations correspond to the displacement and traction transformation between the fluid domain and structure domain, respectively. 

{\setstretch{1.5}
The off-diagonal term $\boldsymbol{A}_{14}$ can be eliminated through the static condensation and Eq.~$(\ref{eqNIFC1})$ can be rewritten as
}
{\setstretch{1.0}
\begin{align}
\left [ \begin{matrix}
\boldsymbol{A}_{11} & 0 & 0 & 0 \\
\boldsymbol{A}_{21} & \boldsymbol{A}_{22} & 0 & 0 \\
0 & \boldsymbol{A}_{32} & \boldsymbol{A}_{33} & 0 \\
0 & 0 & 0 & \tilde{\boldsymbol{A}}_{44} 
\end{matrix}
\right]
\left \{ \begin{matrix}
\Delta \boldsymbol{d}^s \\
\Delta \boldsymbol{d}^I \\
\Delta \boldsymbol{q}^f \\
\Delta \boldsymbol{f}^I 
\end{matrix}
\right \}=
\left \{ \begin{matrix}
\boldsymbol{R}_1 \\
\boldsymbol{R}_2 \\
\boldsymbol{R}_3 \\
\tilde{\boldsymbol{R}}_4 
\end{matrix}
\right \} \label{eqNIFC2} 
\end{align}
}
where $\tilde{\boldsymbol{A}}_{44} =\boldsymbol{A}_{44}-\boldsymbol{A}_{43}\boldsymbol{A}_{33}^{-1}\boldsymbol{A}_{32}\boldsymbol{A}_{22}^{-1}\boldsymbol{A}_{21}\boldsymbol{A}_{11}^{-1}\boldsymbol{A}_{14}$ and $\tilde{\boldsymbol{R}}_4 =\boldsymbol{R}_4-\boldsymbol{A}_{43}\boldsymbol{A}_{33}^{-1}(\boldsymbol{R}_3-\boldsymbol{A}_{32}\boldsymbol{A}_{22}^{-1}(\boldsymbol{R}_2-\boldsymbol{A}_{11}^{-1} \boldsymbol{A}_{21} \boldsymbol{R}_1))$. An iterative scheme is used to correct the forces between the fluid domain and structure domain with a feedback process. The formulation of nonlinear iterative force correction is given as
\begin{align}
\boldsymbol{f}^I_{(k+1)}=\boldsymbol{f}^I_{(k)}+\tilde{\boldsymbol{A}}_{44}^{-1}\tilde{\boldsymbol{R}}_{4(k)}   \label{eqNIFC3} 
\end{align}
where $k$ denotes the nonlinear sub-iteration with a  single time step. The current nonlinear iterative force correction process can be considered as a generalization of Aitken's extrapolation while updating a dynamic stabilization parameter, which could transform a divergent fixed-point iteration to a convergent and stable force correction \cite{jaiman2016stable}. Finally, a numerical stability can be achieved with the aid of the NIFC scheme for the partitioned fluid-flexible multibody system coupling.

\subsection{Fluid-flexible multibody coupling procedure}
\begin{figure}
	\centering
	\includegraphics[width=0.8\textwidth]{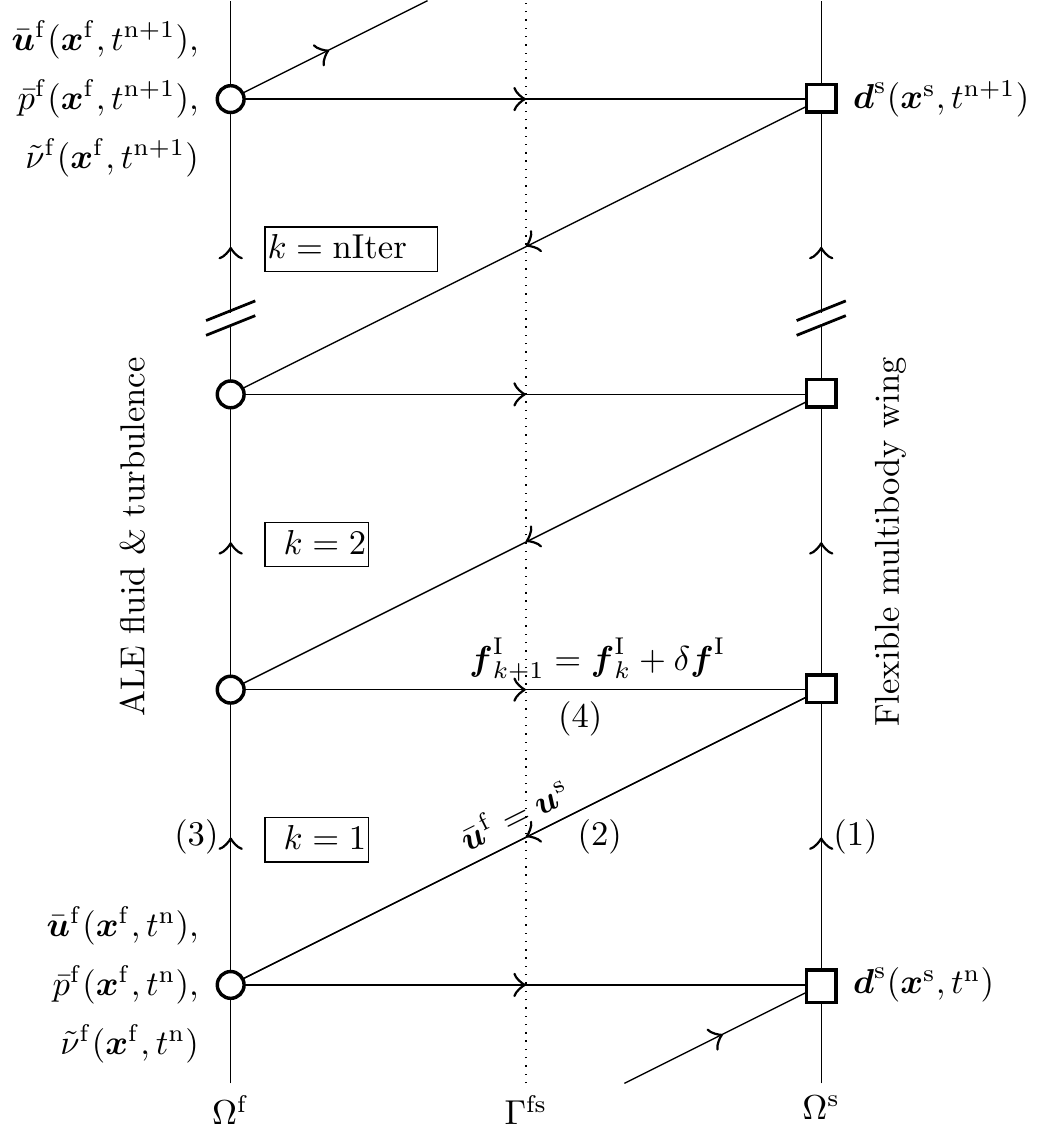}
	\caption{A schematic of predictor-corrector procedure for ALE fluid solver and flexible multibody solver coupling via nonlinear iterative force correction. (1) Solve flexible multibody system with constraints, (2) Transfer predicted structural displacement, (3) Solve ALE fluid and turbulence equations, (4) Correct forces via NIFC filter. nIter represents the maximum number of nonlinear iterations to achieve a desired convergence tolerance in a time step at $t \in [t^n,t^{n+1}]$. }
	\label{nifcschematic}
\end{figure}

The partitioned iterative coupling procedure of the fluid solver with the flexible multibody structural solver is briefly summarized in this section. The schematic of aeroelastic coupling procedure  based on a predictor-corrector algorithm is shown in Fig. \ref{nifcschematic}. The predictor displacement is calculated from the flexible multibody equations and the ALE fluid equations can be solved to provide the forces on the fluid-structure interface as a corrector step. The structural displacement caused by the aerodynamic forces at time $t^n$ is defined as $\boldsymbol{d}^s(\boldsymbol{x}^s,t^n)$. Firstly, the structural displacements of each component in the flexible multibody system is solved via the time discontinuous Galerkin approximation under the loads from the fluid domain at time $t^n$. Then the structural displacements obtained from the previous step are transferred to the fluid solver and it is satisfied the ALE compatibility and the velocity continuity on the interface $\Gamma^{fs}$.  To satisfy the consistency of the non-matching fluid and structural mesh configurations, the mesh displacements are set equal to the structural displacements along the interface as
\begin{align}
\boldsymbol{d}^{m,n+1}=\boldsymbol{d}^s  \quad  \text{on} \quad \Gamma^{fs}  \label{eqbou} 
\end{align}
where $\boldsymbol{d}^{m,n+1}$ represents the mesh displacement at time $t^{n+1}$. Meanwhile, the velocity continuity on interface $\Gamma^{fs}$ is satisfied when the fluid velocity equals to the mesh velocity.
\begin{align}
\overline{\boldsymbol{u}}^{f,n+\alpha^f} = \boldsymbol{u}^{m,n+\alpha^f} = \frac{\boldsymbol{d}^{m,n+1}-\boldsymbol{d}^{m,n}}{\Delta t}   \quad  \text{on} \quad \Gamma^{fs}   \label{eqbou2} 
\end{align}

In the third step, the Navier-Stokes equations under the ALE framework with the turbulence model equations are solved and compute the fluid forces. Finally, the obtained fluid forces are corrected based on NIFC scheme to achieve the numerical stability then transferred to the flexible multibody solver. One aeroelastic sub-iteration is finished and the sub-iteration mentioned above will be executed continuously until the convergence criterion has achieved. Subsequently, all the variables in this aeroelastic solver are updated for the next time step $t^{n+1}$. \\[-0.2cm]

The Generalized Minimal RESidual (GMRES) algorithm is applied in the current framework to compute the velocity, pressure and mesh displacement in fluid equations discretized with finite element method. This algorithm relies on the Krylov subspace iteration and the modified Gram-Schmidt orthogonalization. In order to minimize the linearization error at each time step, the Newton-Raphson scheme is considered for the aeroelastic framework. Furthermore, the discretized algebraic equations based on the flexible multibody system with the co-rotational framework are solved via a classical skyline solver.

\section{Mesh convergence and validation}
Before we demonstrate our multibody aeroelastic framework for a bat-like wing, a mesh convergence study is first conducted to ensure a sufficient mesh resolution employed on both the fluid and structure domains. We validate our multibody aeroelastic solver against the experiment work of \cite{wu2010experimental} by simulating a flexible flapping wing with an anisotropic material, where the structural responses are compared to the experimental data, and the vortex patterns are analyzed. 

\subsection{Mesh convergence}
\begin{figure}
	\centering
	\subfloat[][]{\includegraphics[width=0.85\textwidth]{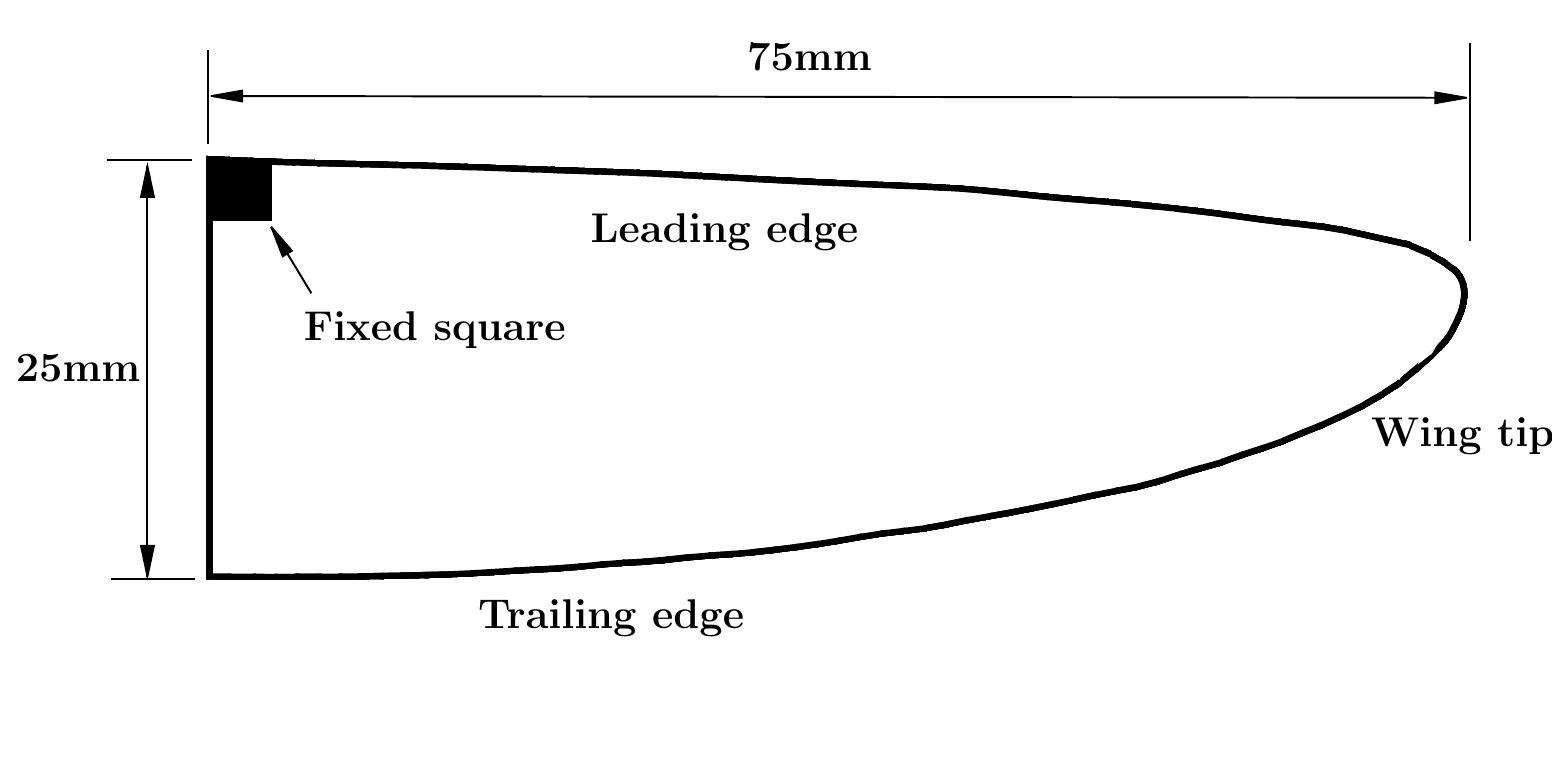} \label{wingDiagram}} \\
	\subfloat[][]{\includegraphics[width=0.85\textwidth]{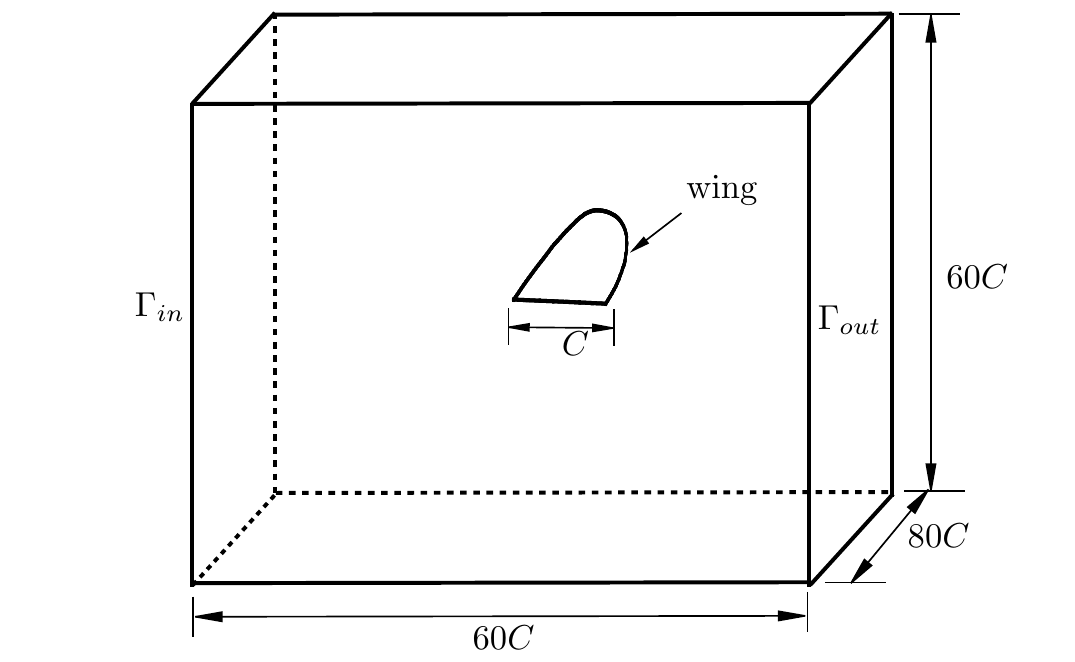} \label{wingDomain}} 
	\caption{Flow past a flapping wing: (a) geometry of isotropic wing, (b) schematic of computational setup.}
	\label{wing}
\end{figure}

The mesh convergence study is conducted by simulating a wing with isotropic material using three different mesh resolutions on the fluid and structure domains. 
The wing simulated is adopted from an experimental study \cite{wu2010experimental}, and its properties are summarized in Table \ref{paraiso}.  
It is an isotropic aluminum wing with Zimmerman shape, which is designed to investigate the aerodynamic characteristics under flapping flight condition. A schematic diagram of the geometry of the wing and its surrounding aeroelastic computational domain are shown in Fig. \ref{wing}. 
The root chord length $C$ is 25 mm and the span length is 75 mm, which results in an aspect ratio of 7.65. The flapping condition is hovering motion with 10 Hz flapping frequency and $21^{\circ}$ flapping amplitude. The reference flow velocity is the velocity measured at the wing tip and the freestream velocity is given as zero. The distances of the inlet $(\Gamma_{in})$ and outlet $(\Gamma_{out})$ boundaries from the leading root of the wing are both 30$C$. The distance between the side walls $(\Gamma_{slip})$ on top and bottom is 60$C$ and the distance increases to 80$C$ for the side walls on both sides. The components of the flow velocity are defined as $\overline{\textbf{u}}^f=(u^f,v^f,w^f)$. The freestream velocity along the $X$-axis at the inlet boundary $\Gamma_{in}$ is given as $u^f=U$. Slip boundary condition is applied on the top and bottom boundaries $(\Gamma_{slip})$, where $\frac{\partial u^f}{\partial z}=0$ and $w^f=0$. Slip boundary condition is applied on both side boundaries $(\Gamma_{slip})$, where $\frac{\partial u^f}{\partial y}=0$ and $v^f=0$. 
A traction-free boundary condition is defined at the outlet boundary $\Gamma_{out}$, where $\sigma_{xx}=\sigma_{yx}=\sigma_{zx}=0$. The boundary condition on flapping wing surfaces is no-slip boundary condition.

\begin{figure}
	\centering
	\subfloat[][ Fluid space mesh]{\includegraphics[width=0.9\textwidth]{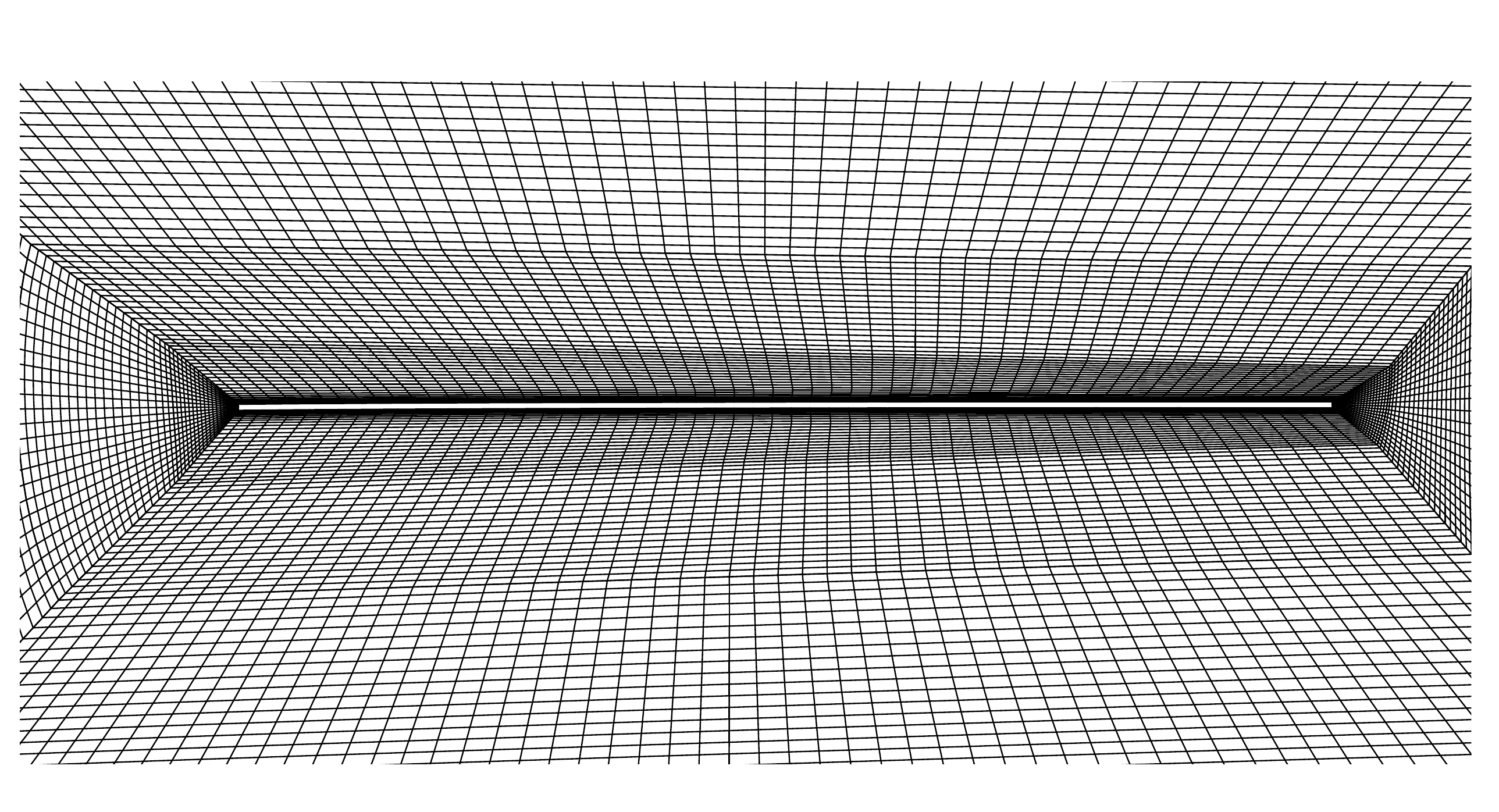} \label{meshfs}} \\ 
	\subfloat[][ Fluid surface mesh]{\includegraphics[width=0.45\textwidth]{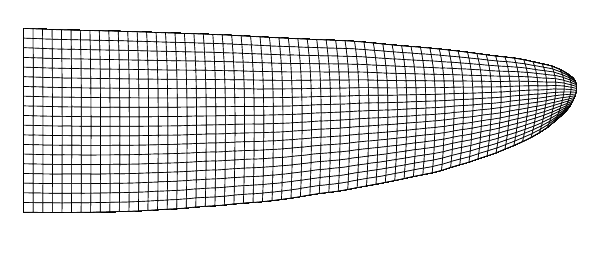} \label{meshfsr}} 
	\quad
	\subfloat[][ Structural surface mesh]{\includegraphics[width=0.45\textwidth]{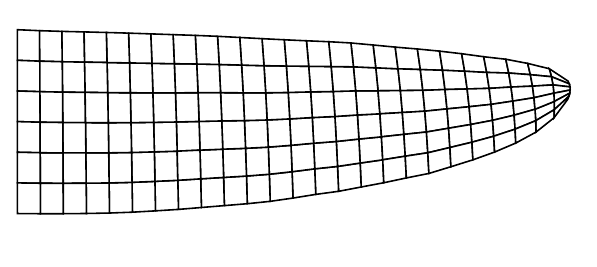} \label{meshss}}  
	\caption{Schematic of mesh characteristics: (a) in space for flapping wing in the fluid domain (M2),  (b) on surface for flapping wing in the fluid domain (M2), (c) on surface for flapping wing in the structural domain (M2).}
	\label{mesh}
\end{figure}

The three-dimensional fluid computational domain is discretized by unstructured eight-node brick finite element mesh and the structural computational domain is discretized by structured four-node rectangular finite element mesh. A summary of three different mesh resolutions is shown in Table \ref{meshsize}. 
A boundary layer with $y^+ \approx 0.5$ in the wall-normal direction is maintained, with a stretching ratio, $\Delta y_{i+1} / \Delta y_i$ of 1.2. The discretization along chord direction, span-wise direction and outside the boundary layer is varied. The non-dimensional time step size, $\Delta t U_{ref}/C$ is chosen as 0.01. Details of mesh statistics in space and on surface for the fluid domain with M2 are shown in Fig. \ref{meshfs} and \ref{meshfsr}, respectively.
A finite element mesh on structural surface with M2 is shown in Fig. \ref{meshss}.


\begin{table}
\centering
\caption{Aeroelastic parameters for an isotropic aluminum wing under hovering motion}
\begin{tabular}{ccc}
\toprule  
Parameters & Value\\
\midrule  
Semi-span at quarter chord& 0.075 m\\
Chord length at wing root& 0.025 m\\
Structural thickness & 0.0004 m\\
Poisson’s ratio& 0.3\\
Material density& 2700 ${\rm{kg/m^3}}$\\
Young’s modulus of material & 70.0 GPa\\
Reference flow velocity (hovering) & 1.0995 m/s\\
Air density & 1.209 ${\rm{kg/m^3}}$\\
Mean chord-based Reynolds number& 2605\\
Flapping frequency& 10 Hz\\
Flapping amplitude& $21^{\circ}$\\
Aspect ratio& 7.65\\
\bottomrule 
\end{tabular}
\label{paraiso}
\end{table}

\begin{table}
\centering
\caption{Mesh statistics for an isotropic wing under hovering motion}
\begin{tabular}{ccccc}
\toprule  
Mesh & Fluid nodes & Fluid elements  & Structural nodes  &  Structural elements \\
\midrule  
M1& 509,082 & 492,630  & 114 & 90\\
M2& 816,312 & 795,614  & 182  & 150\\
M3 & 1,311,120 & 1,284,226 & 506 & 450  \\
\bottomrule 
\end{tabular}
\label{meshsize}
\end{table}

\begin{figure}
	\centering
	\subfloat[][]{\includegraphics[width=0.5\textwidth]{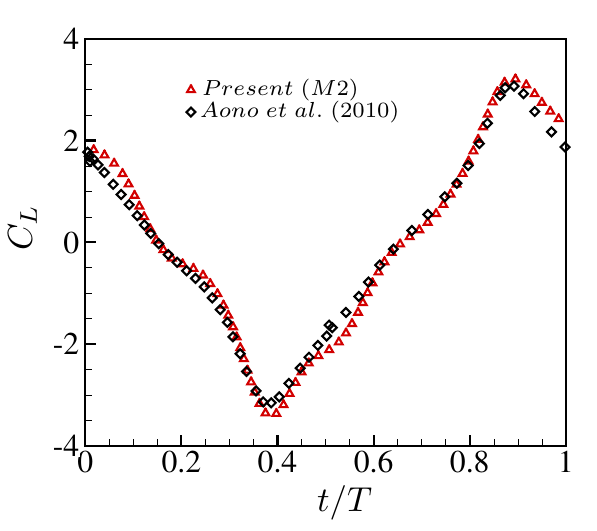} \label{responsea}} 
	\subfloat[][]{\includegraphics[width=0.5\textwidth]{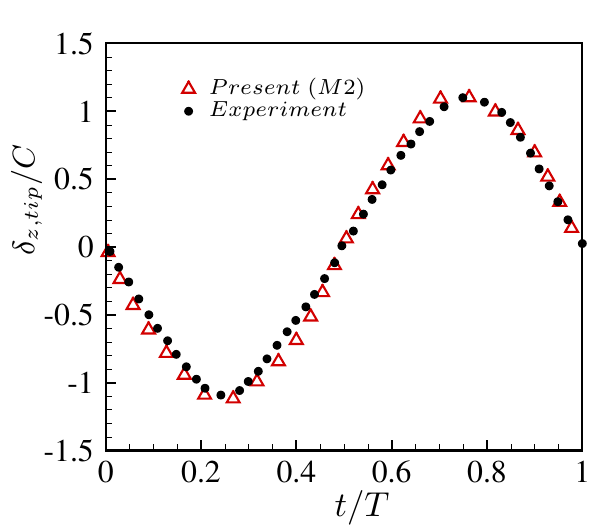} \label{responseb}} 
	\caption{Comparisons of time traces of the isotropic flapping wing: (a) lift coefficient: ({\color{red}$\triangle$}) Present simulation, ({\color{black}$\diamond$}) Aono et al. \cite{aono2010computational}, (b) normalized displacement at wing tip: ({\color{red}$\triangle$}) Present simulation, ({\color{black}$\bullet$}) Experiment \cite{wu2010experimental}.}
	\label{response}
\end{figure}

For the purpose of mesh convergence investigation, the amplitude of lift coefficients for three different meshes are calculated and compared with literature data, which are shown in Table \ref{meshcon}. The results indicate that the gap between M1 and M3 is within 1$\%$ and it reduces to 0.15$\%$ for M2 and M3. It is concluded that M2 has achieved mesh convergence and it can be used as the reference case to compare with experiment data. The comparison of lift coefficient histories in one cycle is displayed in Fig. \ref{responsea} and the simulated result for M2 shows a good comparison with the result of \cite{aono2010computational}. The normalized displacement at wing tip is compared in Fig. \ref{responseb} with that of the experiment \cite{wu2010experimental}. It can be seen that the normalized displacement at wing tip in one non-dimensional cycle has a good match with the experimental measurements. 

\begin{figure}
	\centering
	\subfloat[][ $t/T=0.3$]{\includegraphics[width=0.45\textwidth]{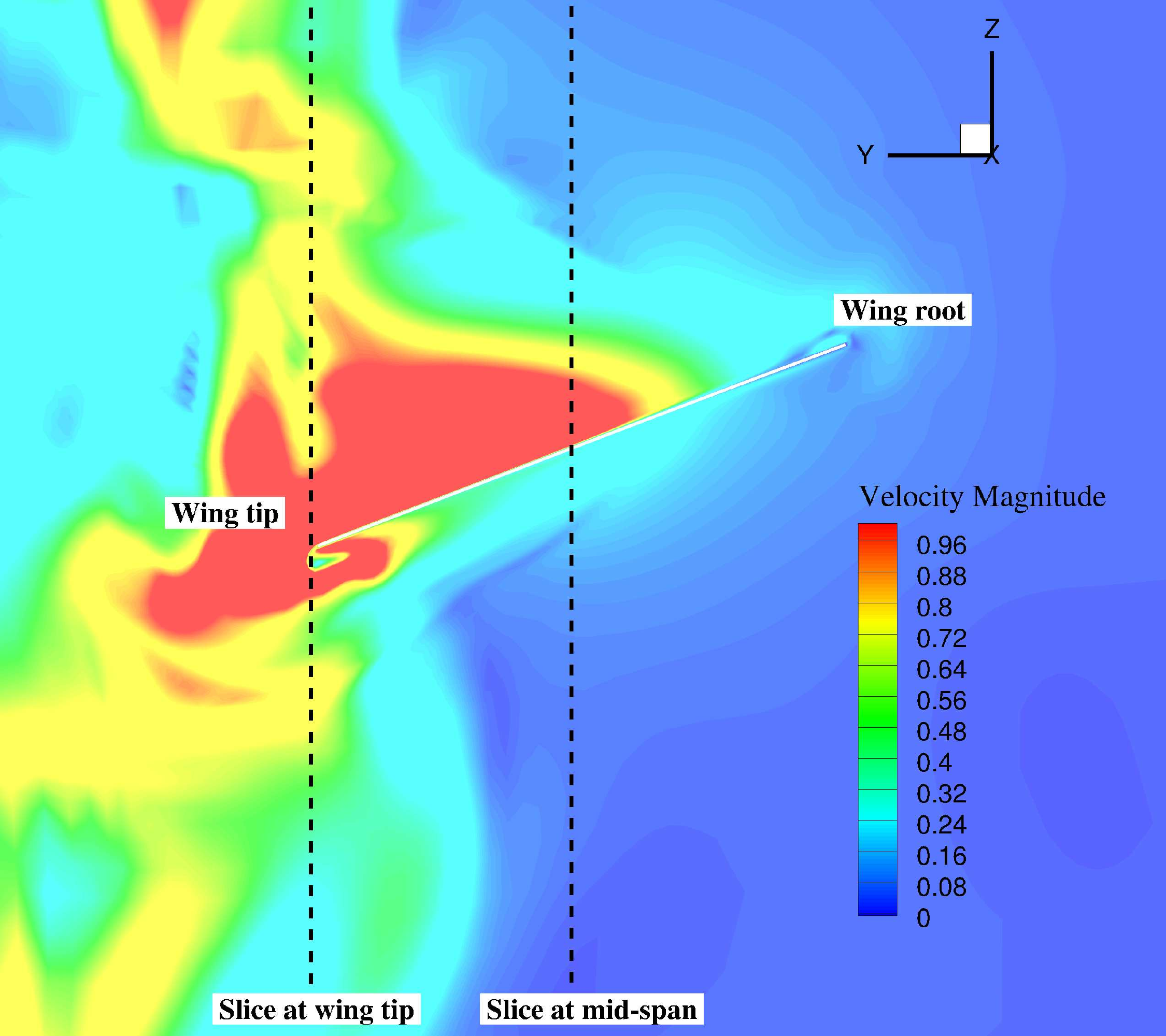}}
	\quad
	\subfloat[][ $t/T=0.48$]{\includegraphics[width=0.45\textwidth]{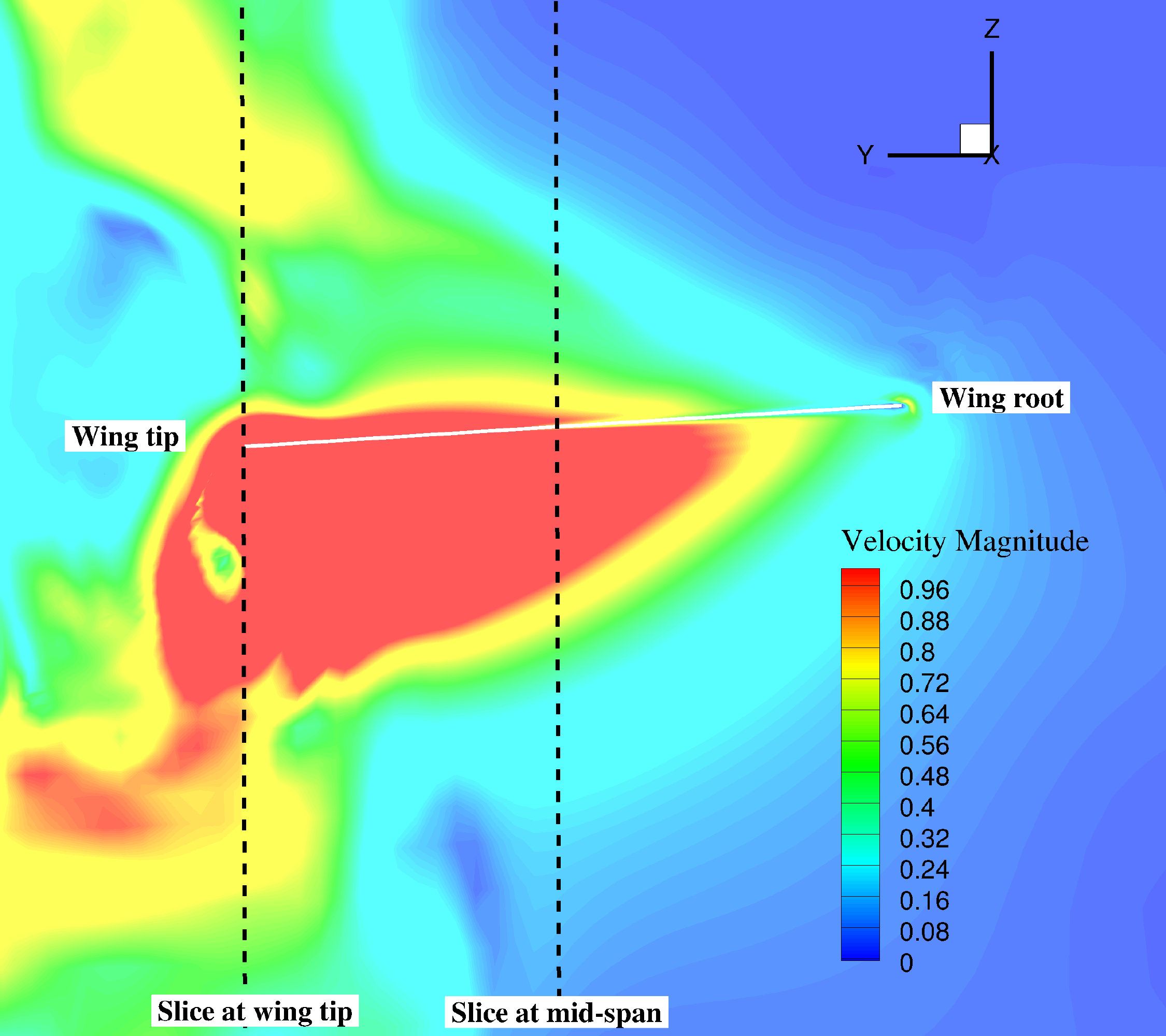}}
	\\
	\subfloat[][ $t/T=0.3$]{\includegraphics[width=0.45\textwidth]{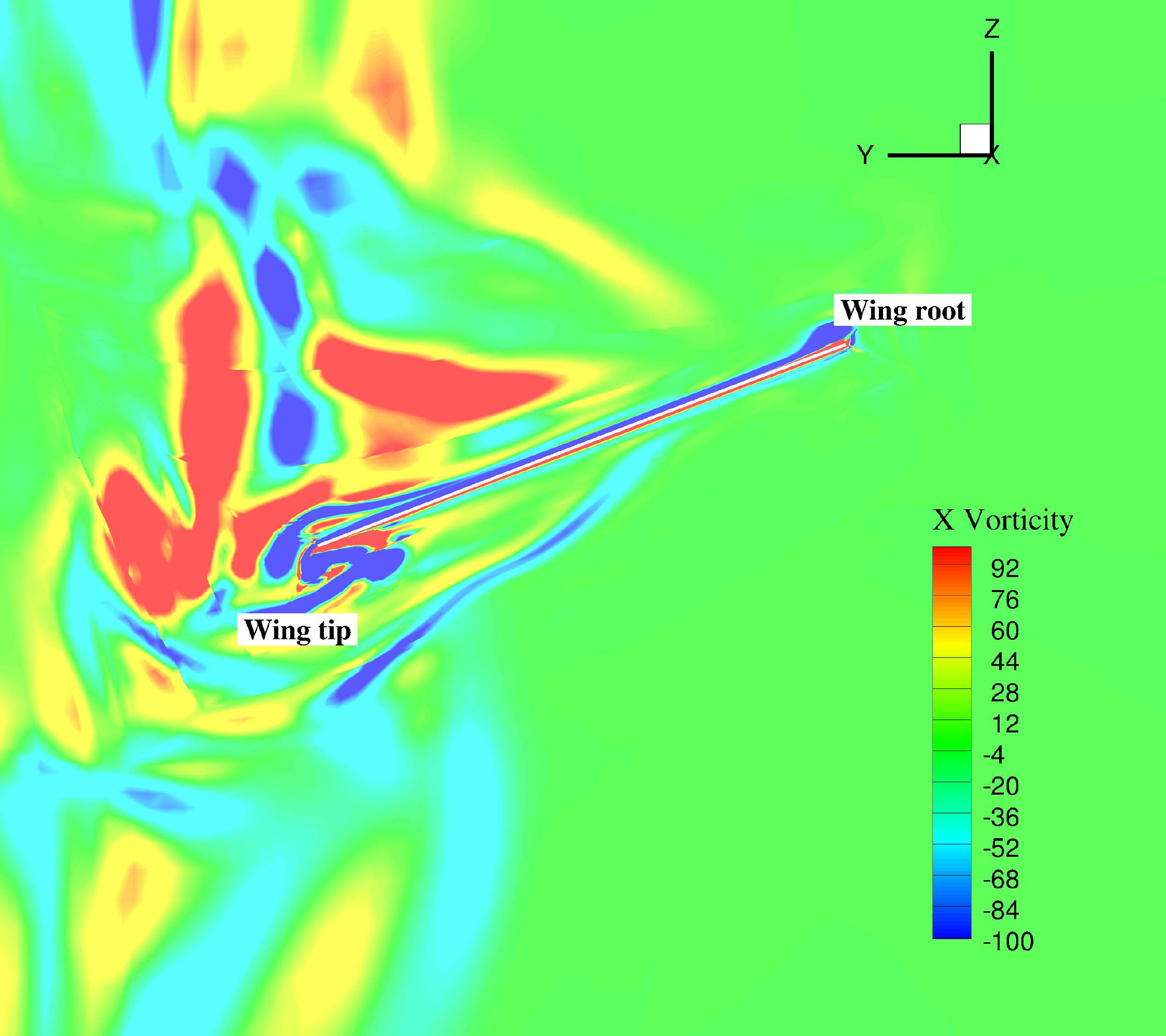}}
	\quad
	\subfloat[][ $t/T=0.48$]{\includegraphics[width=0.45\textwidth]{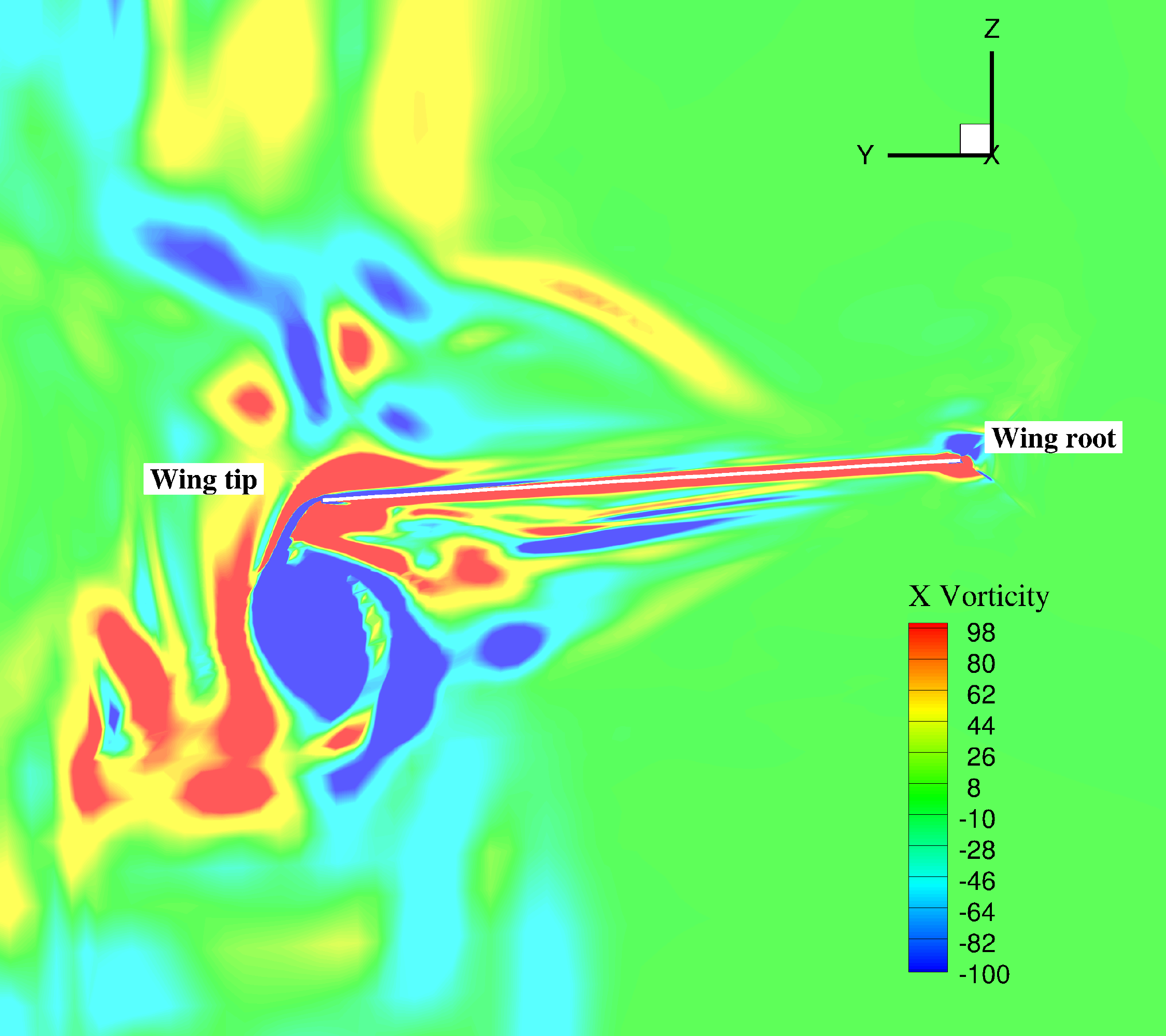}}\\
	\subfloat[][ $t/T=0.3$]{\includegraphics[width=0.45\textwidth]{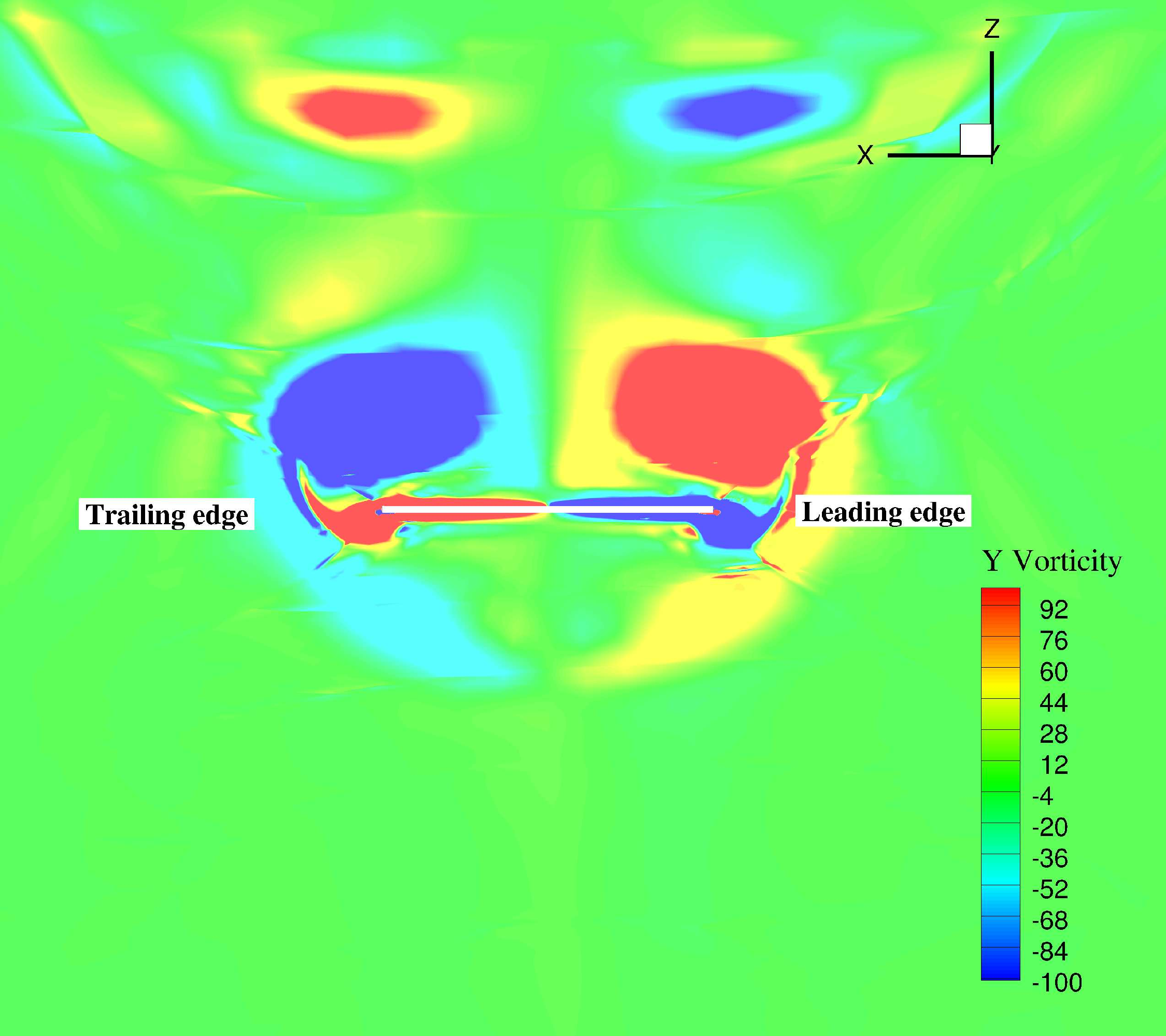}}
	\quad
	\subfloat[][ $t/T=0.48$]{\includegraphics[width=0.45\textwidth]{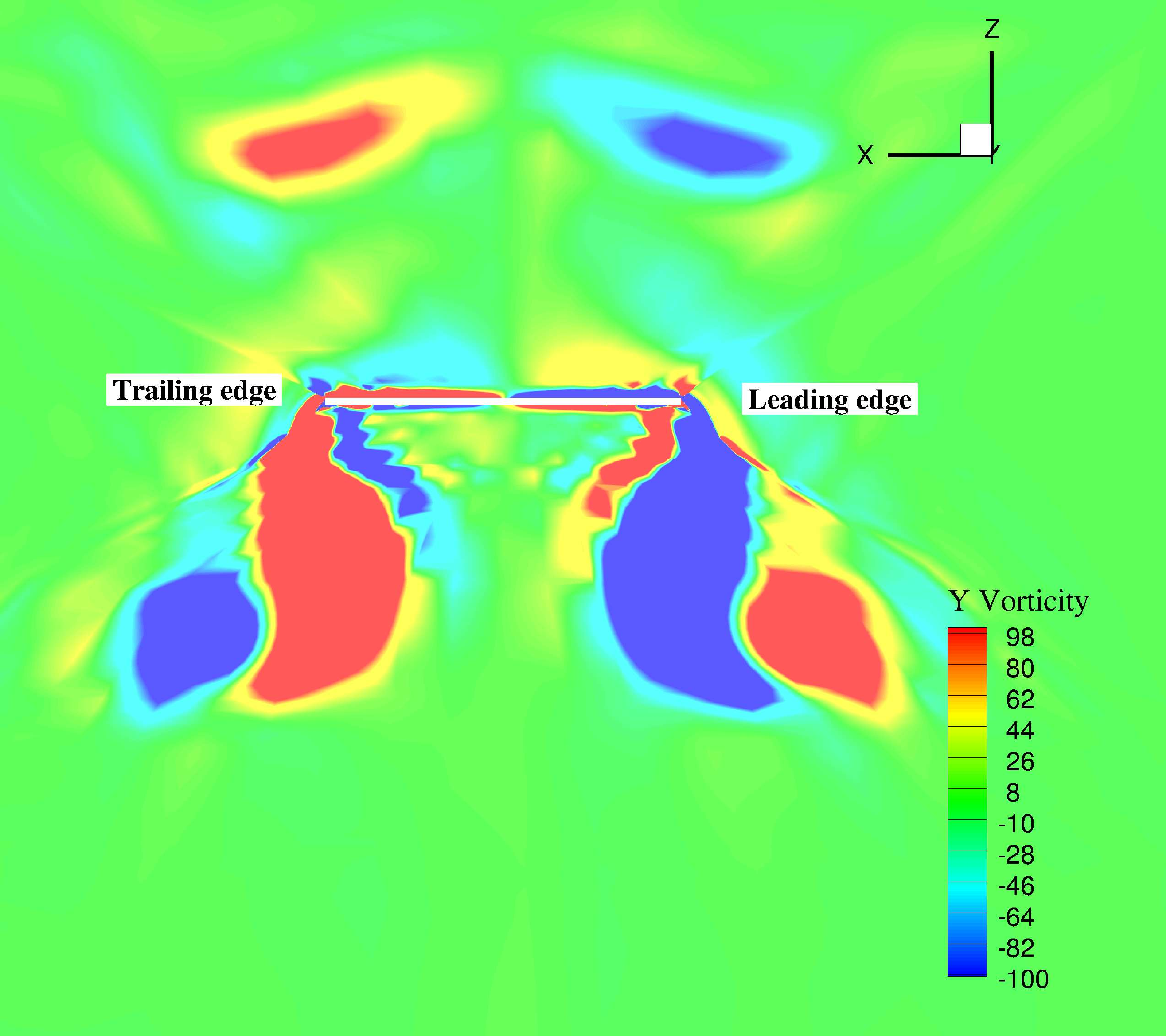}}
	\caption{Flow fields around the isotropic flapping wing in uniform flow: velocity magnitude at (a)  $t/T=0.3$, (b) $t/T=0.48$,  $X$-vorticity at (c) 
$t/T=0.3$, (d) $t/T=0.48$, and  $Y$-vorticity contours at (e) $t/T=0.3$, (f) $t/T=0.48$.}
	\label{fluid03}
\end{figure}

\begin{figure}
	\centering
	\subfloat[][ $t/T=0.3$ at mid-span]{\includegraphics[width=0.5\textwidth]{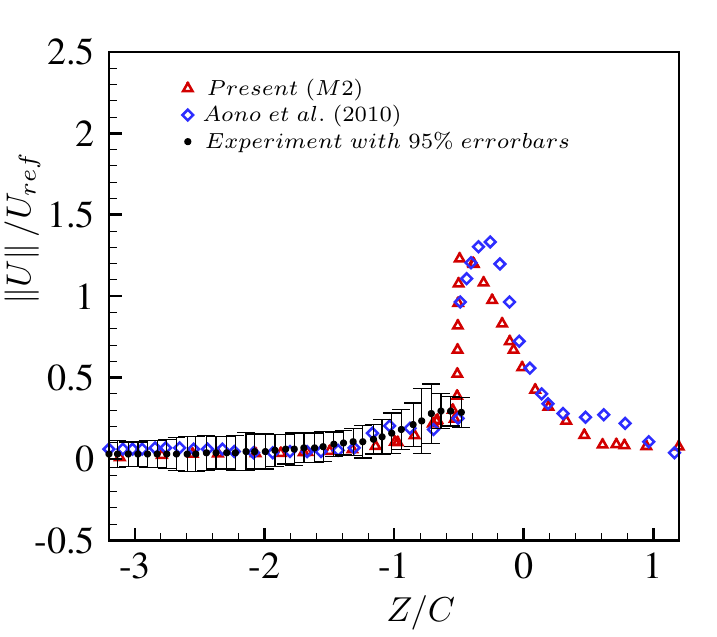}}
	\subfloat[][ $t/T=0.3$ at wing tip]{\includegraphics[width=0.5\textwidth]{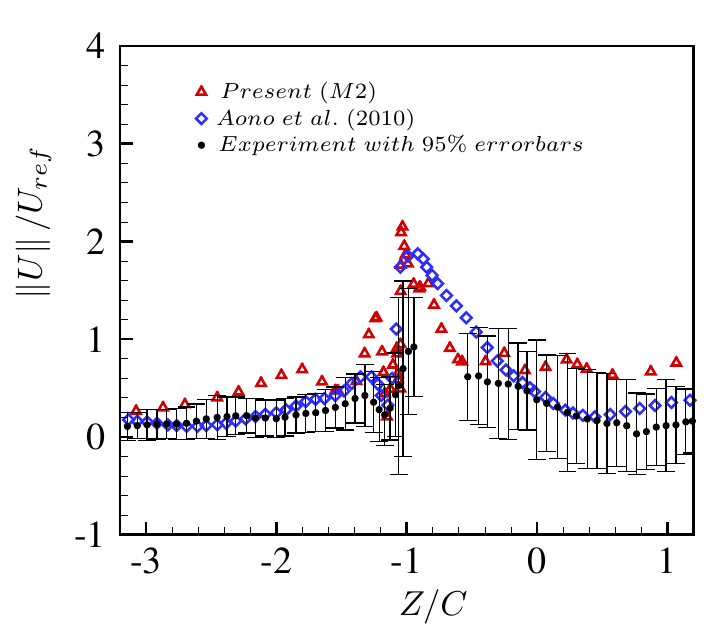}} \\
	\subfloat[][ $t/T=0.48$ at mid-span]{\includegraphics[width=0.5\textwidth]{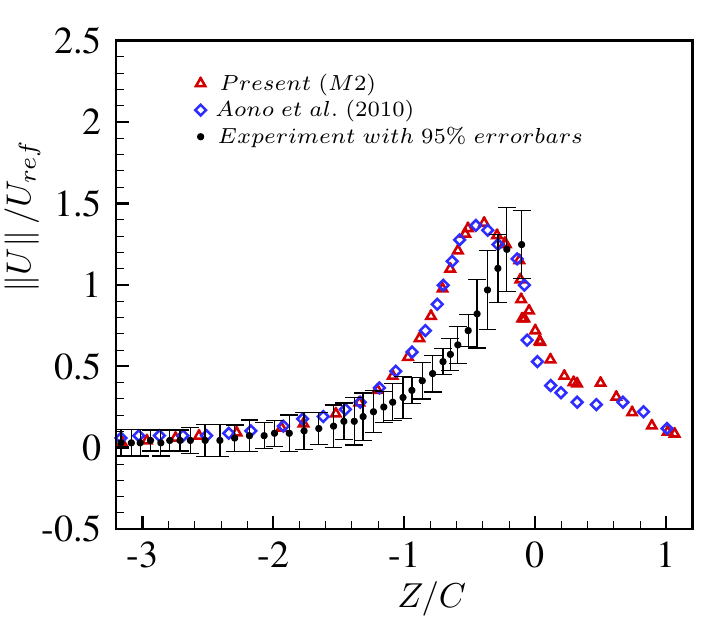}}
	\subfloat[][ $t/T=0.48$ at wing tip]{\includegraphics[width=0.5\textwidth]{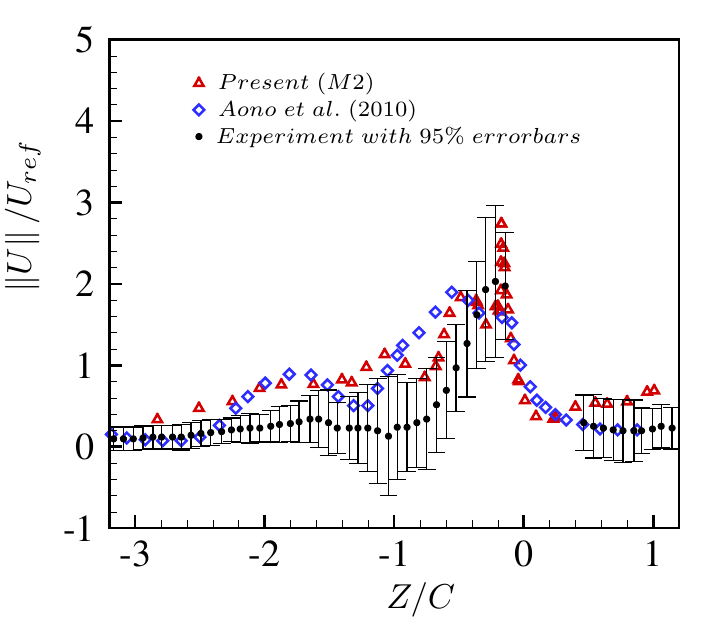}}
	\caption{Flow past an isotropic flappig wing: Comparison of instantaneous velocity magnitude for isotropic wing at various time instants: (a) $t/T=0.3$ at mid-span, (b) $t/T=0.3$ at wing tip, (c) $t/T=0.48$ at mid-span, (d) $t/T=0.48$ at wing tip. ({\color{red}$\triangle$}) Present simulation, ({\color{blue}$\diamond$}) Aono et al. \cite{aono2010computational}, ({\color{black}$\bullet$}) Experiment with 95$\%$ errorbars \cite{wu2010experimental}. }
	\label{velocity}
\end{figure}

The velocity magnitude contour, $X$-vorticity contour for a slice at the quarter position along chord-wise and $Y$-vorticity contour for a slice at the middle position along span-wise at both $t/T=0.3$ and $t/T=0.48$ are shown in Fig. \ref{fluid03}, respectively. It can be seen from the $Y$-vorticity contours that a pair of main vortices is generated at the leading edge and trailing edge during the hovering motion. The normalized velocity magnitude distributions along vertical direction of the flapping wing at wing tip and mid-span for both time instants $t/T=0.3$ and $t/T=0.48$ are extracted from the flow field and compared with those obtained from \cite{wu2010experimental} and \cite{aono2010computational}, which are shown in Fig. \ref{velocity}. The experiment result is displayed with 95$\%$ errorbars of the instantaneous values. The normalized velocity magnitude distributions simulated via our aeroelastic framework show similar trends with those in literature and the experiment. 

\begin{figure}
	\centering
	\subfloat[][ $t/T=0$]{\includegraphics[width=0.45\textwidth,angle=-90]{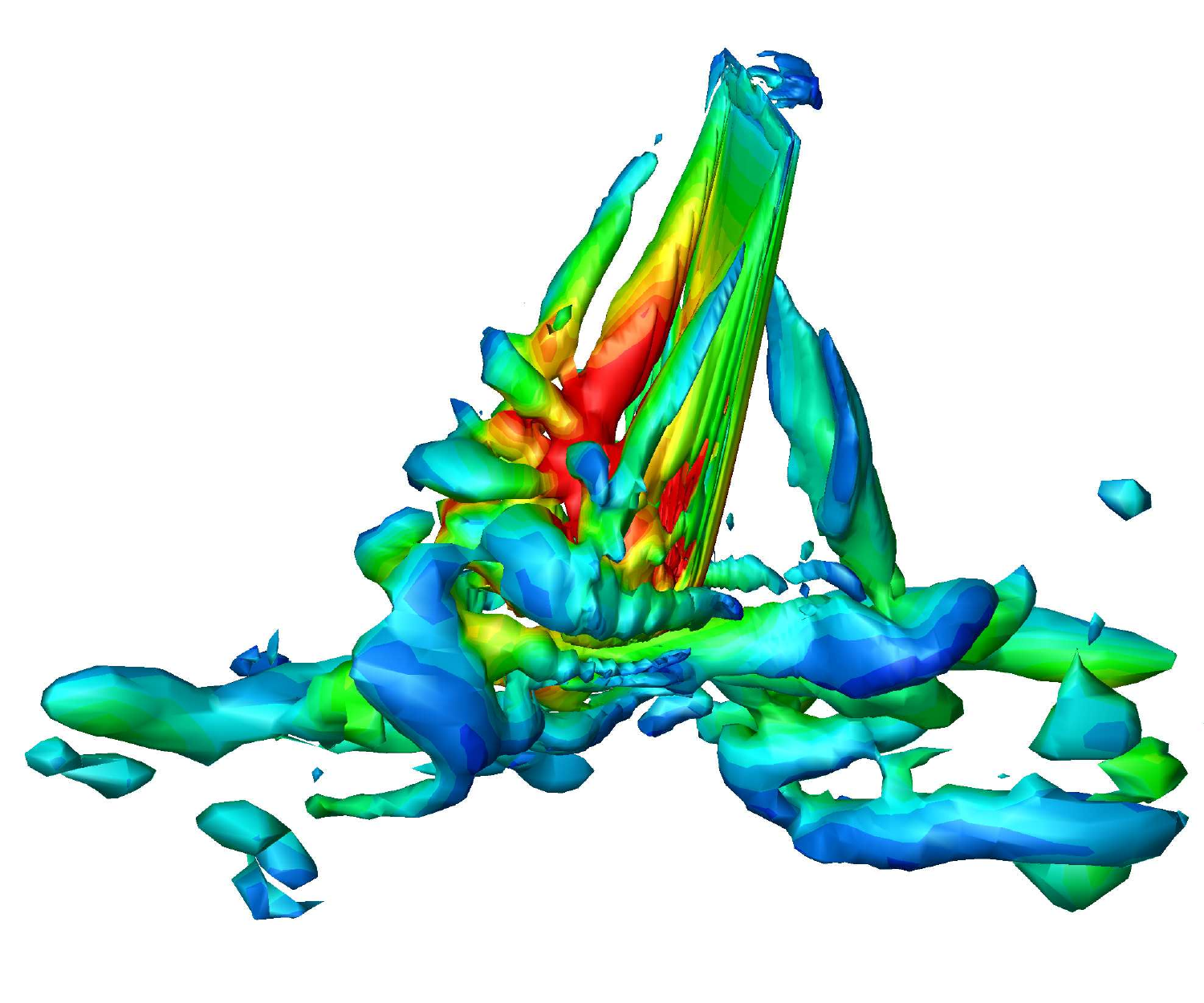}}
	\quad
	\subfloat[][ $t/T=0.25$]{\includegraphics[width=0.45\textwidth,angle=-90]{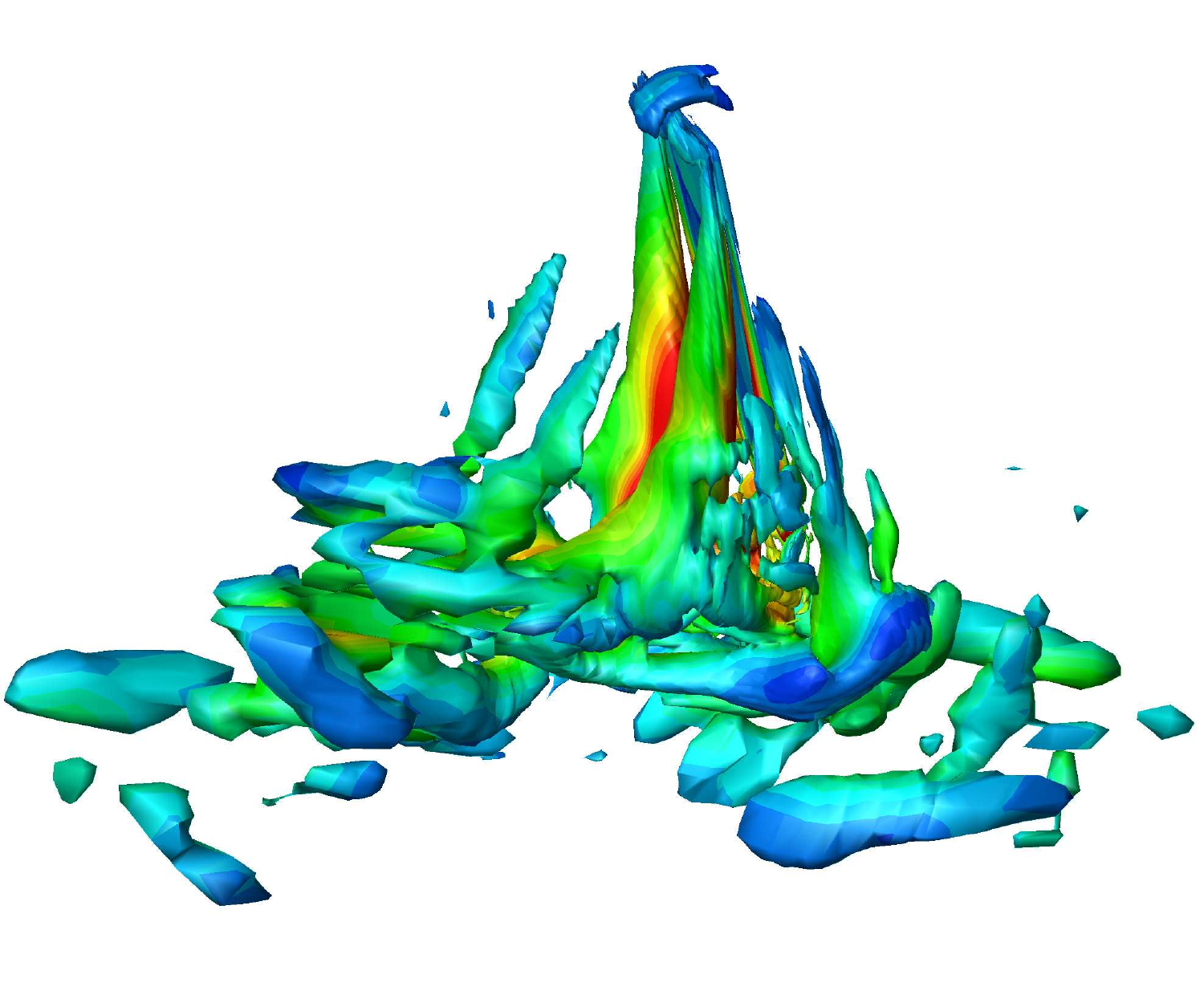}} \\
	\subfloat[][ $t/T=0.5$]{\includegraphics[width=0.45\textwidth,angle=-90]{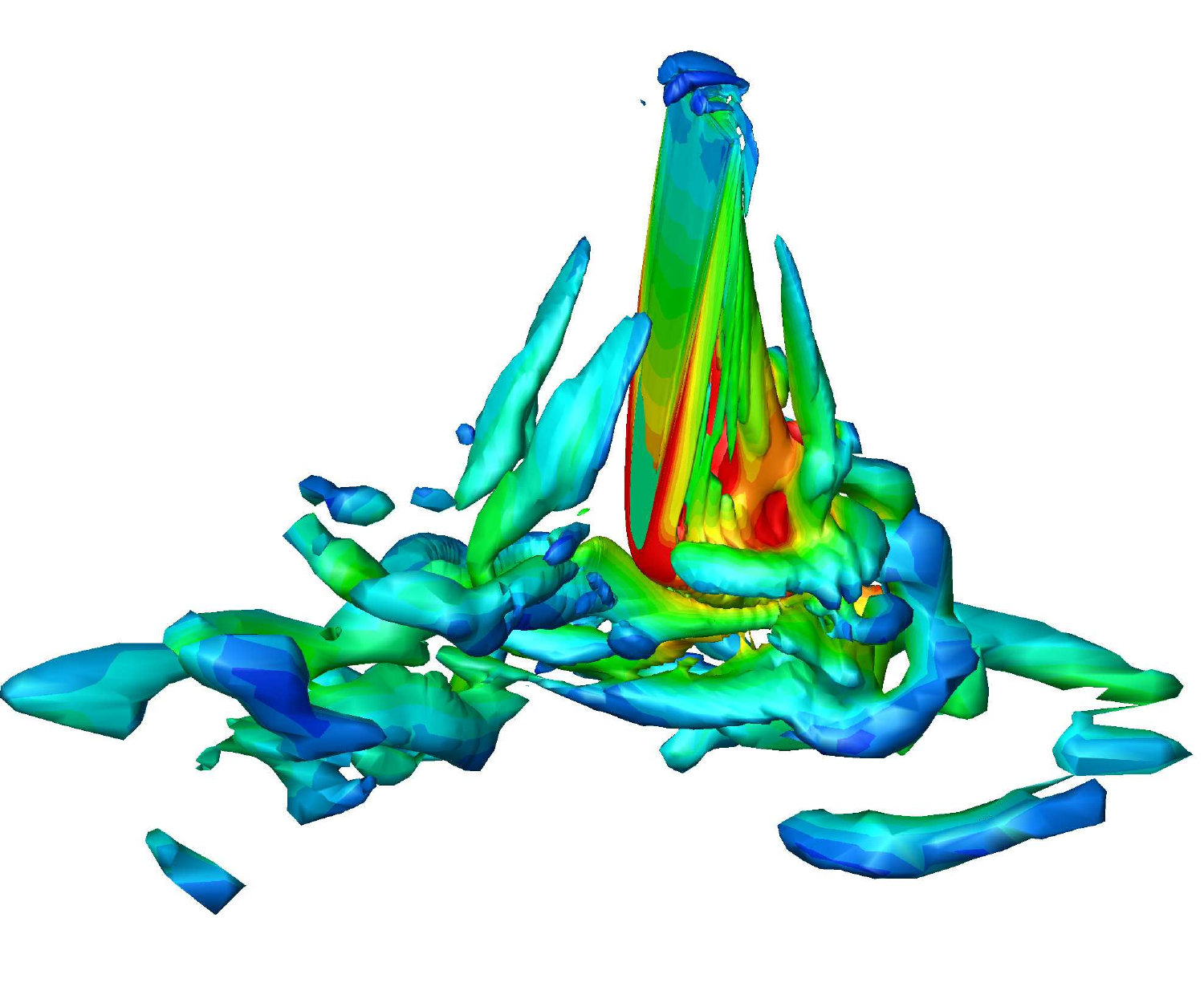}}
	\quad
	\subfloat[][ $t/T=0.75$]{\includegraphics[width=0.45\textwidth,angle=-90]{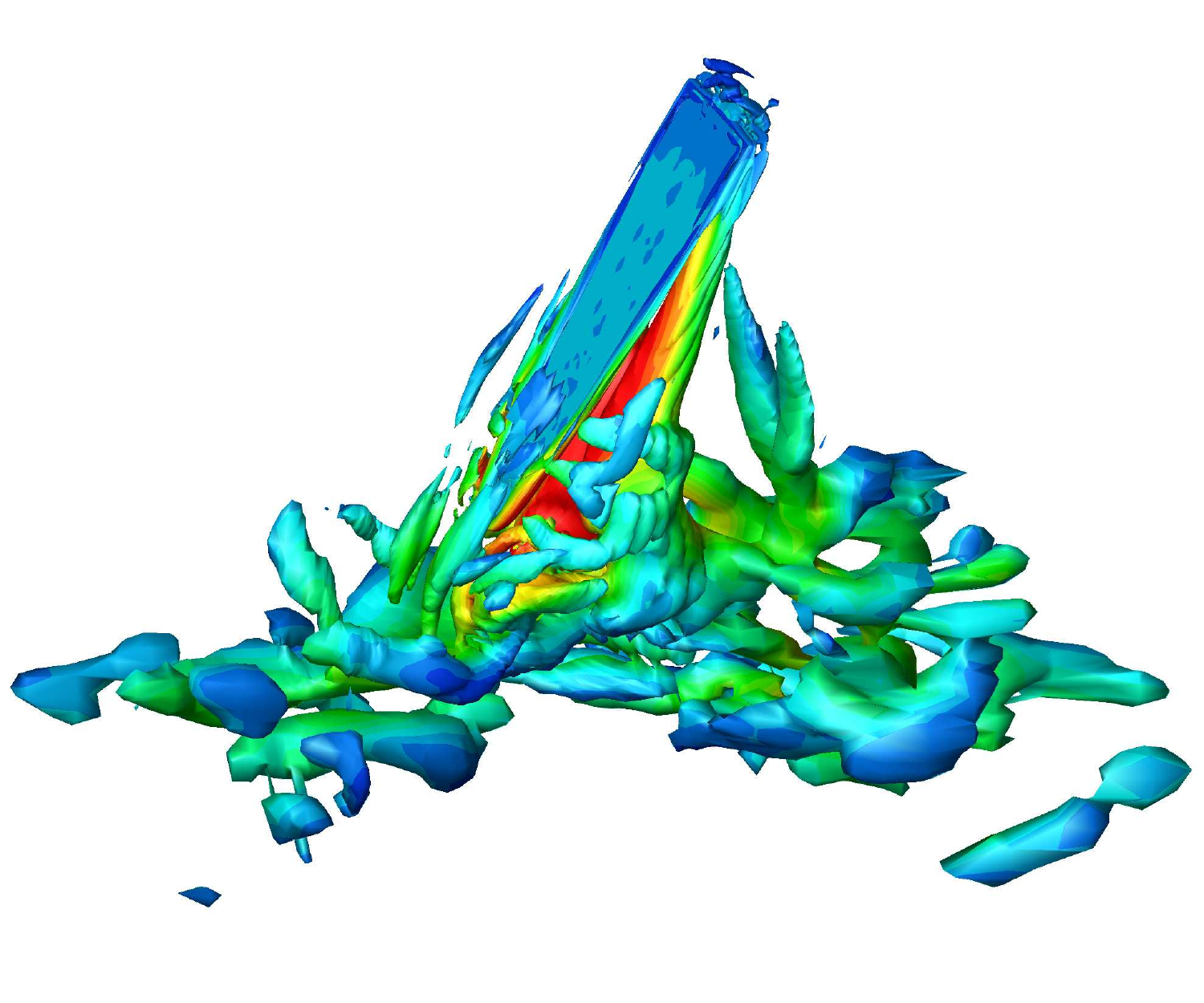}}
	\\
	\includegraphics[width=0.65\textwidth]{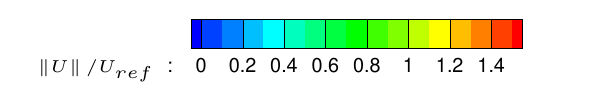}
	\caption{Flow past an isotropic flapping wing: Wake structures based on the instantaneous iso-surfaces of $Q(=-\frac{1}{2} \frac{\partial u^f_i}{\partial x_j} \frac{\partial u^f_j}{\partial x_i})$ value at (a) $t/T=0$, (b) $t/T=0.25$, (c) $t/T=0.5$, (d) $t/T=0.75$. Iso-surfaces of non-dimensional $Q^+ \equiv Q(C/U_{ref})^2=1$ are colored by the normalized velocity magnitude $\left \| U \right \| /U_{ref}$.}
	\label{Qwake}
\end{figure}

The instantaneous turbulent wake fields of the complex three-dimensional flow around the flapping wing at four various time instances with $t/T=$ 0, 0.25, 0.5 and 0.75 are displayed in Fig. \ref{Qwake}. The non-dimensional Q-criterion at isosurface of 1 is used to depict the detailed and complex vortex structures around flapping wing and it is colored by the normalized velocity magnitude during hovering motion. 

\begin{table}
\centering
\caption{Mesh convergence study for the lift coefficient $C_L$. The percentage differences are computed based on M3 result.}
\begin{tabular}{ccc}
\toprule  
Results & Amplitude of lift coefficient $C_L$\\
\midrule  
Present (M1)& 6.65 (0.61$\%$) \\
Present (M2)& 6.6 (-0.15$\%$)\\
Present (M3)& 6.61 \\
Aono et al.& 6.24\\
\bottomrule 
\end{tabular}
\label{meshcon}
\end{table}

\subsection{Validation of multibody aeroelastic framework}


%

\begin{figure}
	\centering
	\subfloat[][]{\includegraphics[width=0.5\textwidth]{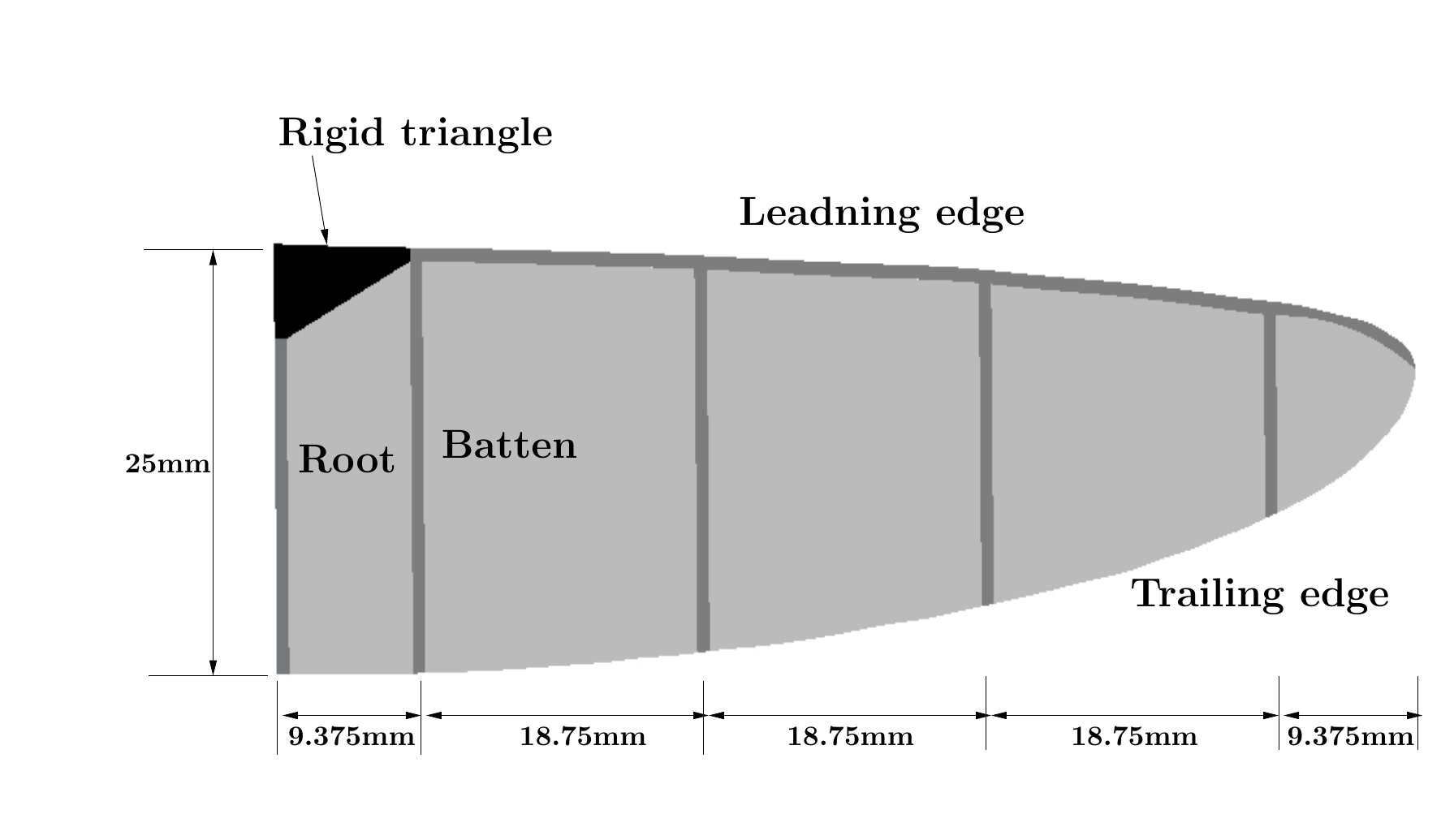} \label{aniwinga}}
	\subfloat[][]{\includegraphics[width=0.5\textwidth]{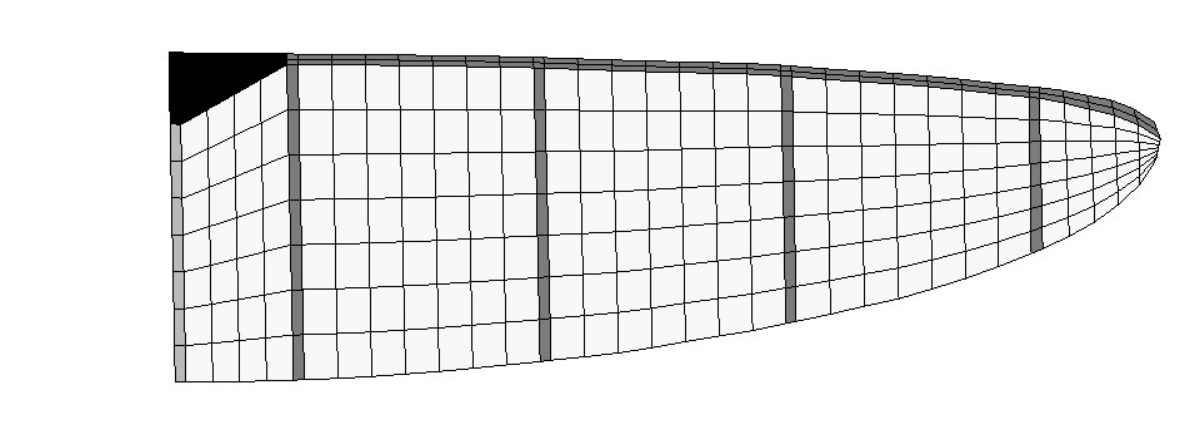}\label{aniwingb}}
	\\
	\subfloat[][]{\includegraphics[width=0.5\textwidth]{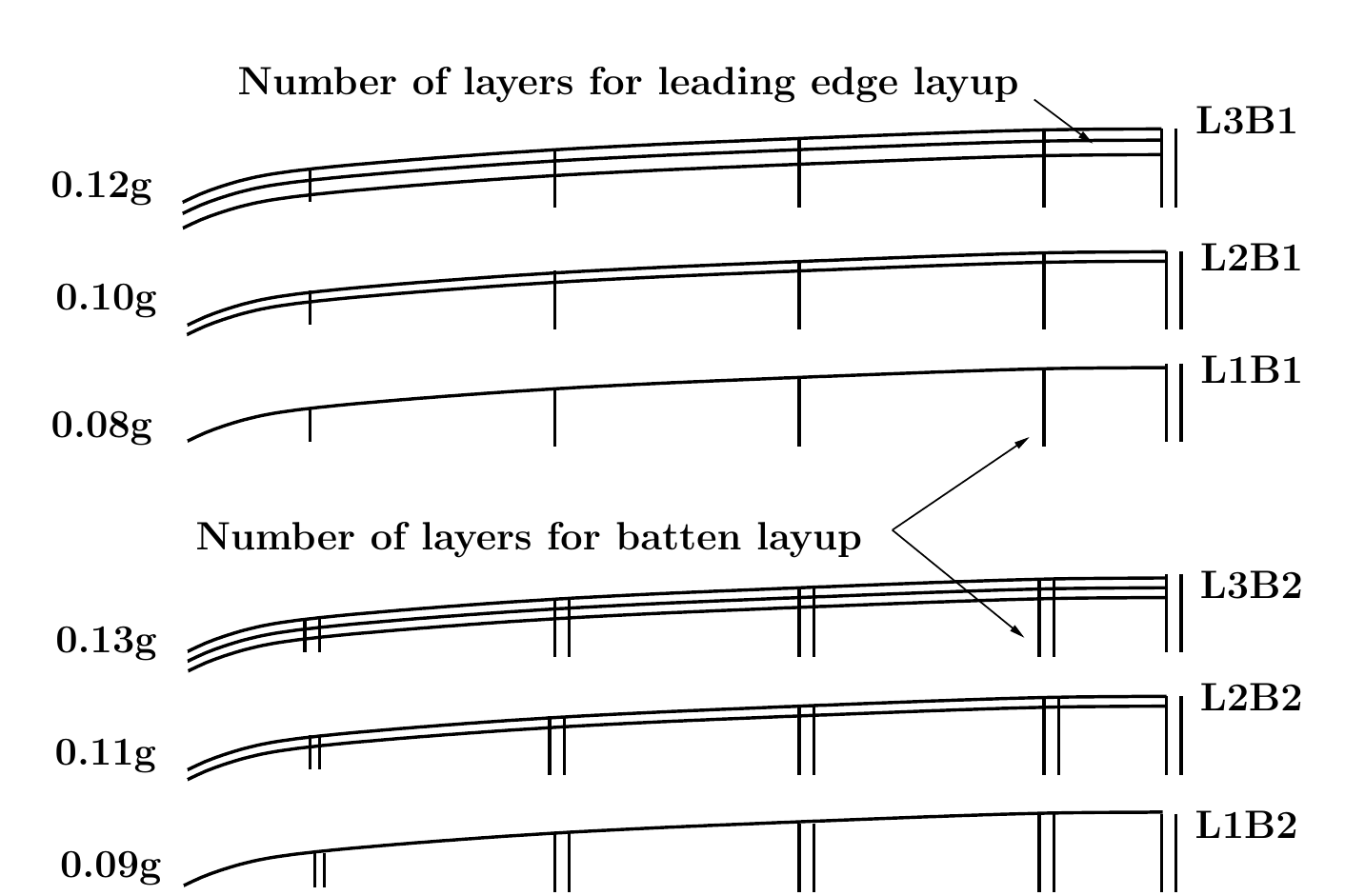} \label{aniwingc}}
	\subfloat[][]{\includegraphics[width=0.5\textwidth]{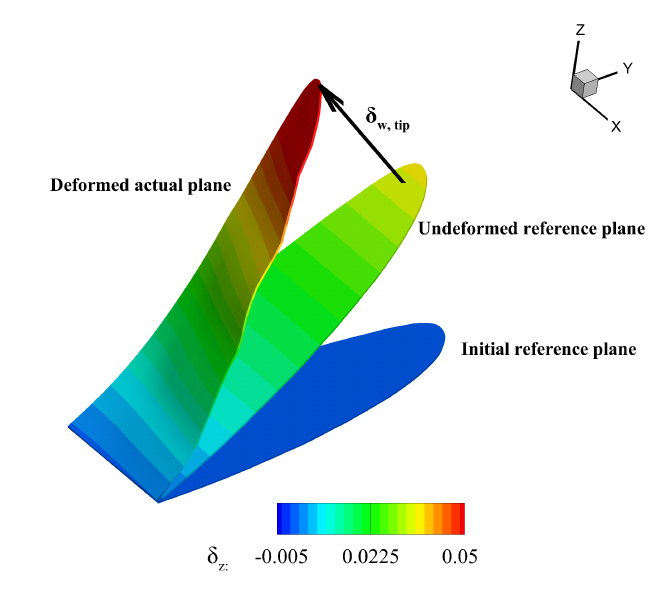} \label{aniwingd}}
	\caption{Problem set-up for anisotropic wing configuration: (a) geometry information, (b) finite element representation, (c) the topological layout \cite{wu2010experimental}, (d) initial reference plane, undeformed reference plane and deformed actual plane.}
	\label{aniwing}
\end{figure}

A multibody flexible wing adopted from the experiment \cite{wu2010experimental} with composite material flapping in a vacuum is first simulated to validate the flexible multibody structural solver. The wing flapping in air flow is considered for our validation purpose of the proposed aeroelastic solver. This wing is designed to mimic a real hummingbird wing based on Zimmerman planform and the membrane with Capran material is supported by several skeletons made of unidirectional carbon fibers. The schematic of this anisotropic wing for a detailed geometric information is depicted in Fig. \ref{aniwinga}. A rigid triangle made of three layers of bidirectional carbon fiber is used to mount the wing during the flapping deformation and it is inserted at the corner of the wing root and the leading edge. A schematic of the topological layout is presented in Fig. \ref{aniwingc}. This anisotropic wing is reinforced at the leading edge, the wing root and the surface of membrane with different layers of carbon fiber strips of 0.8 $mm$ width. Different wings with specific layout schemes are termed L$i$B$j$, where L represents the leading edge, B stands for the batten and $i$, $j$ denotes the number of layers in the leading edge and the batten, respectively. The wing root is reinforced with 2 carbon fiber layers for all wings.  

L2B1 and L3B1 models are considered to validate the high-fidelity flexible multibody solver in our proposed aeroelastic framework. For L2B1 and L3B1 models, the leading edge is reinforced with two and three layers, respectively. Each batten is reinforced with 1 layer for both models. Details of material properties of the composite skeleton and the wing membrane for the anisotropic wing with L2B1 and L3B1 models are summarized in Table \ref{paraaniso}. Some structural material parameters are adjusted to satisfy the structural mode frequencies and shapes obtained from experiments according to the correction work in \cite{gogulapati2011nonlinear}. Geometrically exact co-rotational shell elements with anisotropic material are employed for skeletons and shell elements with homogeneous isotropic material are used for the wing membrane. Different finite elements are connected with their adjacent elements via kinematic constraints. The structural model is discretized by 315 structured four-node rectangular finite elements and the detailed mesh characteristic is given in Fig. \ref{aniwingb}. 

The first six order natural frequencies are analyzed for L2B1 and L3B1 models with specific material properties. The comparison with frequencies obtained from the experiment \cite{wu2010experimental} and literature \cite{gogulapati2011nonlinear} shows that the present results have good agreement, which is given in Table \ref{freaniso}. In the experiment, this anisotropic wing is given a prescribed rotation flapping motion along wing root direction at the mounting rigid triangle in a vacuum. The flapping amplitude is $35^{\circ}$ and the flapping frequency is set to 25 Hz. The time transient structural responses at wing tip is measured and compared with those obtained from experiment \cite{wu2010experimental}. Here $\delta_{z,tip}$ denotes the actual location in the vertical direction of wing tip at computational coordinate and $\delta_{w,tip}$ represents the distance between deformed actual wing tip and the undeformed reference plane. The comparison of the initial reference plane, the undeformed reference plane and the deformed actual plane is depicted in Fig. \ref{aniwingd}.

\begin{figure}
	\centering
	\subfloat[][L2B1]{\includegraphics[width=0.5\textwidth]{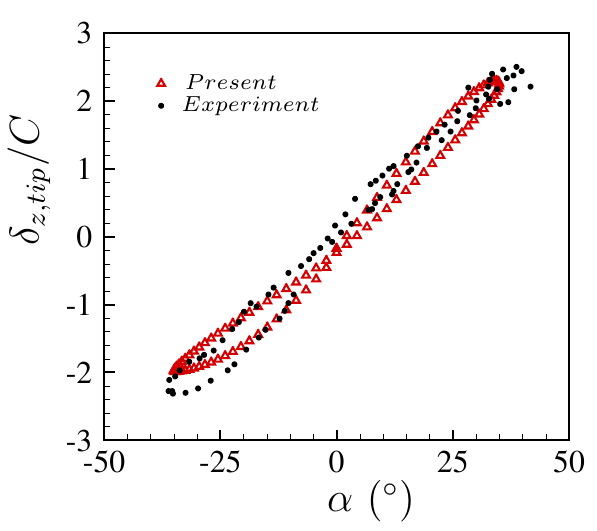}}
	\subfloat[][L2B1]{\includegraphics[width=0.5\textwidth]{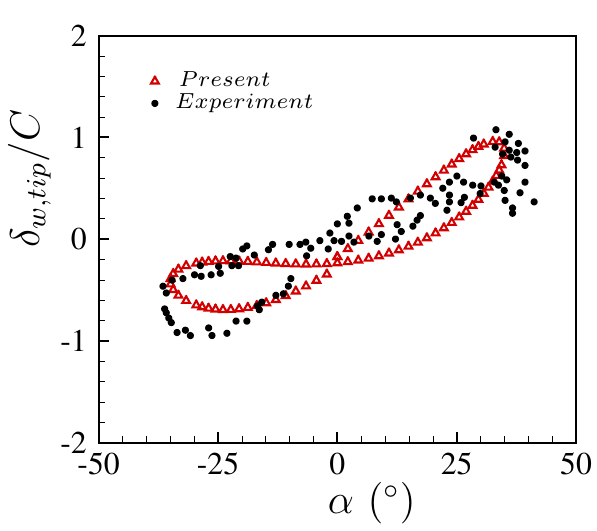}} 
	\\
	\subfloat[][L3B1]{\includegraphics[width=0.5\textwidth]{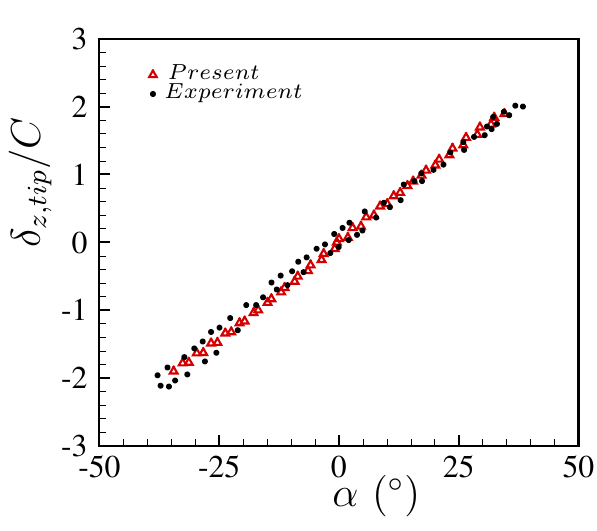}}
	\subfloat[][L3B1]{\includegraphics[width=0.5\textwidth]{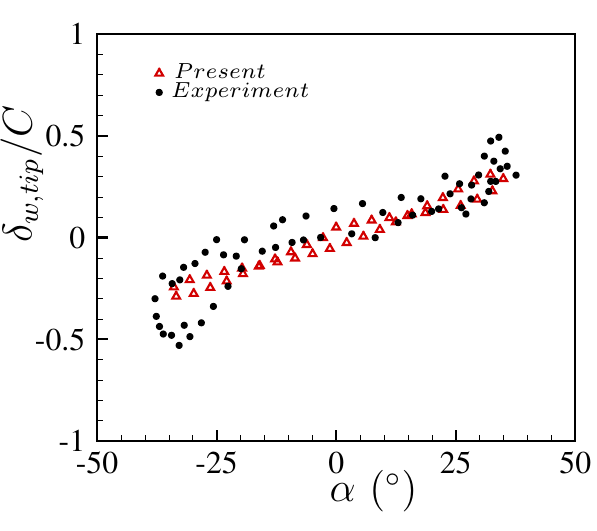}} 
	\caption{Comparison of structural responses for anisotropic wing: (a) normalized location in vertical direction of L2B1 model at wing tip, (b) normalized displacement of L2B1 model at wing tip, (c) normalized location of L3B1 model in vertical direction at wing tip, (d) normalized displacement of L3B1 model at wing tip. ({\color{red}$\triangle$}) Present simulation, ({\color{black}$\bullet$}) Experiment \cite{wu2010experimental}.}
	\label{anire}
\end{figure}

\begin{figure}
	\centering
	\subfloat[][]{\includegraphics[width=0.26\textwidth,angle=-90]{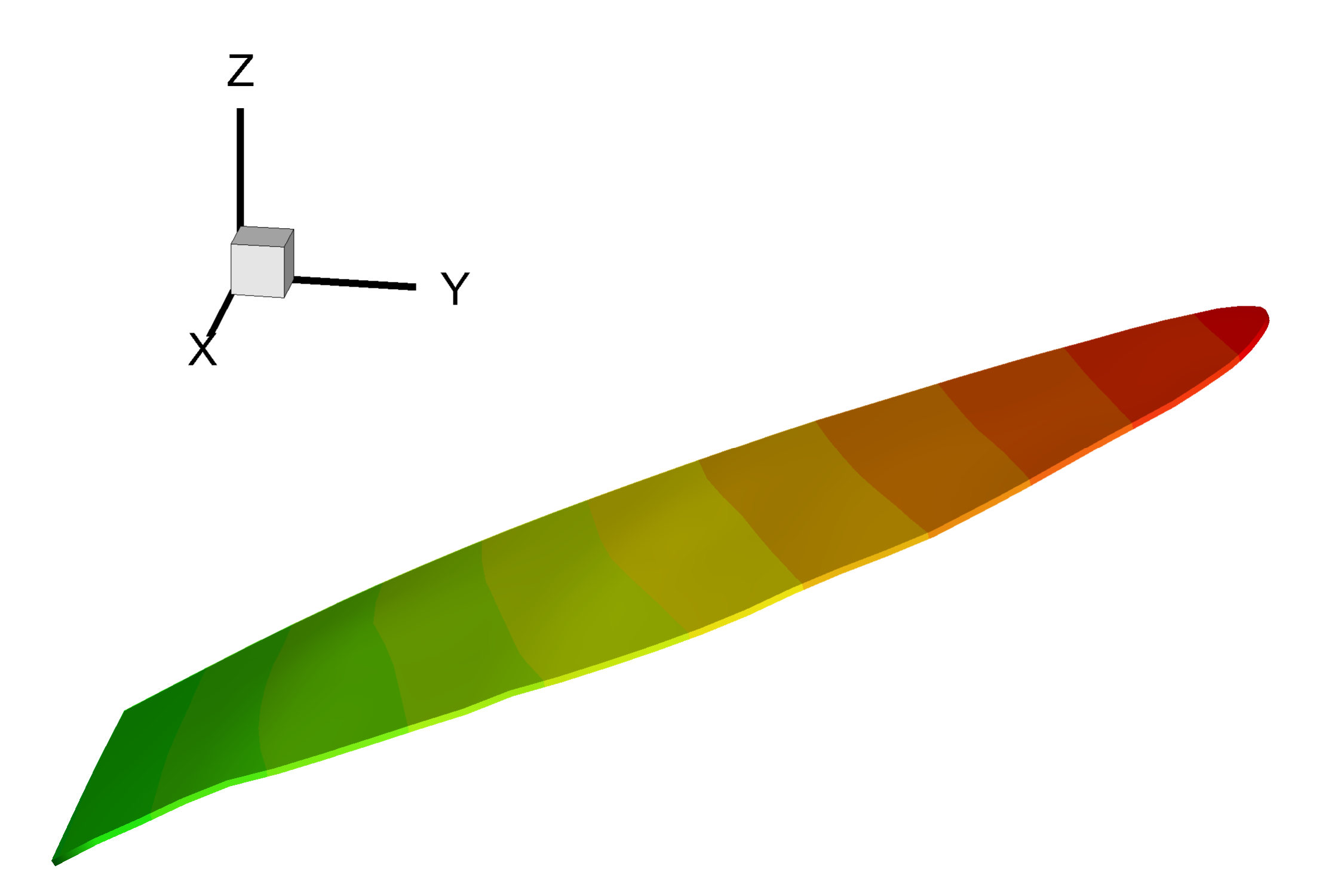}}
	\quad
	\subfloat[][]{\includegraphics[width=0.26\textwidth,angle=-90]{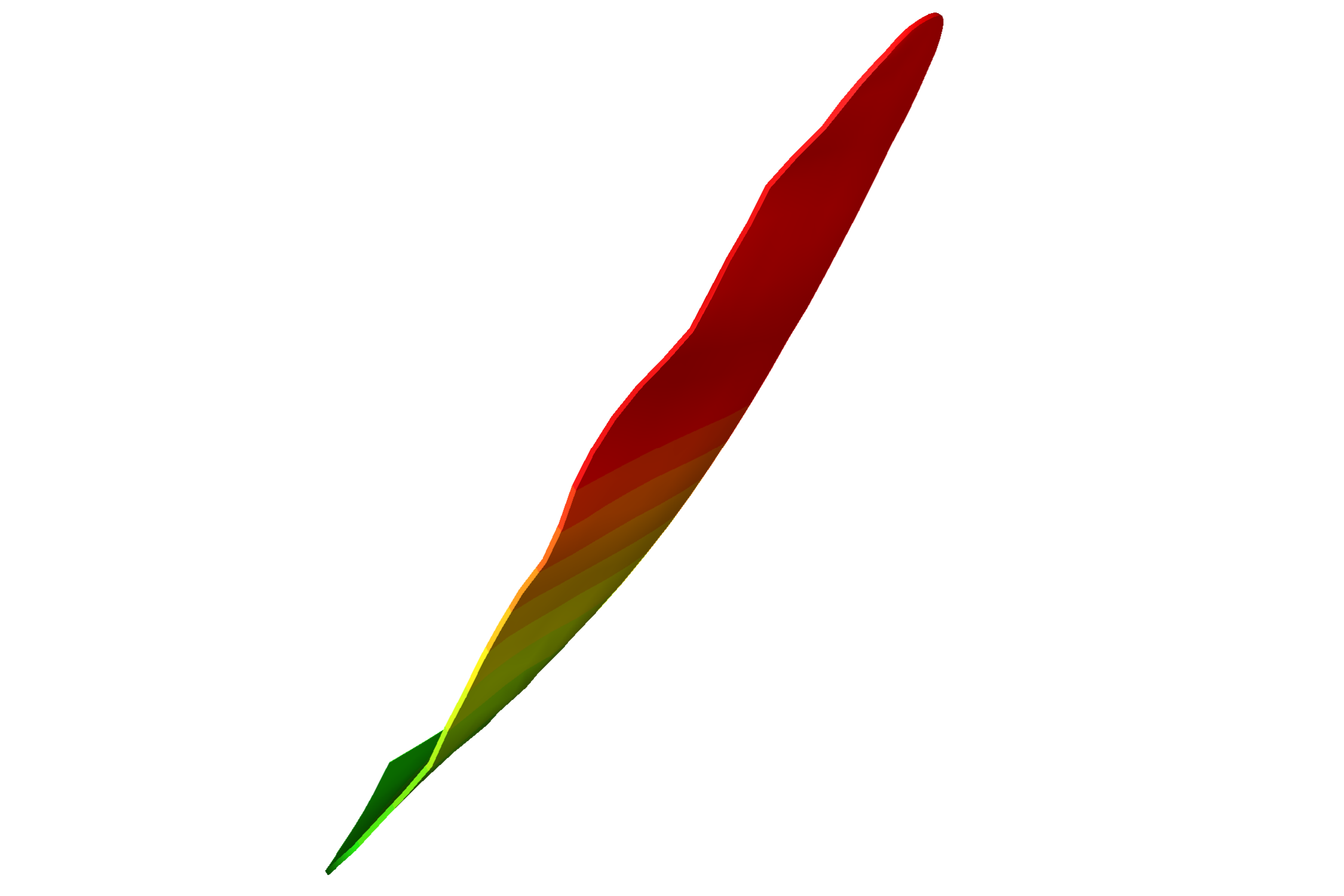}} \\ 
	\subfloat[][]{\includegraphics[width=0.26\textwidth,angle=-90]{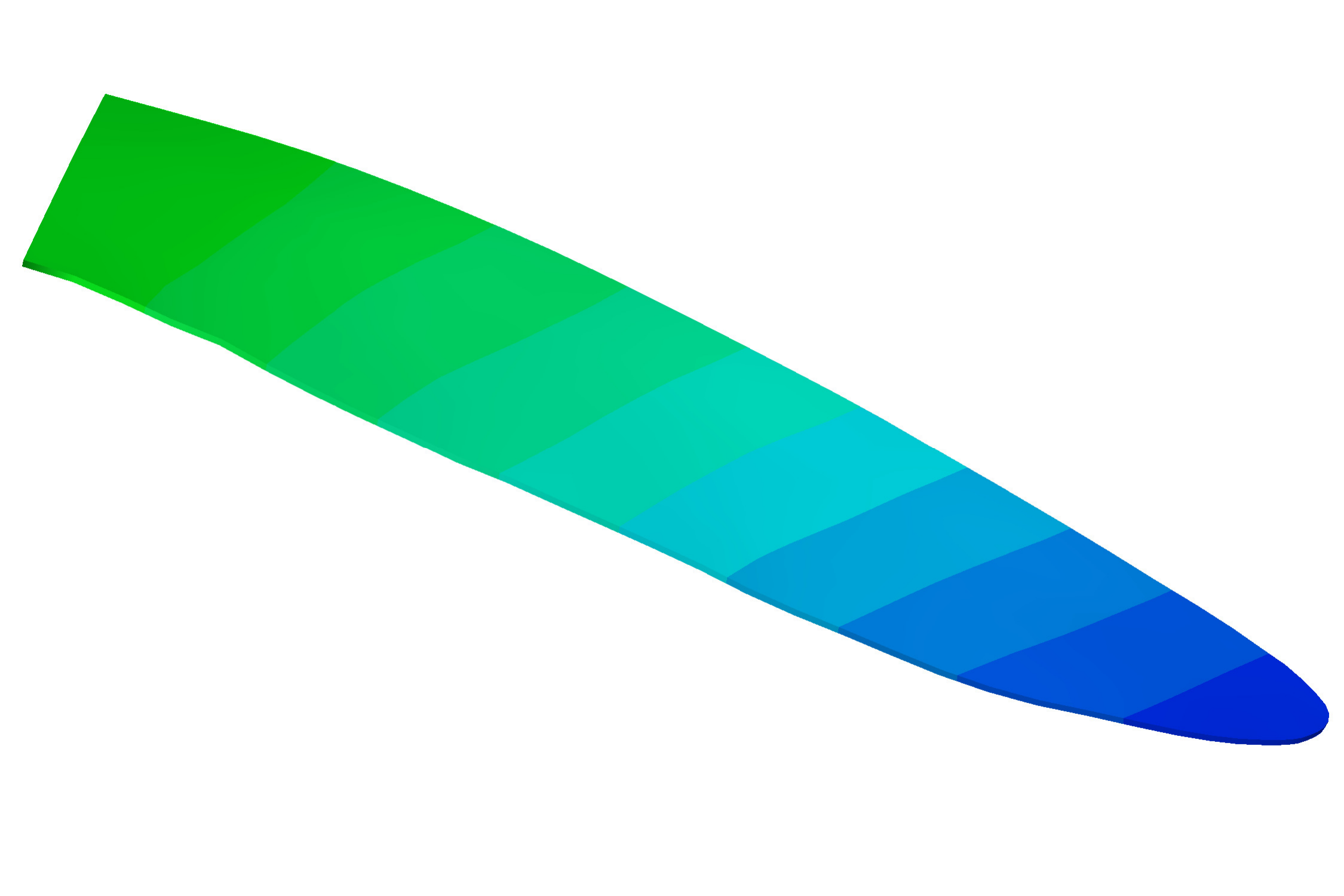}}
	\quad
	\subfloat[][]{\includegraphics[width=0.26\textwidth,angle=-90]{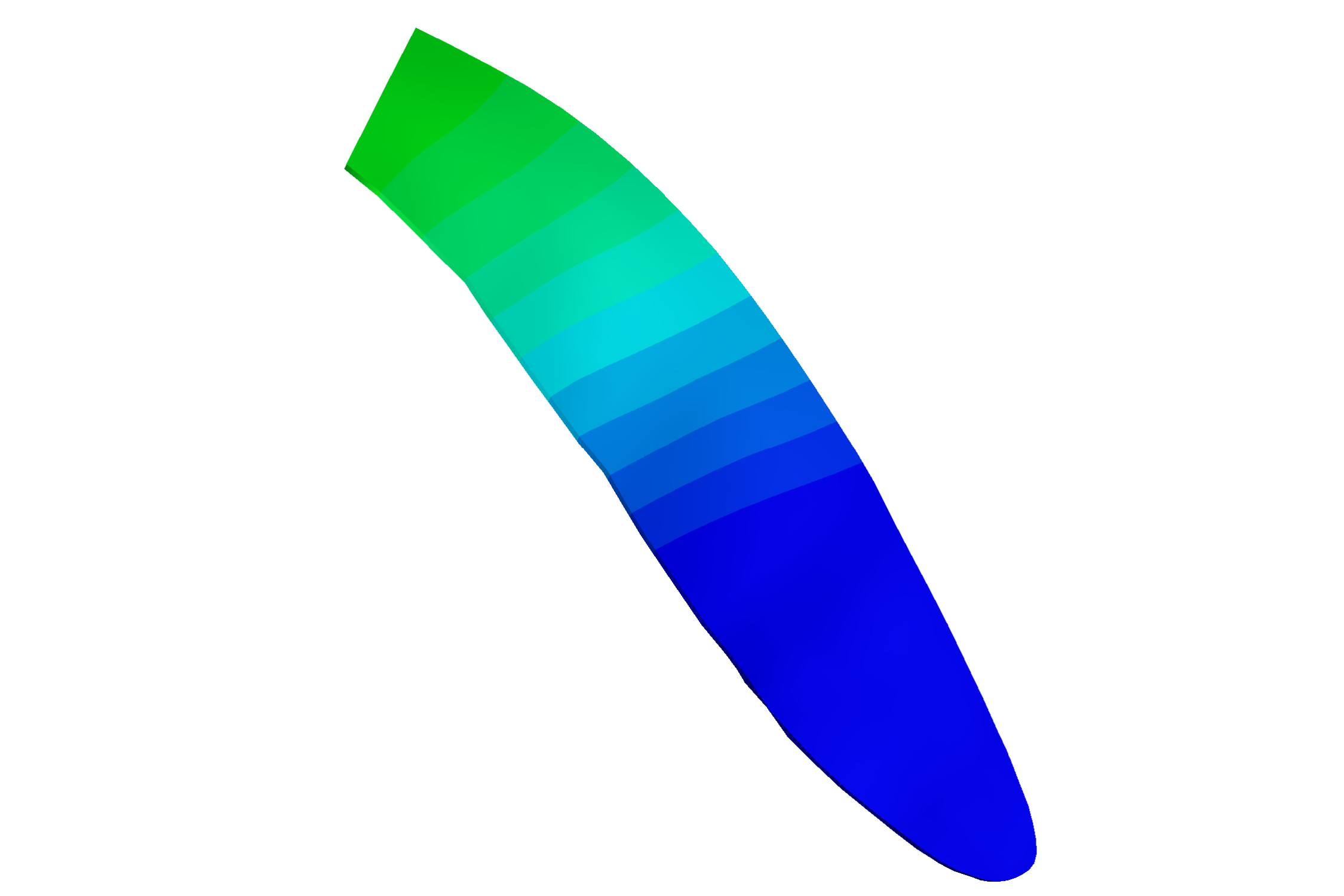}}
	\\
	\includegraphics[width=0.7\textwidth]{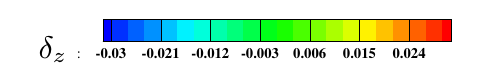}
	\caption{Structural displacement contours of L2B1 model at various time instants: (a) $t/T=0.19$ with flapping angle $32.54^\circ$, (b) $t/T=0.34$ with flapping angle $29.55^\circ$,  (c) $t/T=0.55$ with flapping angle $-10.50^\circ$, (d) $t/T=0.69$ with flapping angle $-32.82^\circ$.}
	\label{anisodis}
\end{figure}

Comparisons of time histories for the normalized location in the vertical direction and  the normalized displacement at the wing tip for L2B1 and L3B1 models undergoing prescribed rotational motion in a vacuum are shown in Fig. \ref{anire},  respectively. The results indicate that the high-fidelity flexible multibody solver can simulate the nonlinear structural responses of the anisotropic wing with multiple components and shows reasonable agreement with the date obtained from the experiment. In our simulation results, the wing tip deformation is varied in different cycles, which is similar to the conclusion obtained in \cite{gogulapati2011nonlinear}. The structural responses are averaged over several cycles, according to the periodic assumption in experimental measurements. One possible reason causing discrepancies in the comparison is some uncertain factors connected with the actuation mechanism \cite{wu2010experimental}. The instantaneous structural displacement contours of L2B1 model for four time instants in a whole flapping cycle are given in Fig. \ref{anisodis}. The anisotropic wing shows a relatively large elastic deformation and the wing twist due to the high flexibility of the wing material.

\begin{table}
\centering
\caption{Material properties of the composite skeleton and wing membrane for the anisotropic wing (L2B1 and L3B1 models) \cite{gogulapati2011nonlinear}}
\begin{tabular}{cc}
\toprule  
Component & Material properties\\
\midrule  
\multirow{9}{4cm}{Carbon fiber prepreg (Properties of one layer)} & $E_{11}$=233 GPa \\
& $E_{22}$=23.7 GPa \\
& $G_{12}$=10.5 GPa \\
& $G_{23}=G_{31}$=1.7 GPa \\
& $v_{12}$=0.05 \\
& $v_{23}=v_{31}$=0.32\\
& $\rho$=1740 ${\rm{kg/m^3}}$\\
& Thickness=0.1 mm \\
\midrule  
\multirow{4}{4cm}{Capran membrane
(From experiments)} & $E$=2.76 GPa \\
& $v_{12}$=0.489 \\
& $\rho$=1384 ${\rm{kg/m^3}}$ \\
& Thickness=0.015 mm \\
\bottomrule 
\end{tabular}
\label{paraaniso}
\end{table}

\begin{table}
\centering
\caption{Comparison of the natural frequencies for anisotropic wing}
\begin{tabular}{ccccccccc}
\toprule  
\multirow{2}*{Model} & \multirow{2}*{Result} & \multicolumn{6}{c}{Natural frequencies (Hz)}\\
\cline{3-8} 
& & $1^{st}$ & $2^{nd}$ & $3^{rd}$ & $4^{th}$ & $5^{th}$ & $6^{th}$\\
\midrule  
\multirow{3}*{L2B1} & Present & 45.03 & 73.16 & 76.39 & 94.45 & 106.35 & 116.31 \\
& Gogulapati A. & 47.00 & 72.00 & 76.50 & 88.00 & 109.00 & 118.80\\
& Experiment & 42.00 &  &   & 84.00 &  & 126.00 \\
\midrule  
\multirow{3}*{L3B1} & Present & 62.71 & 76.13 & 79.55 & 106.31 & 111.00 & 119.39 \\
& Gogulapati A. & 65.00 & 75.50 & 76.80 & 107.00 & 109.00 & 120.00\\
& Experiment & 59.00 &  &   & 104.00 &  & 138.00 \\
\bottomrule 
\end{tabular}
\label{freaniso}
\end{table}

%

In the experiment, the anisotropic wing flapping in air condition is also performed to investigate the aeroelastic effect on the flapping motion. Therefore, an anisotropic wing with L3B1 model is simulated by using the proposed aeroelastic framework for validation purpose. In the experiment, the material properties of the anisotropic wing are summarized in Table \ref{paraaniso} and a hovering motion is employed with zero freestream velocity. Detailed simulation parameters for the aeroelastic simulation are given in Table \ref{paraani}. Considering the similar simulation condition as the isotropic flapping wing, a similar aeroelastic computational domain is  shown in Fig. \ref{wingDomain}. Identical boundary conditions for the computational domain of anisotropic wing are applied  as those of the isotropic wing configuration. A similar mesh distribution is adopted for the three-dimensional fluid computational domain with 492,630 unstructured finite elements. The structural mesh is same as shown in Fig. \ref{aniwingb} with 315 structured finite elements. The distance between the first grid point in the boundary layer and wing surface is set as 9.28$\times 10^{-4}C$ with $y^+ \approx 0.5$ and the number of divisions along the vertical direction of the wing surface is 25 and stretching ratio, $\Delta y_{i+1} / \Delta y_i$ is 1.2. The non-dimensional time step size, $\Delta t U_{ref}/C$ is set to 0.0018. 

\begin{figure}
	\centering
	\subfloat[][]{\includegraphics[width=0.5\textwidth]{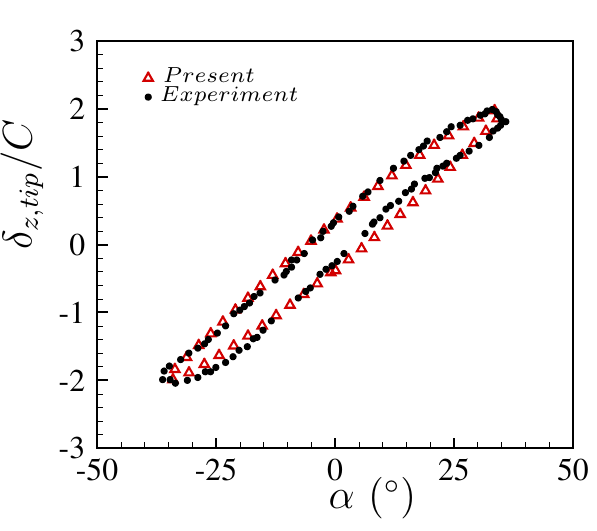}}
	\subfloat[][]{\includegraphics[width=0.5\textwidth]{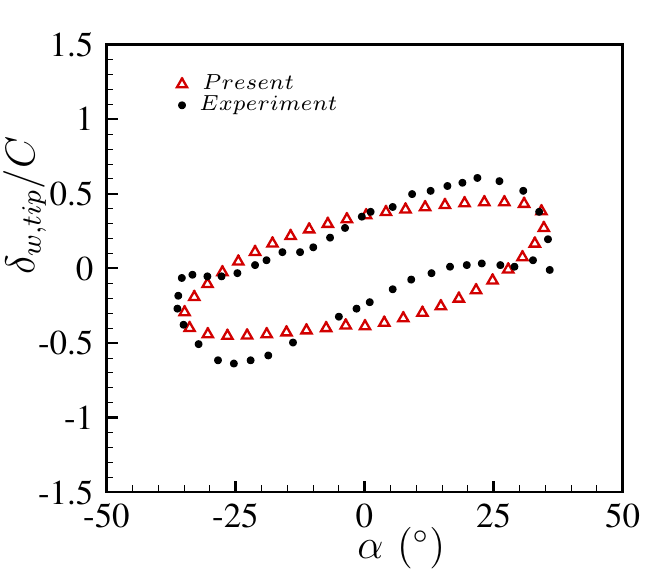}} 
	\caption{Comparison of coupled aeroelastic responses for anisotropic flapping wing: (a) normalized location of L3B1 model in vertical direction at wing tip, (b) normalized displacement of L3B1 model at wing tip. ({\color{red}$\triangle$}) Present simulation, ({\color{black}$\bullet$}) Experiment \cite{wu2010experimental}.}
	\label{aniair}
\end{figure}

\begin{figure}
	\centering
	\subfloat[][$t/T=0$]{\includegraphics[width=0.45\textwidth,angle=-90]{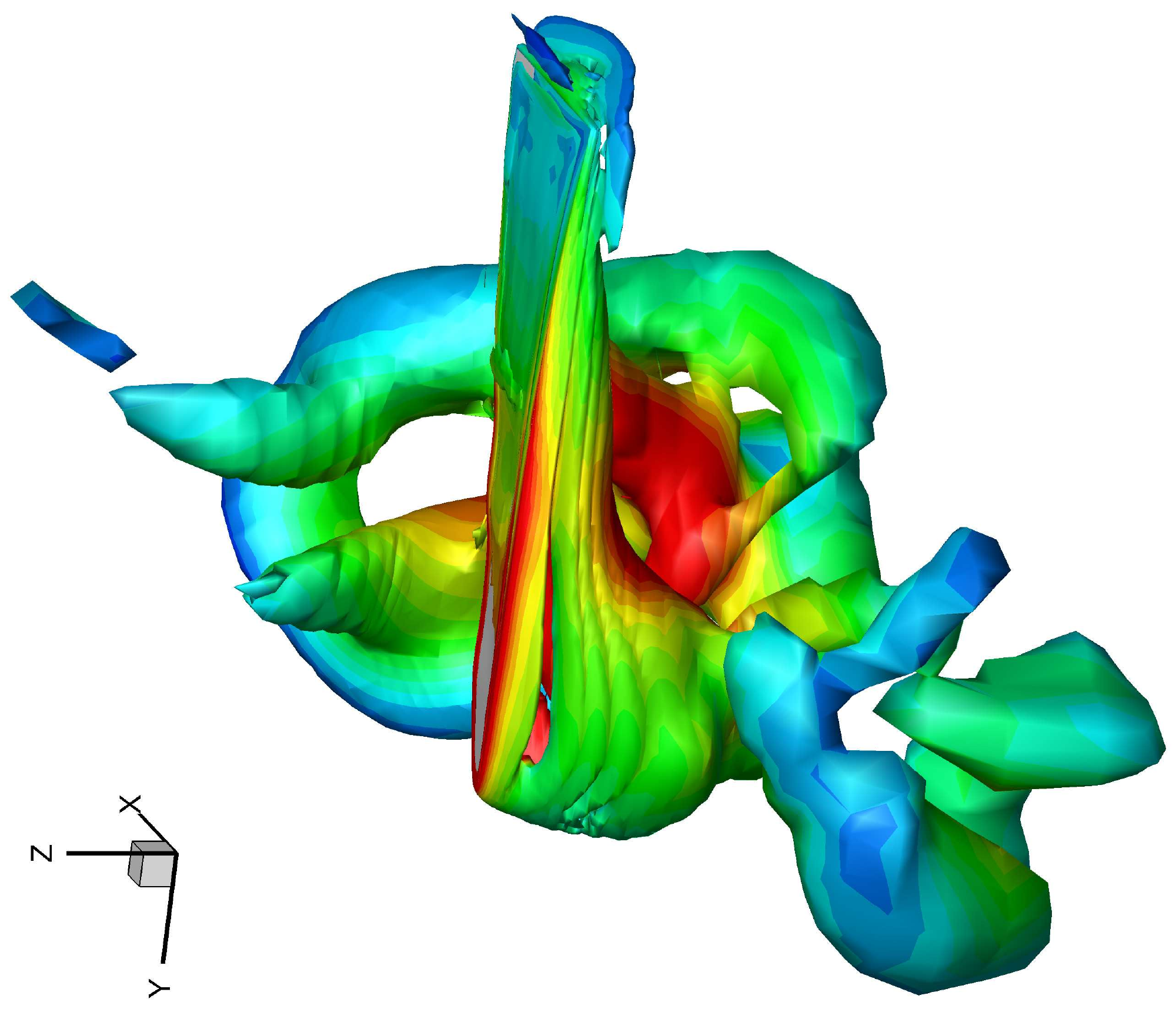}}
	\quad
	\subfloat[][$t/T=0.25$]{\includegraphics[width=0.45\textwidth,angle=-90]{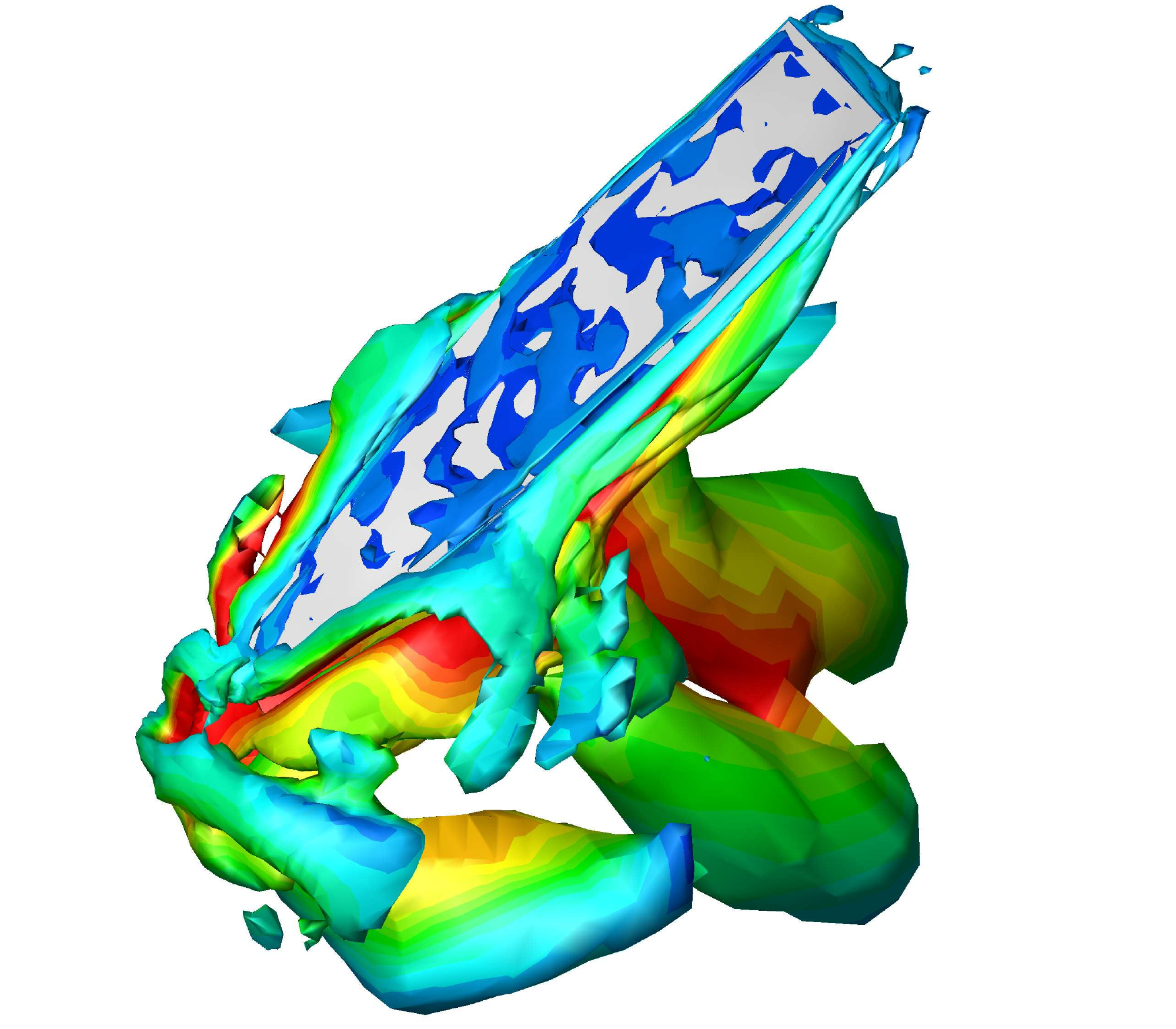}} \\
	\subfloat[][$t/T=0.5$]{\includegraphics[width=0.45\textwidth,angle=-90]{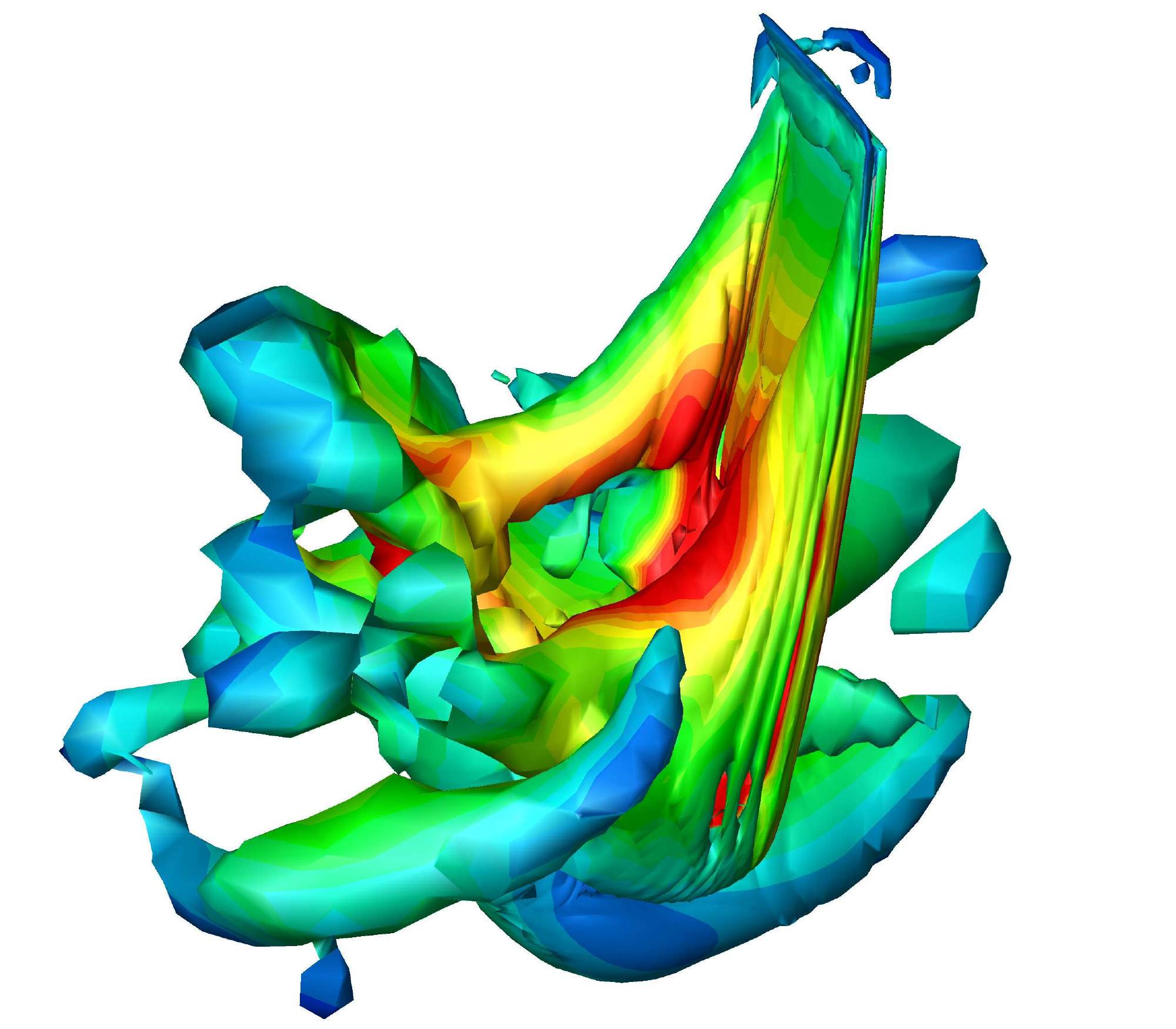}}
	\quad
	\subfloat[][$t/T=0.75$]{\includegraphics[width=0.45\textwidth,angle=-90]{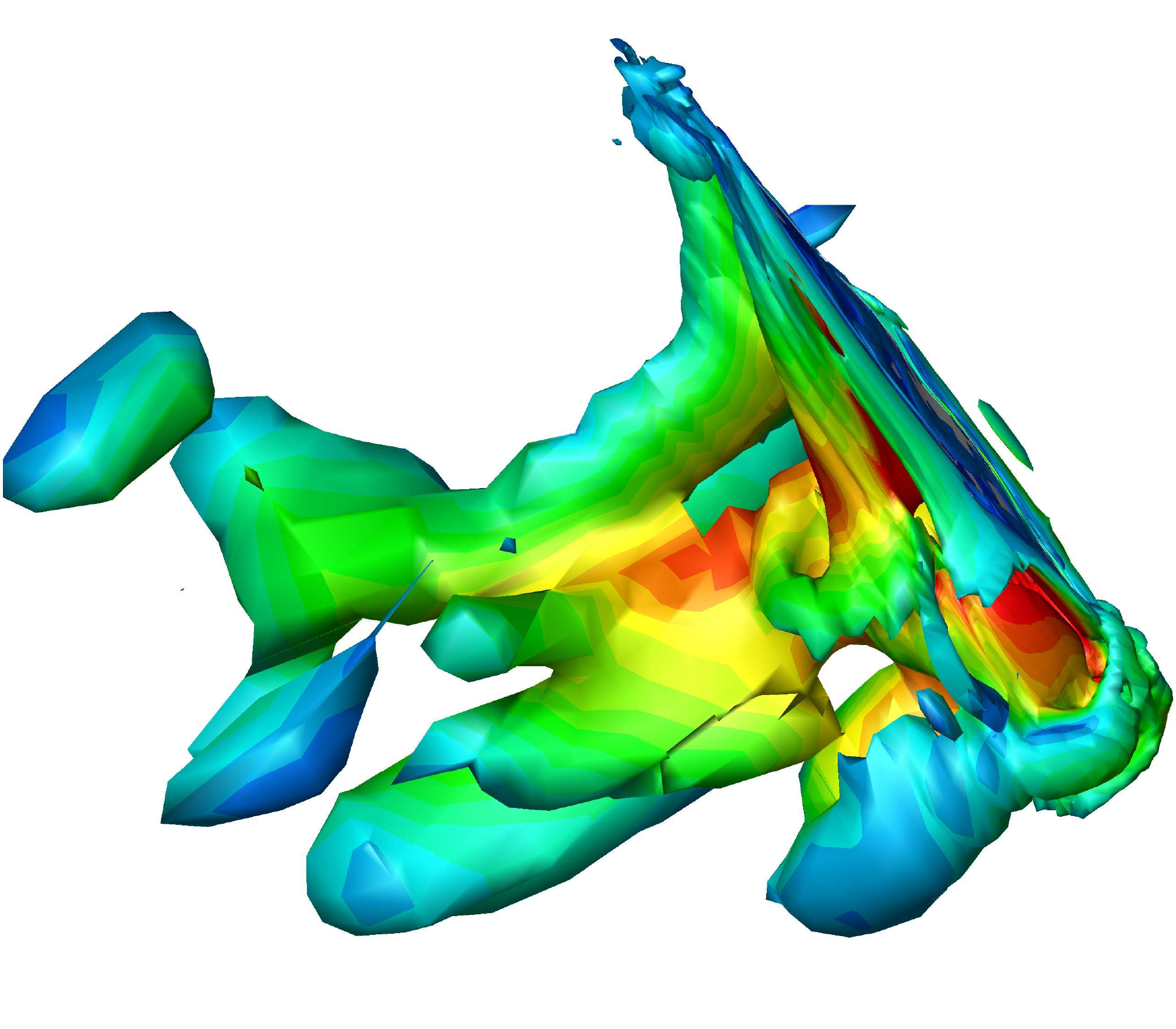}}
	\\
	\includegraphics[width=0.65\textwidth]{./vorleng.pdf}
	\caption{Flow past an anisotropic flapping wing: Wake structures of L3B1 model based on the instantaneous iso-surfaces of $Q(=-\frac{1}{2} \frac{\partial u^f_i}{\partial x_j} \frac{\partial u^f_j}{\partial x_i})$ value at (a) $t/T=0$, (b) $t/T=0.25$, (c) $t/T=0.5$, (d) $t/T=0.75$. Iso-surfaces of non-dimensional $Q^+ \equiv Q(C/U_{ref})^2=1$ are colored by the normalized velocity magnitude $\left \| U \right \| /U_{ref}$.}
	\label{Qaniwake}
\end{figure}

For the purpose of validation, the comparisons of normalized location in vertical direction and normalized displacement at the wing tip are given in Fig. \ref{aniair}, respectively. Results indicate that the overall trends of the flapping wing are well predicted, compared with the experimental data. Fig. \ref{Qaniwake} graphically shows the iso-surface of the non-dimensional Q-criterion of 1 for L3B1 model colored by the normalized velocity magnitude during a flapping motion period. Detailed turbulent wake structures and their evolution process can be observed.  
Overall, the proposed fluid-flexible multibody solver is able to simulate flexible flapping wing with multibody components. The detailed aerodynamic characteristics around the flapping wing and nonlinear structural responses are captured accurately, and compared well with the available experimental data. 

\begin{table}
	\centering
	\caption{Aeroelastic parameters for an anisotropic wing under hovering motion}
	\begin{tabular}{ccc}
		\toprule  
		Parameters & Value\\
		\midrule  
		Reference velocity (hovering) & 4.58 m/s\\
		Air density & 1.206 ${\rm{kg/m^3}}$\\
		Reynolds number& 7304\\
		Flapping frequency& 25  Hz\\
		Flapping amplitude& $35^{\circ}$\\
		\bottomrule 
	\end{tabular}
	\label{paraani}
\end{table}

\section{Application to bat-like flapping dynamics}
As mentioned earlier, the flight patterns and the mechanism of bats are quite different from birds and insects. The lack of understanding about how bats fly efficiently prompts researchers to investigate via experiments and CFD simulations. However, the complex flight kinematics, the special wing structures comprising skeletons with multi-degree of freedom and highly flexible wing-membrane skin as well as the significant aeroelastic phenomena during flight prevent the research progresses for bat flight. The presently available results about the bat flapping dynamics are quite limited, especially from fully-coupled computational modeling. For the first time, we demonstrate the capability of our fully-coupled multibody aeroelastic framework to simulate a bat-like flexible wing. The effects of flexibility and aerodynamic load are investigated to explore their impact on the dynamics of the wing.



\subsection{Structural model setup}
\begin{figure}
	\centering
	\subfloat[][]{\includegraphics[width=0.5\textwidth]{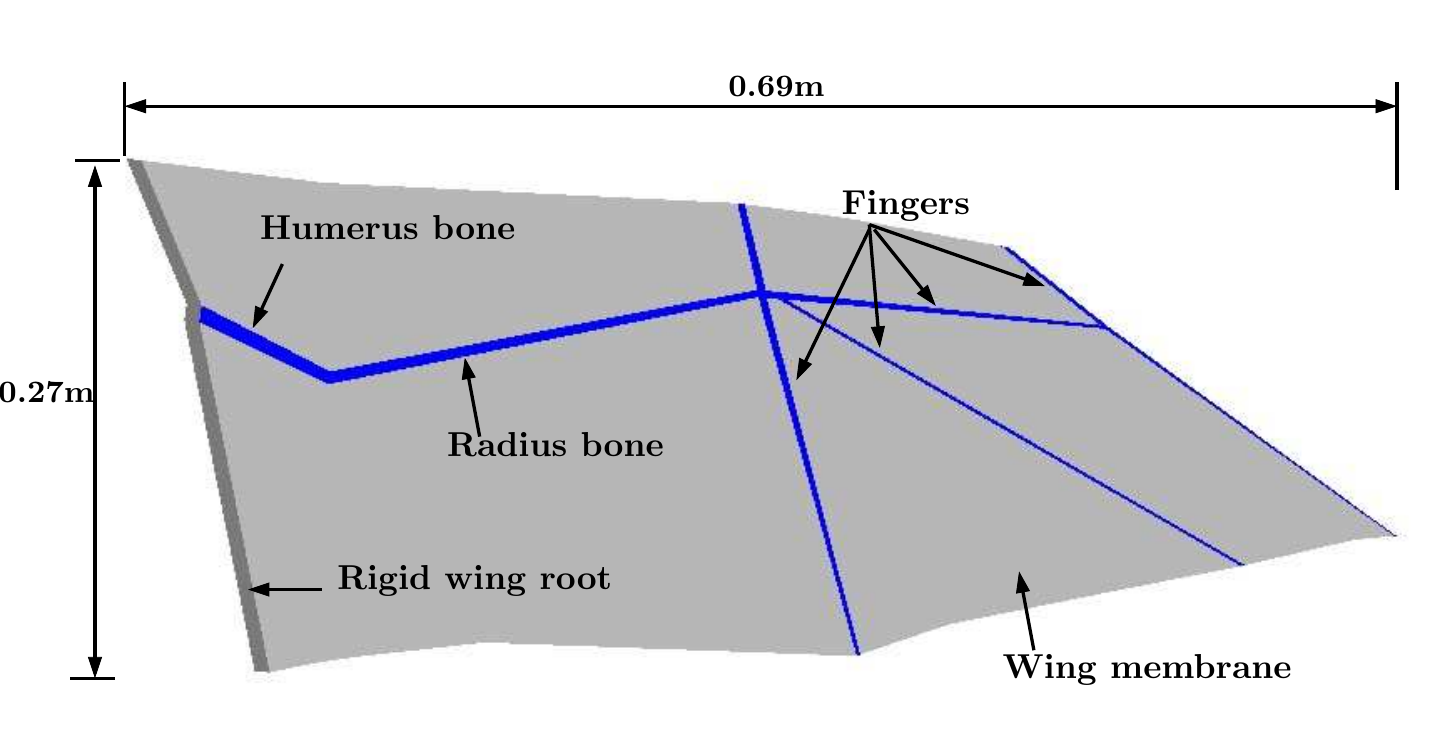} \label{batwinga}}
	\subfloat[][]{\includegraphics[width=0.5\textwidth]{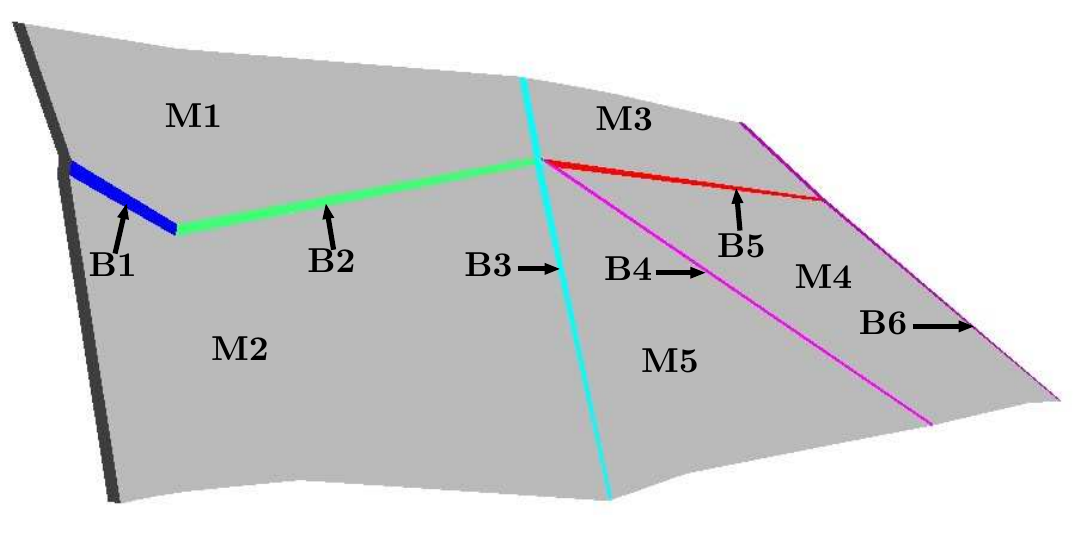} \label{batwingb}}
	\caption{Schematic of bat-like wing: (a) geometric information, (b) indication of different components.}
	\label{batwing}
\end{figure}

A bat-like wing made of several wing skin-like membranes and reinforced skeletons is constructed to investigate the aeroelastic responses with different flight patterns. We consider the bat-like wing but ignore its body and other components. A schematic of the constructed bat-like wing is shown in Fig. \ref{batwinga}. The basic, simplified shape of the bat-like wing is referred to a flapping bat robot made in Brown University \cite{bahlman2013design}. In our simulation, the chord at the wing root is 0.27 m and the span is 0.69 m, resulting in a wing area of 0.124 ${\rm{m^2}}$. 

A rigid strip is applied at the wing root to fix the wing like the bat body. Two strips near to the wing root represent the humerus bone and radius bone, respectively. Several strips in the outer region stand for bat fingers. All these components are used to support the covered wing membranes. The detailed geometric sizes of these bones and fingers are designed to approach those of a real bat, based on the research work of P. Watts \cite{watts2001computational}. The widths and thicknesses of these bones and fingers are varied along their axial direction in order to adjust and achieve reasonable stiffness distributions to support the wing membranes. Meanwhile, the thickness of the wing membrane becomes thinner in the outer region than that near to the wing root, which allows a relative larger elastic deformation in the outer region. The indication of six bones and five membranes in this wing is provided in Fig. \ref{batwingb}. All the components in this bat-like wing are assumed to be made of homogeneous, isotropic material. Material properties of the skeletons and wing membranes in different wing regions are summarized in Table \ref{parabat}.

\begin{table}
	\centering
	\caption{Material properties of the skeletons and wing membrane for the bat-like wing ($\rightarrow$ represents the variation of thickness along axial direction)}
	\begin{tabular}{ccccccc}
		\toprule  
		\multirow{2}*{Material properties} & \multicolumn{6}{c}{Components}\\
		\cline{2-7} 
		& M1/M2/M5 & M3/M4  & B1/B2 & B3/B4 & B5 & B6\\
		\midrule  
		Young's modulus (MPa) & 4 & 3  & 5000 & 3000 & 2500 & 2000 \\
		Poisson ratio   &  0.3 & 0.3 & 0.3 & 0.3 & 0.3 & 0.3\\
		Density (${\rm{kg/m^3}}$)  & 1100 & 1100 & 2200   & 2200  & 2200 & 2200 \\
		Thickness (mm)  & 0.8/0.6/0.4 & 0.2 & 5/4   & 2$\rightarrow$1.5  & 3$\rightarrow$2 & 2 \\
		\bottomrule 
	\end{tabular}
	\label{parabat}
\end{table}

\subsection{Effect of flexibility}
\label{flex}
\begin{figure}
	\centering
	\subfloat[][]{\includegraphics[width=0.85\textwidth]{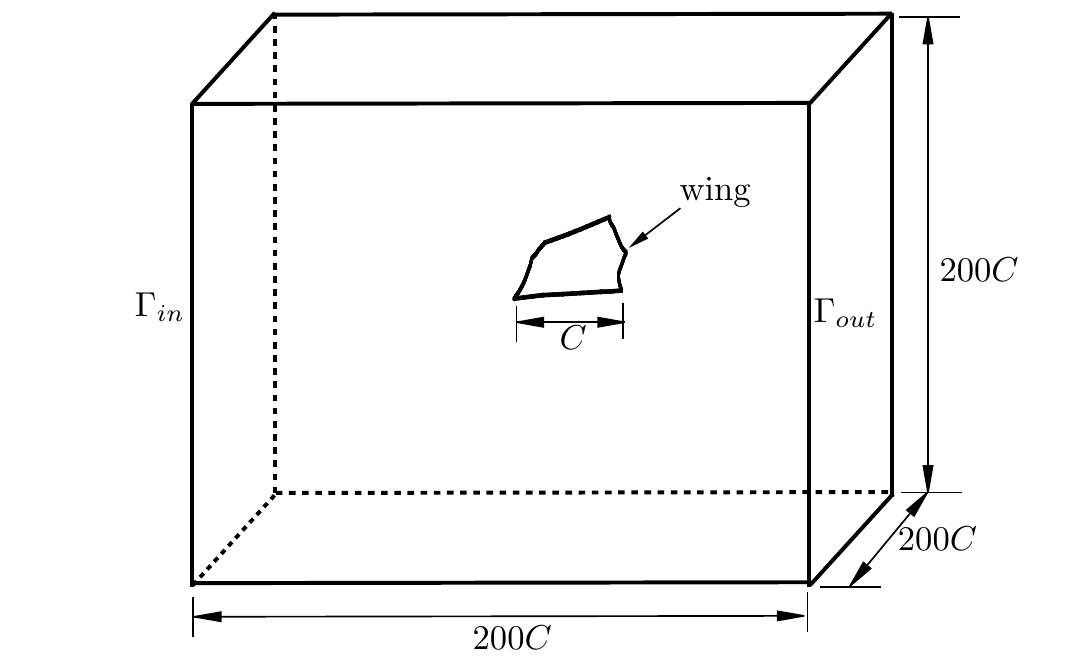} \label{batpitchwinga}}
	\quad
	\subfloat[][]{\includegraphics[width=0.85\textwidth]{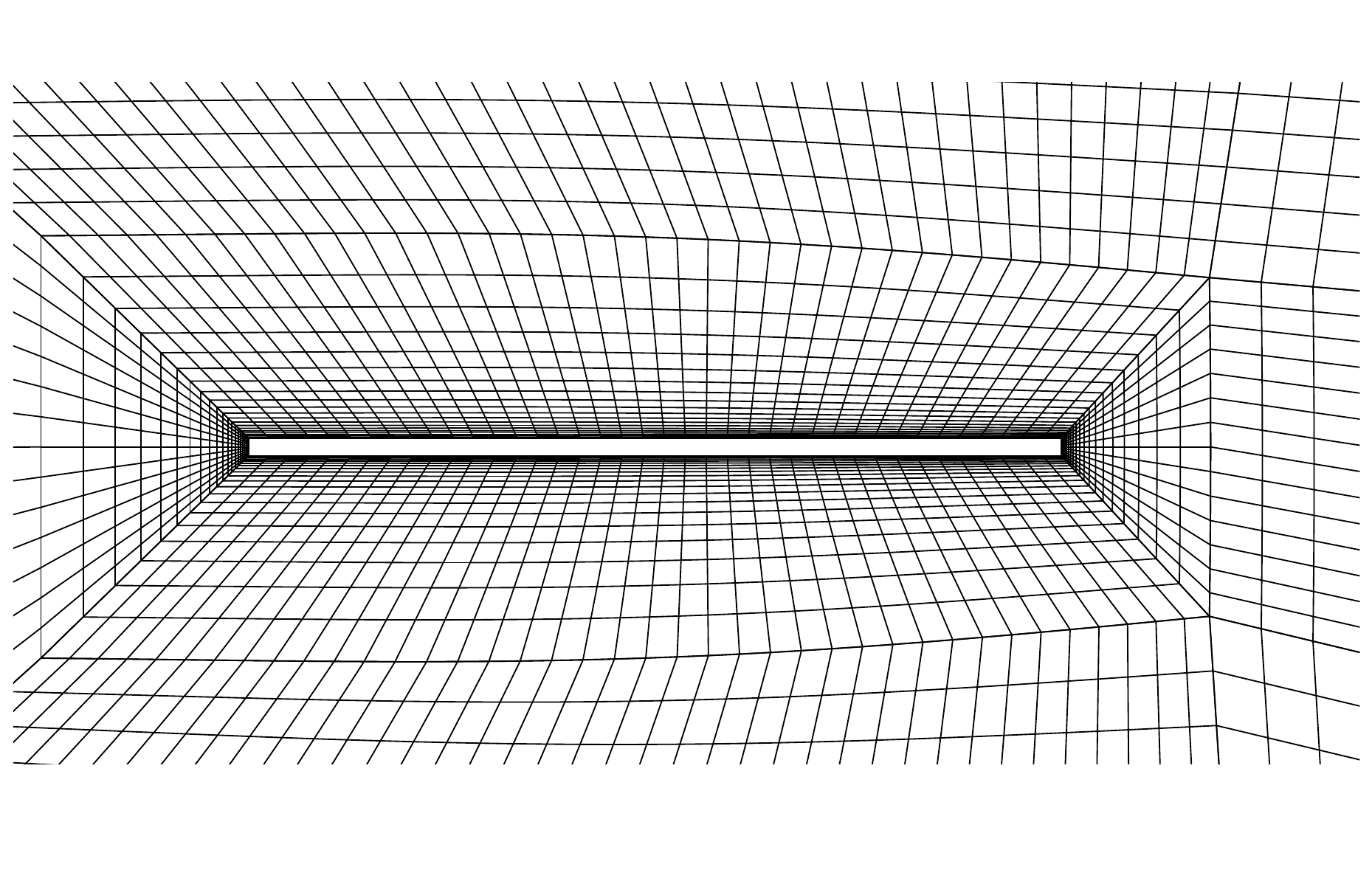} \label{batpitchwingb}} 
	\caption{Flow past a bat-like flapping wing: (a) schematic of computational setup, (b) mesh distribution slice in the fluid domain at mid-span.}
	\label{batpitchwing}
\end{figure}

To investigate the effect of flexibility on the dynamics of a bat-like flexible wing, a rigid wing with identical geometry is simulated and compared to its flexible counterpart. To eliminate the effect of flapping motion, the bat-like wing in a gliding flight is considered. The gliding flight is simulated by applying a uniform flow to the wing while constraining its root in all six degree-of-freedom. Structural deformation due to aerodynamic load is allowed for the flexible wing, but not allowed for the rigid wing. The detailed parameters for the simulation are summarized in Table \ref{parabatfsi}. A schematic diagram of the bat-like wing setup is depicted in Fig. \ref{batpitchwinga}. The length of chord at the wing root is defined as $C$. The distances between the inlet boundary $(\Gamma_{in})$ and the outlet boundary $(\Gamma_{out})$, the top and bottom boundaries $(\Gamma_{slip})$ and the boundaries on both sides are the same of $200C$. The no-slip Dirichlet boundary condition is applied on the surface of bat-like wing. The freestream velocity along the X-axis at the inlet boundary $\Gamma_{in}$ is defined as $u^f=U$, where $u^f$ is the X-component of the flow velocity $\overline{\textbf{u}}^f=(u^f,v^f,w^f)$. The slip boundary condition is applied on the top and bottom as well as the side boundaries. A traction-free boundary condition is defined at the outlet boundary $\Gamma_{out}$, where $\sigma_{xx}=\sigma_{yx}=\sigma_{zx}=0$.  

\begin{figure}
	\centering
	\subfloat[][]{\includegraphics[width=0.45\textwidth]{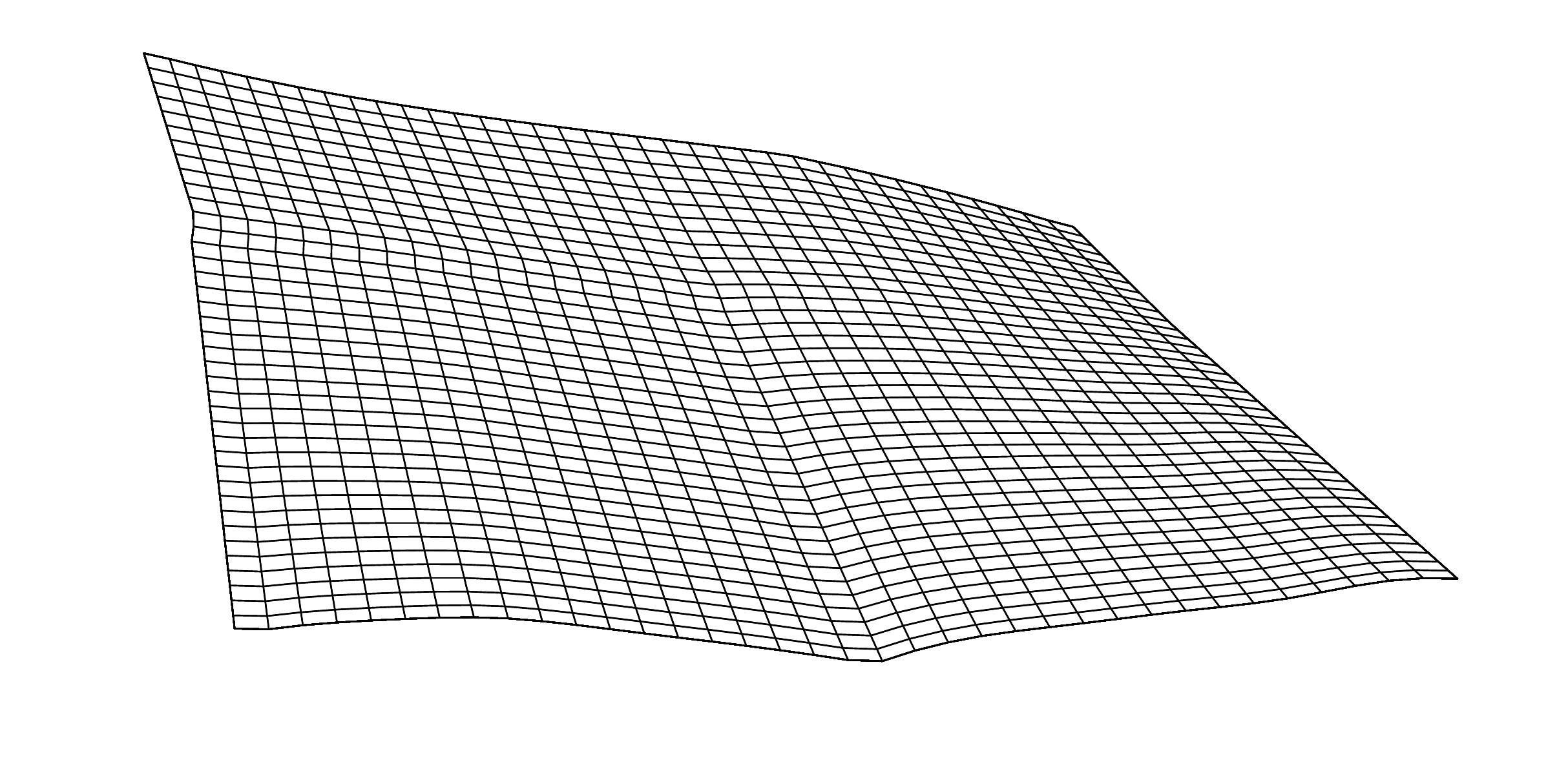} \label{batwingmesha}}
	\quad
	\subfloat[][]{\includegraphics[width=0.45\textwidth]{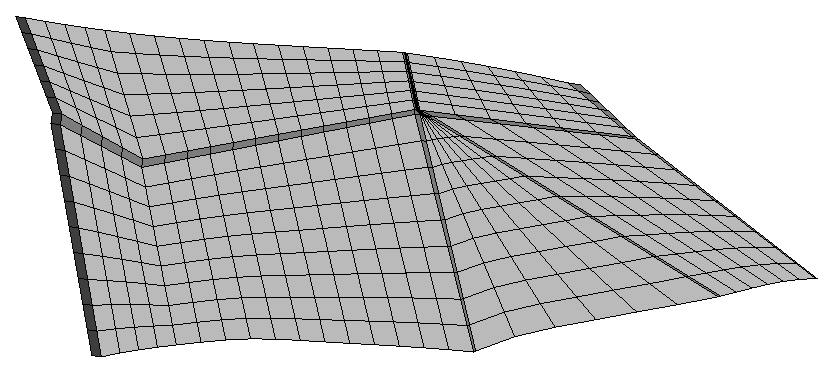} \label{batwingmeshb}}
	\caption{Mesh characteristics on bat-like wing surface in (a) the fluid domain, (b) the structural domain.}
	\label{batwingmesh}
\end{figure}

The three-dimensional fluid computational domain is discretized by 454,258 unstructured eight-node brick finite element meshes. A boundary layer is maintained around the wing such that $y^+ < 1.0$ in the wall-normal direction. Mesh distribution slice in  the fluid domain at mid-span is shown in Fig. \ref{batpitchwingb}. Geometrically exact co-rotational shell elements are employed for wing membrane and bones. The structural model is discretized by 496 structured four-node rectangular finite elements. Mesh characteristics on the wing surface in both the fluid domain and the structural domain are compared and given in Fig. \ref{batwingmesha} and \ref{batwingmeshb}, respectively. The non-dimensional time step size is selected as $\Delta t U/C=0.022$.

\begin{figure}
	\centering
	\subfloat[][]{\includegraphics[width=0.45\textwidth,angle=-90]{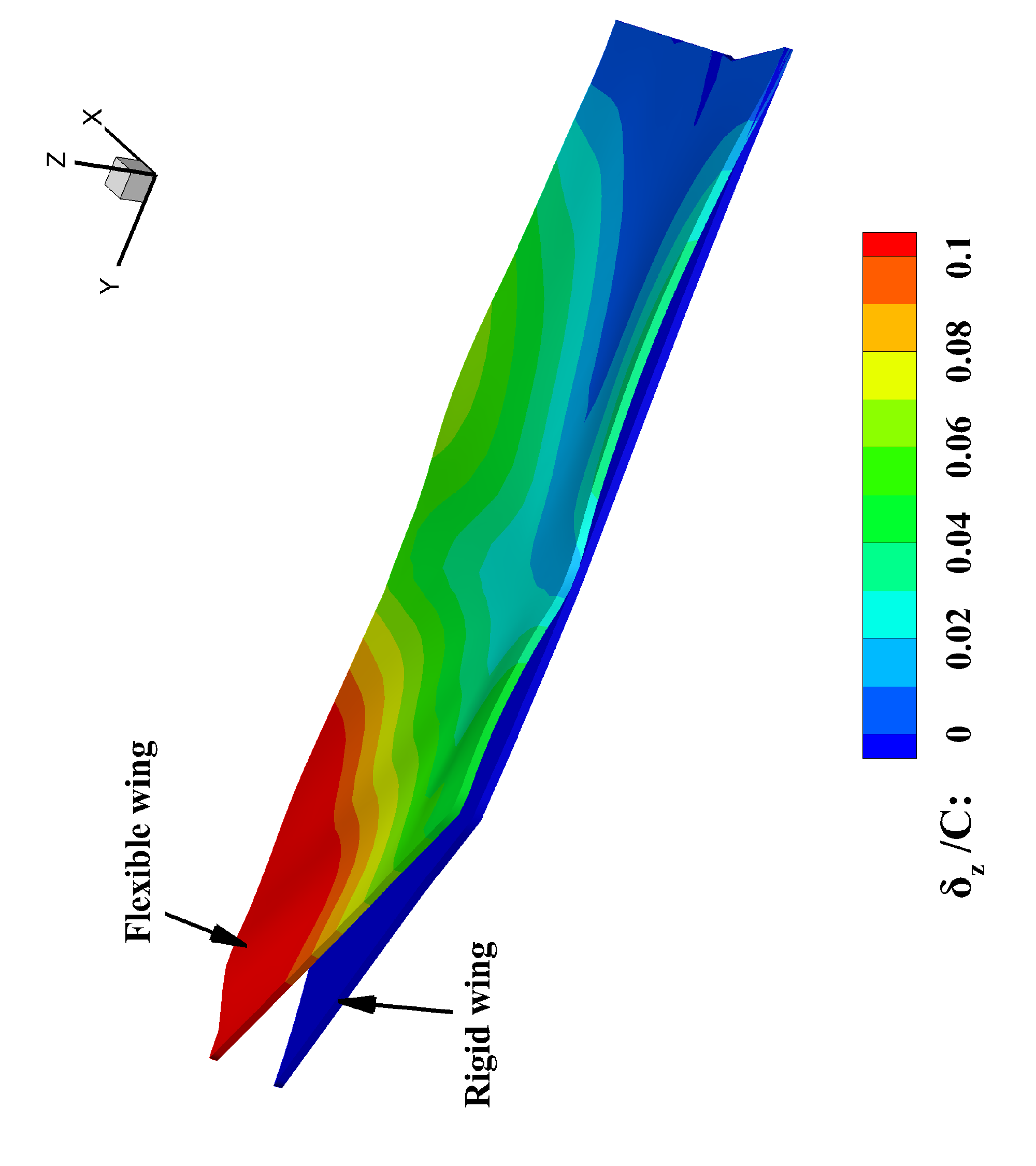} \label{Qbatwakefixa}}
	\subfloat[][]{\includegraphics[width=0.45\textwidth,angle=-90]{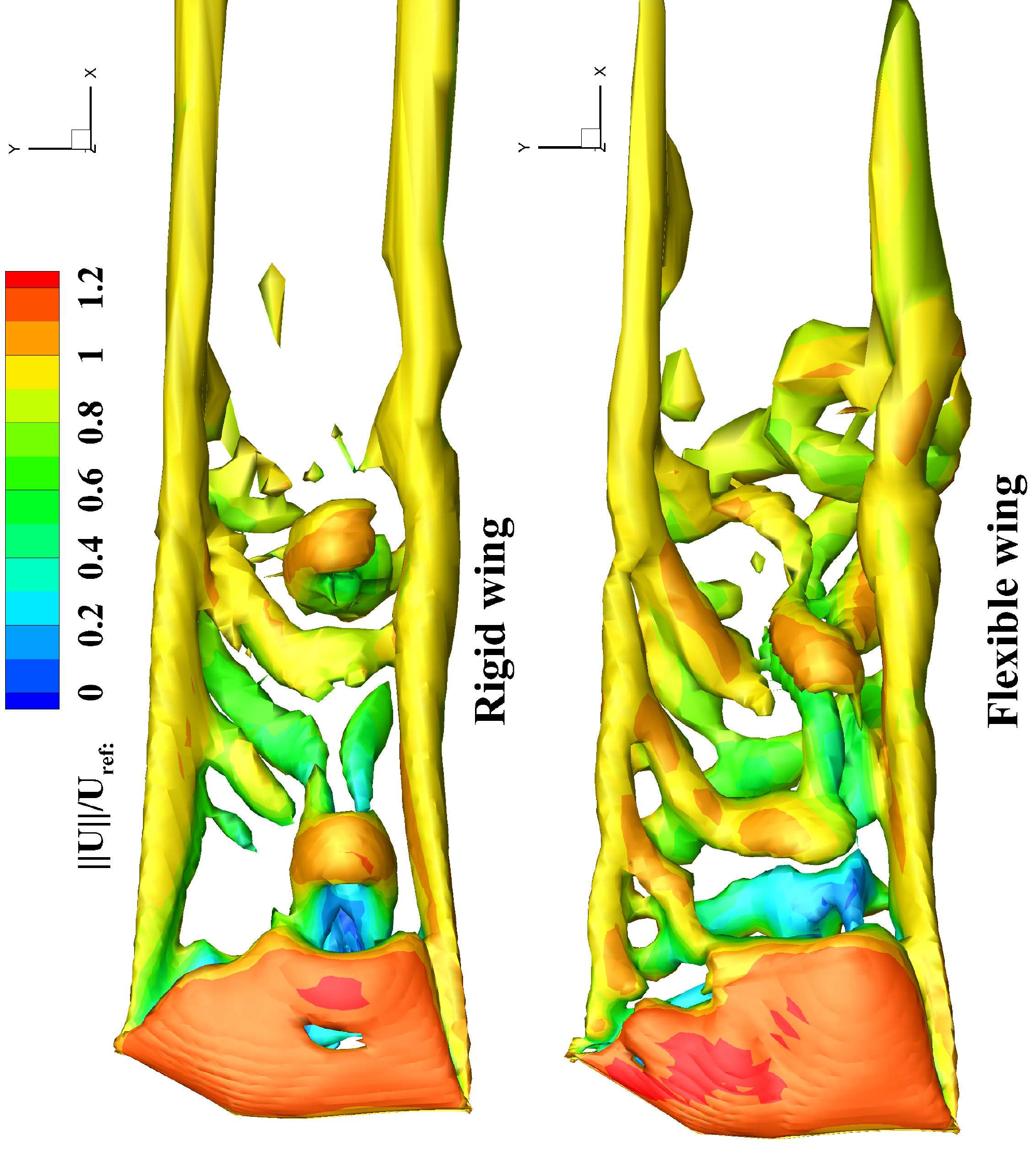} \label{Qbatwakefixb}}
	\caption{(a) Comparison of non-dimensional displacement contours in the vertical direction for both rigid and flexible wings. (b) Wake structures of bat-like wing based on the instantaneous iso-surfaces of $Q(=-\frac{1}{2} \frac{\partial u^f_i}{\partial x_j} \frac{\partial u^f_j}{\partial x_i})$ value for both rigid and flexible wings. Iso-surfaces of non-dimensional $Q^+ \equiv Q(C/U_{ref})^2=0.25$ are colored by the normalized velocity magnitude $\left \| U \right \| /U_{ref}$.}
	\label{Qbatwakefix}
\end{figure}

Comparison of the deformed flexible wing and rigid wing colored by non-dimensional displacement in the vertical direction is shown in Fig. \ref{Qbatwakefixa}. The maximum displacement is observed at the wing tip and the covered membranes have some small wrinkles on the surface. Comparison of the mean lift coefficient ($\bar{C}_L$) and the mean drag coefficient ($\bar{C}_D$) for the rigid and flexible wing is presented in Table \ref{batforce}. $\bar{C}_L$ decreases from 0.6234 to 0.5746 and $\bar{C}_D$ reduces by 13.17$\%$ due to the passive deformation of flexible wing and its compliant membrane components. As a result, the lift-to-drag ratio of flexible wing ($\bar{C}_L$/$\bar{C}_D$) increases by 6.13$\%$, compared with the rigid wing. 

\begin{figure}
	\centering
	\subfloat[][ Rigid wing]{\includegraphics[width=0.4\textwidth,angle=-90]{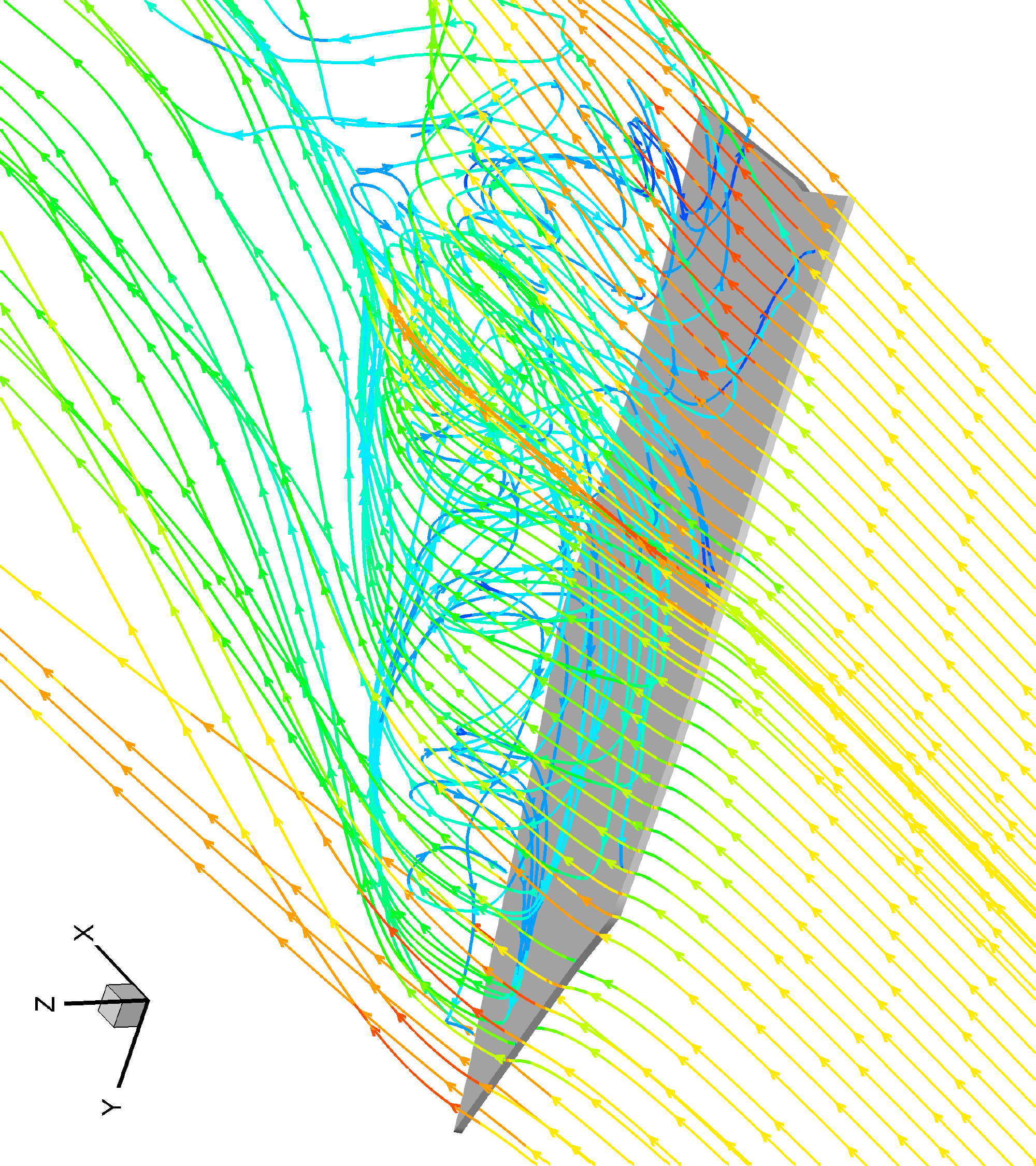}}
	\subfloat[][ Flexible wing]{\includegraphics[width=0.4\textwidth,angle=-90]{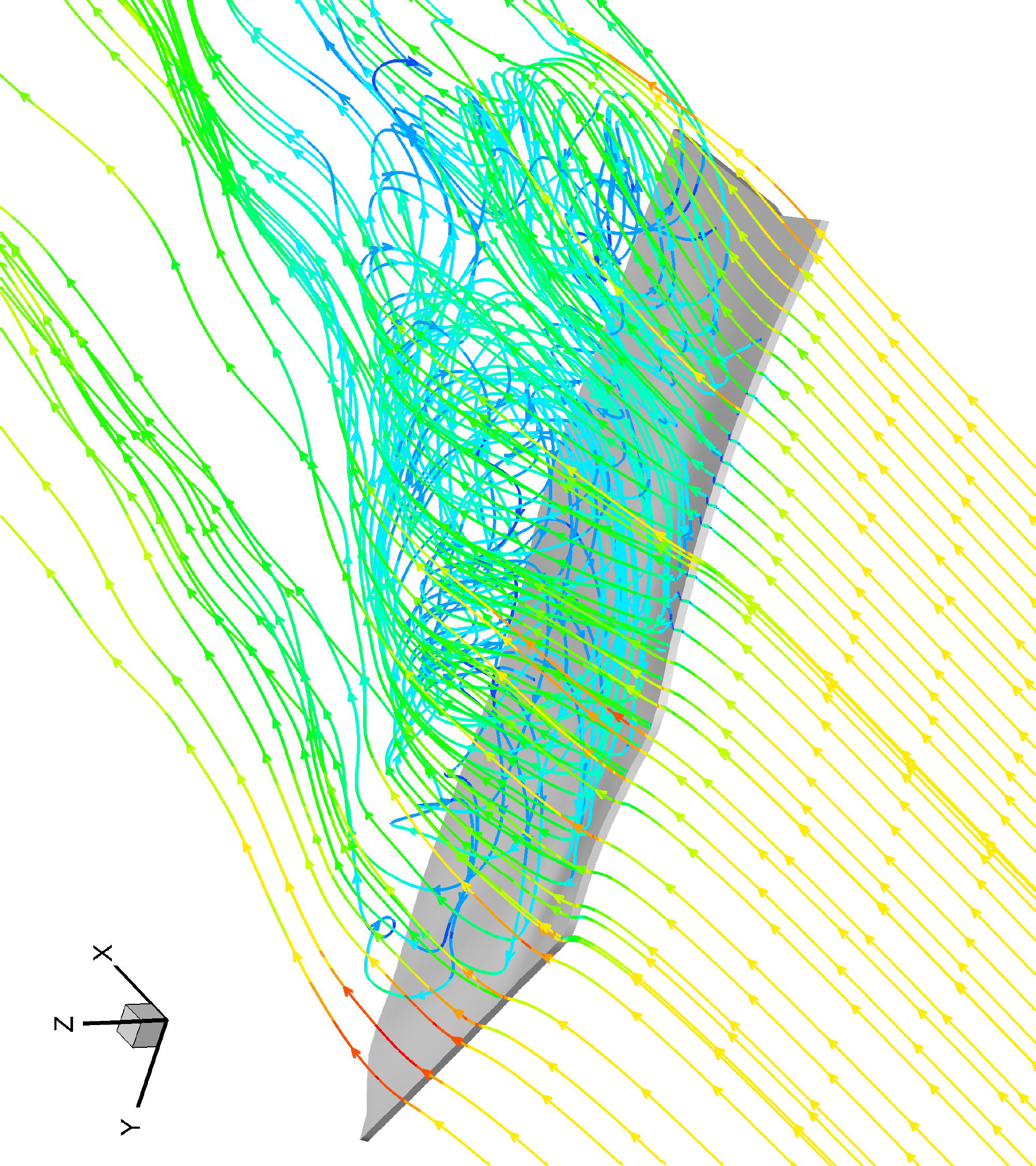}}
	\\
	\includegraphics[width=0.65\textwidth]{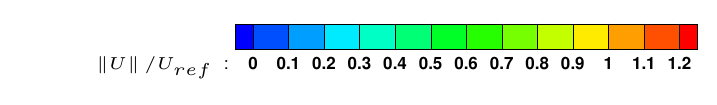}
	\caption{Streamlines colored by the normalized velocity magnitude $\left \| U \right \| /U_{ref}$ around the bat-like wing for (a) rigid wing, (b) flexible wing.}
	\label{pitchshowfix}
\end{figure}

For further investigation, wake structures for both rigid and flexible wings based on instantaneous iso-surfaces of non-dimensional Q-criterion of 0.25 colored by the normalized velocity magnitude are given in Fig. \ref{Qbatwakefixb}. The shedding vortex structures behind the bat-like wing are changed by the wing deformation via aeroelastic coupling. Fig. \ref{pitchshowfix} provides streamlines around the rigid and flexible wings colored by the normalized velocity magnitude. A massive separation flow on the upper surface is observed and the flow patterns are altered by the wing deformation, compared with the rigid wing counterpart. In summary, the flexibility of the wing significantly affects the dynamics of a bat-like wing. Further investigation can be conducted by simulating a range of flexibilities to quantitatively investigate its effect on the aerodynamic performance of the wing.

\begin{table}
	\centering
	\caption{Simulation parameters for a bat-like wing under gliding motion}
	\begin{tabular}{ccc}
		\toprule  
		Parameters & Value\\
		\midrule  
		Semi-span length & 0.69 m\\
		Chord length at wing root& 0.27 m\\
		Wing area & 0.124 ${\rm{m^2}}$\\
		Freestream velocity & 2.0 m/s\\
		Angle of attack& $20^{\circ}$\\
		Air density & 1.225 ${\rm{kg/m^3}}$\\
		Mean chord-based Reynolds number & 24609\\
		\bottomrule 
	\end{tabular}
	\label{parabatfsi}
\end{table}

\begin{table}
	\centering
	\caption{Comparison of mean lift coefficient ($\bar{C}_L$), mean drag coefficient ($\bar{C}_D$) and lift-to-drag ratio ($\bar{C}_L$/$\bar{C}_D$) for rigid and flexible wing}
	\begin{tabular}{cccc}
		\toprule  
		Results & $\bar{C}_L$ & $\bar{C}_D$ & $\bar{C}_L$/$\bar{C}_D$\\
		\midrule  
		Rigid wing & 0.6234 & 0.2391 & 2.61 \\
		Flexible wing & 0.5746 (7.83 $\%$ $\downarrow$) & 0.2076 (13.17$\%$ $\downarrow$) & 2.77 (6.13 $\%$ $\uparrow$) \\
		\bottomrule 
	\end{tabular}
	\label{batforce}
\end{table}

\subsection{Effect of aerodynamic load}
In this section, we investigate the effect of aerodynamic load on the structural responses of the flapping wing. Both uncoupled and coupled aeroelastic cases are simulated herein. The uncoupled case does not consider aerodynamic load in the flapping motion, while the coupled aeroelastic case includes the coupling with surrounding air flow. The angle of attack is assumed as zero and the freestream velocity is set as 1.0 m/s for the flapping flight. The flapping amplitude is given as 20$^\circ$ and the flapping frequency is set as 1 Hz. The identical mesh distributions in Section \ref{flex} for both fluid and structural models are adopted for the flapping wing simulation. The non-dimensional time step size is selected as $\Delta t U/C=0.00926$. 
	
\begin{figure}
	\centering
	\subfloat[][]{\includegraphics[width=0.5\textwidth]{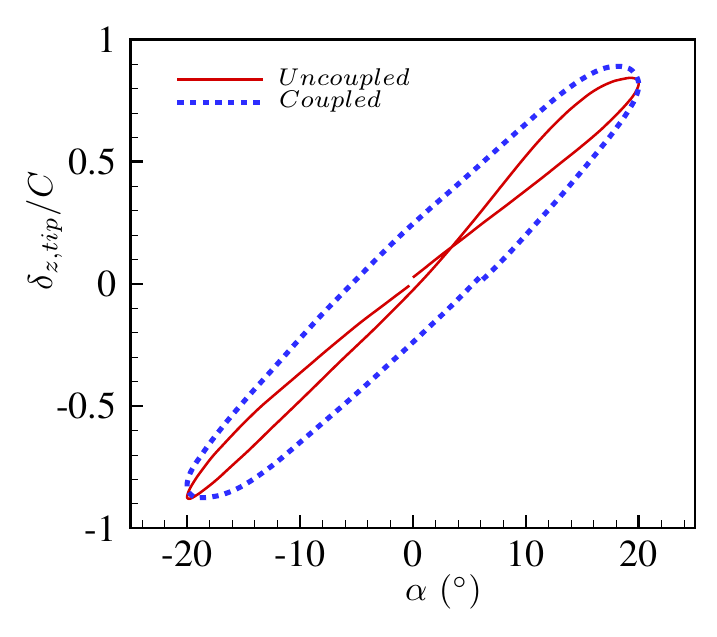}}
	\subfloat[][]{\includegraphics[width=0.5\textwidth]{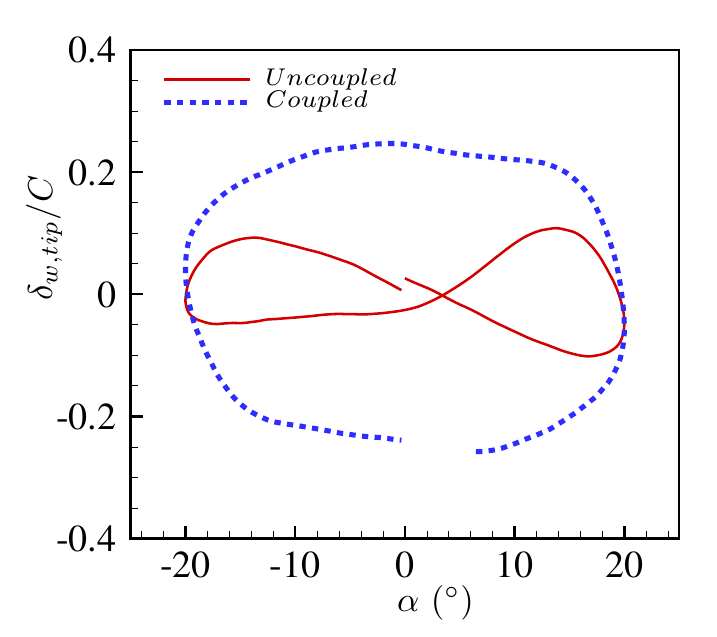}}
	\caption{Comparison of uncoupled and coupled aeroelastic structural responses of the bat-like wing for (a) normalized location in the vertical direction at wing tip, (b) normalized displacement at wing tip.}
	\label{pitchre}
\end{figure}

\begin{figure}
	\centering
	\subfloat[][$t/T=0$]{\includegraphics[width=0.35\textwidth,angle=-90]{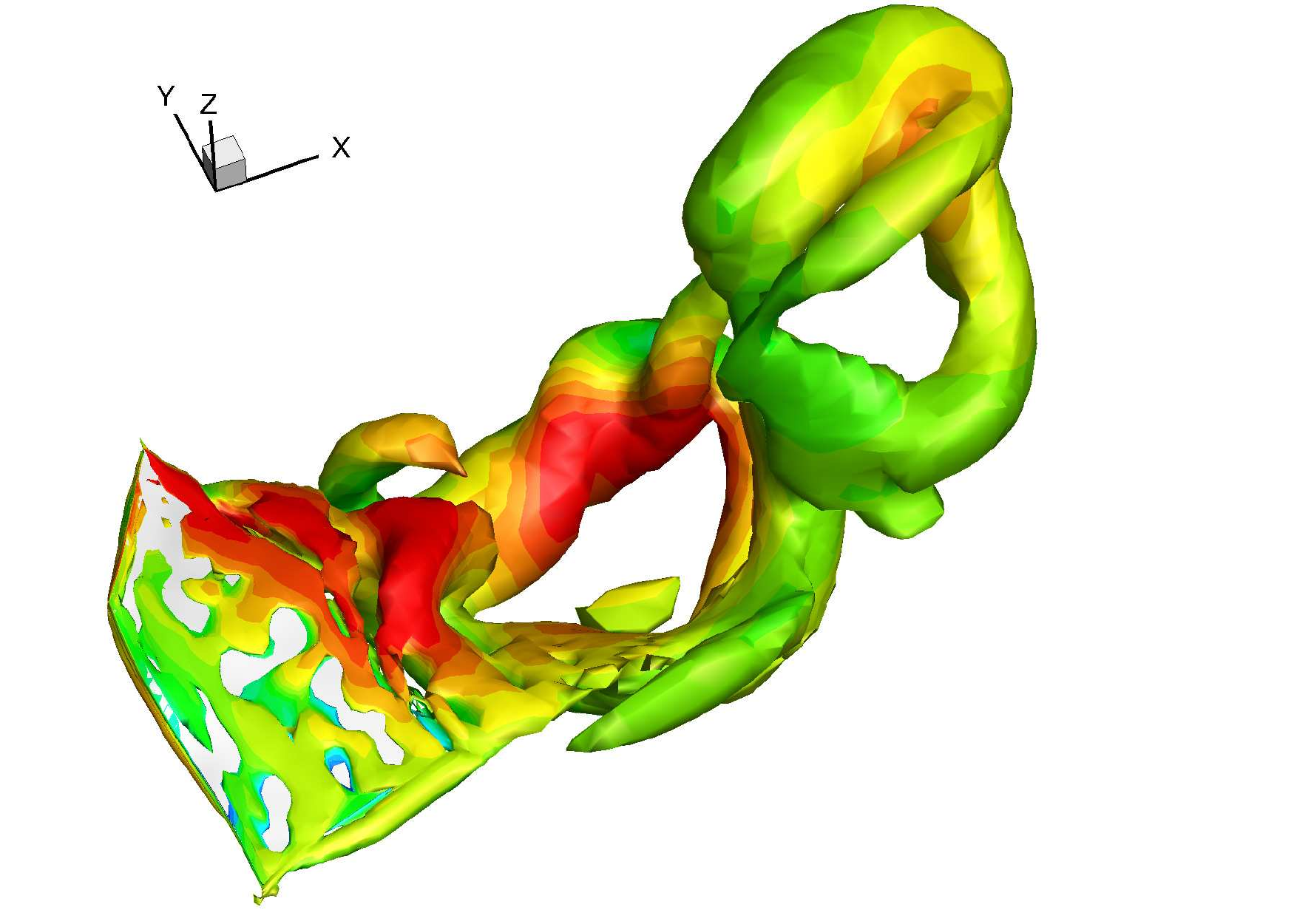}}
	\subfloat[][$t/T=0.25$]{\includegraphics[width=0.35\textwidth,angle=-90]{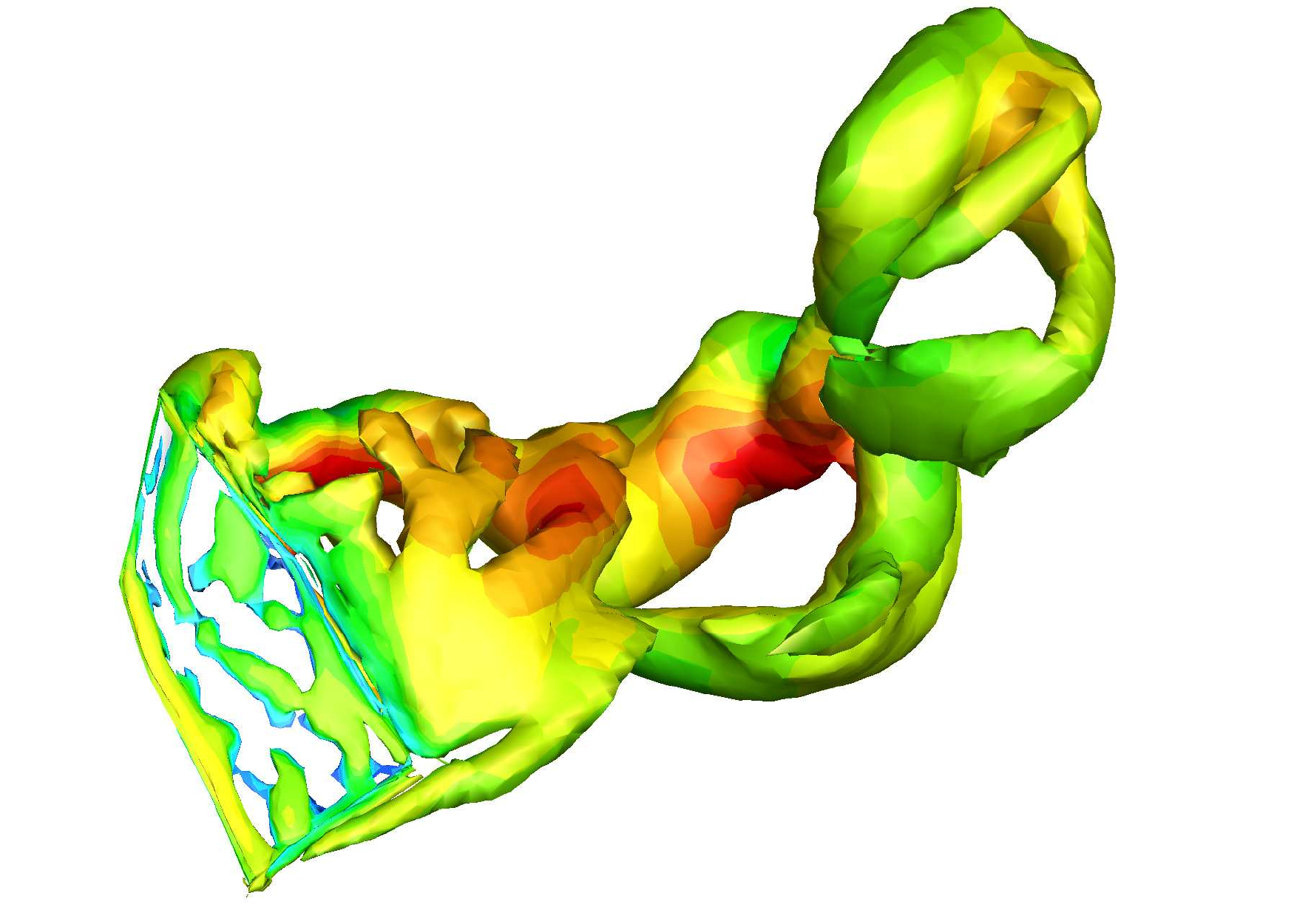}} \\
	\subfloat[][$t/T=0.5$]{\includegraphics[width=0.35\textwidth,angle=-90]{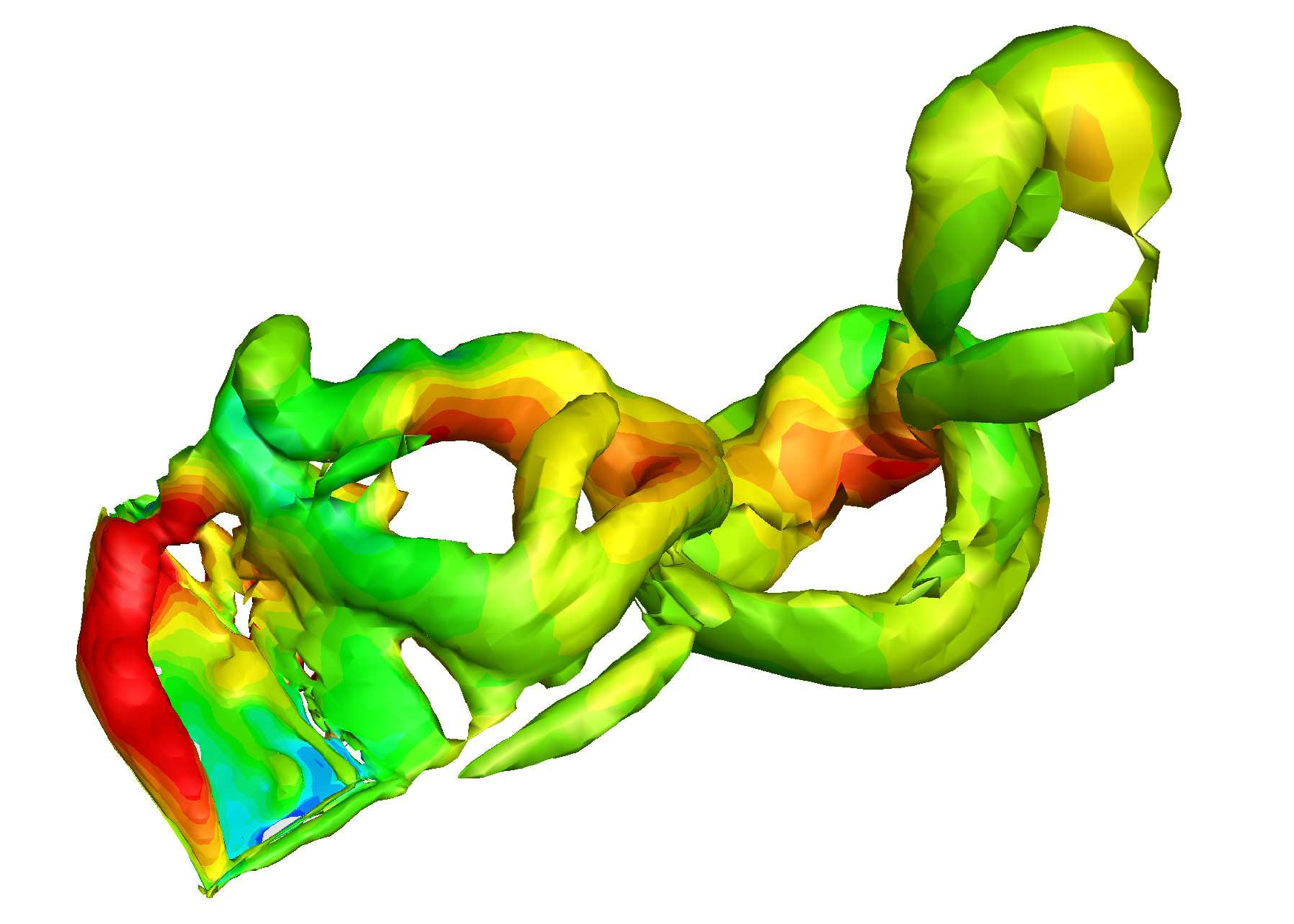}}
	\subfloat[][$t/T=0.75$]{\includegraphics[width=0.35\textwidth,angle=-90]{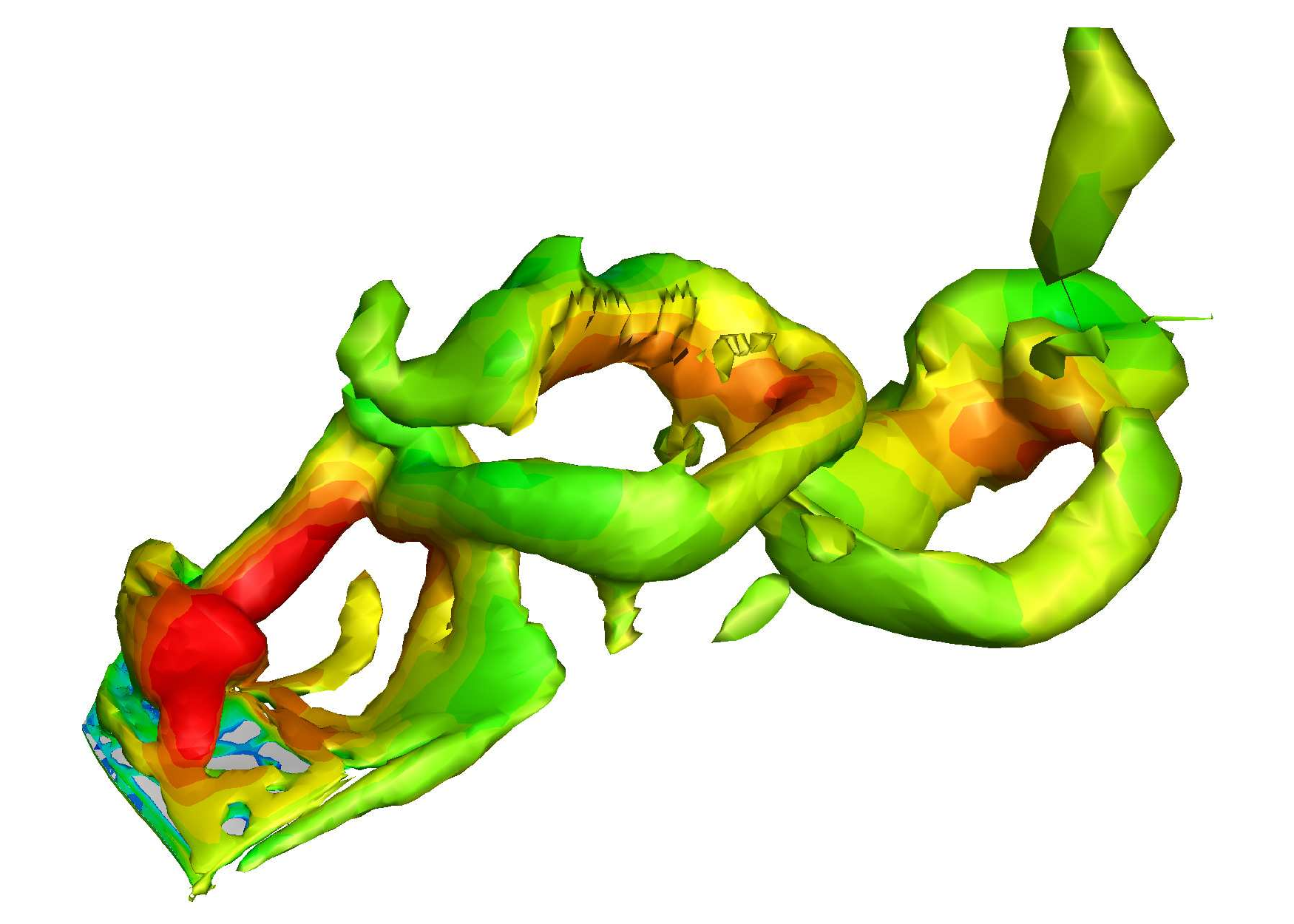}}
	\\
	\includegraphics[width=0.65\textwidth]{./vorleng.pdf}
	\caption{Flow past a bat-like wing: Wake structures of bat-like wing based on the instantaneous iso-surfaces of $Q(=-\frac{1}{2} \frac{\partial u^f_i}{\partial x_j} \frac{\partial u^f_j}{\partial x_i})$ value at (a) $t/T=0$, (b) $t/T=0.25$, (c) $t/T=0.5$, (d) $t/T=0.75$. Iso-surfaces of non-dimensional $Q^+ \equiv Q(C/U_{ref})^2=0.25$ are colored by the normalized velocity magnitude $\left \| U \right \| /U_{ref}$.}
	\label{Qbatwake}
\end{figure}

\begin{figure}
	\centering
	\subfloat[][ $t/T=0.25$]{\includegraphics[width=0.4\textwidth,angle=-90]{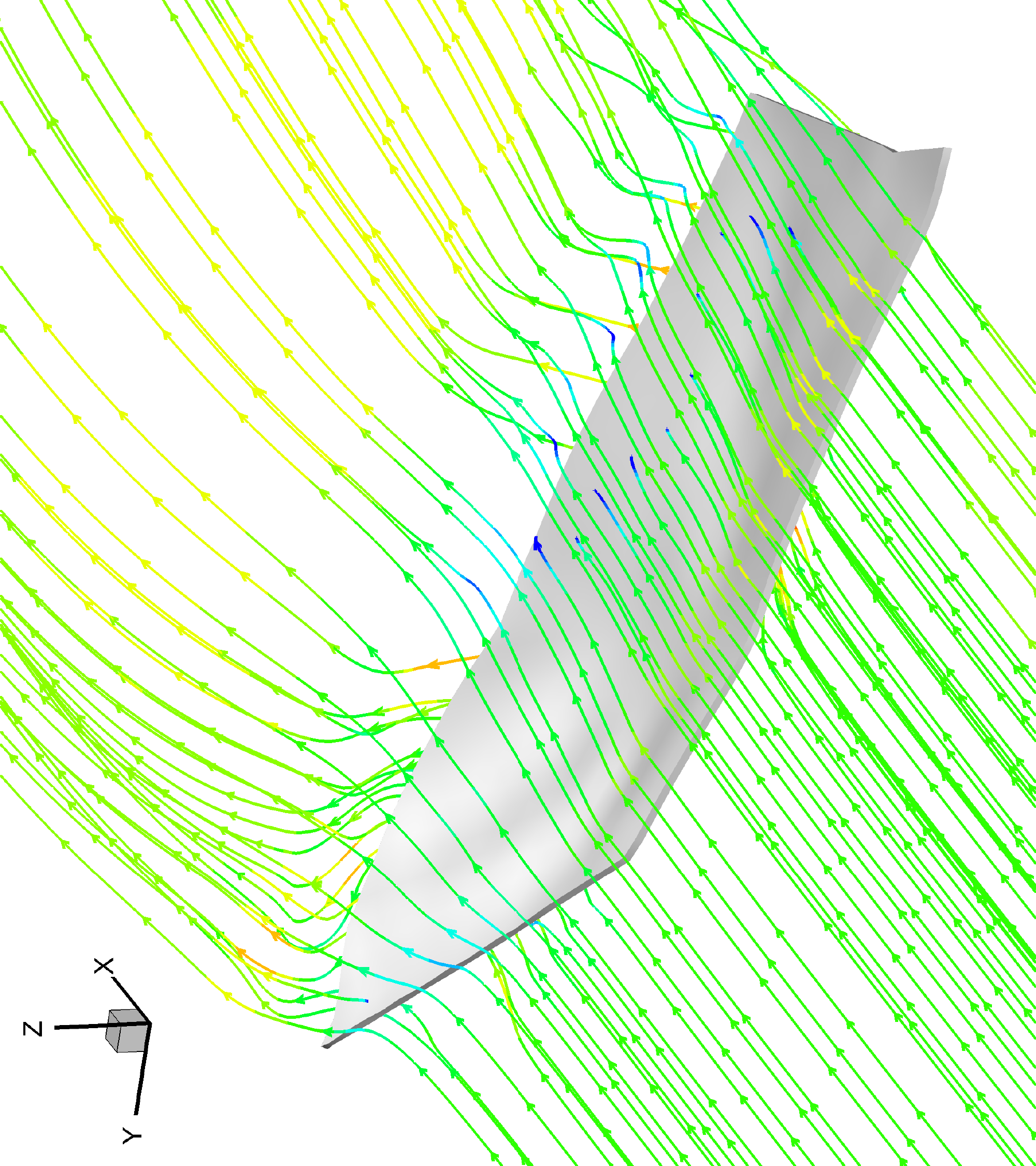}}
	\subfloat[][ $t/T=0.75$]{\includegraphics[width=0.4\textwidth,angle=-90]{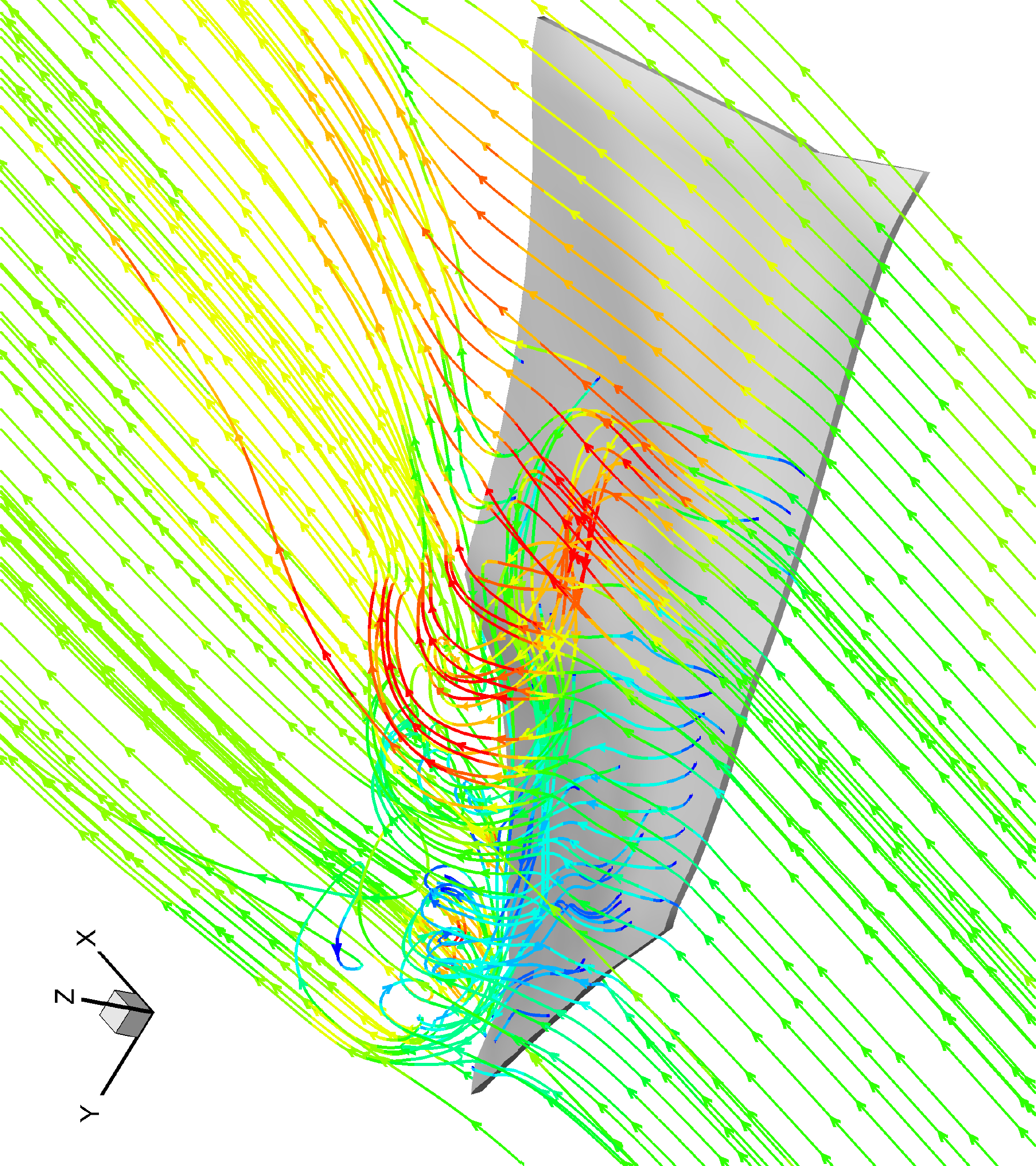}}
	\\
	\includegraphics[width=0.65\textwidth]{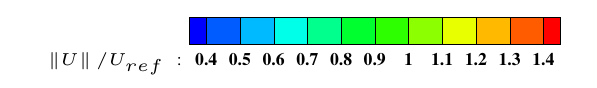}
	\caption{Streamlines colored by the normalized velocity magnitude $\left \| U \right \| /U_{ref}$ around the bat-like wing at (a) $t/T=0.25$, (b) $t/T=0.75$.}
	\label{pitchshow}
\end{figure}

The comparisons of the normalized location and the displacement in the vertical direction at wing tip for both cases are shown in Fig. \ref{pitchre},  respectively. The nonlinear dynamic responses for the flexible bat-like wing have been affected by the aeroelastic coupling effect significantly and the lagging effect of coupled case becomes stronger than the uncoupled one. The iso-surfaces of the non-dimensional Q-criterion of 0.25 for the bat-like wing colored by the normalized velocity magnitude during a whole flapping motion period are given in Fig. \ref{Qbatwake}. The evolution of vortex generated from the leading edge, the trailing edge and the wing tip during flapping flight can be observed. For a detailed analysis, streamlines around this wing at $t/T=0.25$ and $t/T=0.75$ are shown in Fig. \ref{pitchshow}, respectively. A massive separation flow on upper surface near to the wing tip is viewed at the downstroke motion. Such complex, nonlinear aerodynamic phenomena, including leading edge vortex (LEV) generation and trailing edge vortex (TEV) shedding, influence the structural responses via the aeroelastic coupling process. The effect of aerodynamic load on the bat-like flapping wing is quite substantial.

According to the discussion presented above, the proposed multibody aeroelastic solver is able to capture the physics of aeroelastic phenomena of the flexible multibody flapping wing and it can be extended to a more general bat-like wing. Meanwhile, the implementation of RBF method in current framework and combination with NIFC approach have been proved as an effective scheme to simulate flexible flapping wing. Finally, a bat-like wing with different flexibilities for the bones and the membranes and the flapping frequencies should be explored for further mechanism investigation based on the proposed aeroelastic framework.

%
%

\section{Conclusion}
In this paper, a three-dimensional multibody aeroelastic framework is developed by coupling an incompressible turbulent flow solver with DDES method and a flexible multibody structural solver discretized by geometrically exact co-rotational finite elements via partitioned domain decomposition strategy. The nonlinear iterative force correction approach has been implemented and coupled with the RBF interpolation method during the integration of incompressible turbulent flow and a flexible multibody system to achieve a stable and robust partitioned coupling process. The proposed aeroelastic framework provides a feasible and high-fidelity tool to explore the design and optimization of a wide range of flexible flapping wings, such as bionic MAVs and UAVs. We validated the accuracy of our framework by simulating an anisotropic wing made of multiple reinforced battens and membranes. A good agreement to an experiment has been achieved, where the characteristic responses of the flapping wing compared quite well with the experimental data. We further demonstrated the applicability of our framework by investigating the aeroelastic responses of a bat-like wing. The rigid and flexible wings under gliding flight simulations  were carried out and   the effect of flexibility on the dynamics of the bat-like wing was quantified. The lift-to-drag ratio increases by 6.13\% for the flexible wing due to the passive deformation of the wing and its compliant membranes, compared to its rigid counterpart. Subsequently, the effect of aerodynamic load has been studied by simulating a coupled and an uncoupled aeroelastic cases. It is observed that the aerodynamic load enhances the lagging effect of the nonlinear structural responses significantly. In future, various aspects of a bat flight can be investigated via the developed framework, which may include structural optimization, acoustic reduction, flight control, and among others.

\section*{Acknowledgements}
The authors wish to acknowledge supports from the National University of Singapore
and the Ministry of Education, Singapore.

\bibliographystyle{elsarticle-num}
\bibliography{reference}

\end{document}